\documentstyle[12pt]{article}

\catcode`\@=11
\def\marginnote#1{}
\newcount\hour
\newcount\minute
\newtoks\amorpm
\hour=\time\divide\hour by60
\minute=\time{\multiply\hour by60 \global\advance\minute by-\hour}
\edef\standardtime{{\ifnum\hour<12 \global\amorpm={am}%
        \else\global\amorpm={pm}\advance\hour by-12 \fi
        \ifnum\hour=0 \hour=12 \fi
        \number\hour:\ifnum\minute<10 0\fi\number\minute\the\amorpm}}
\edef\militarytime{\number\hour:\ifnum\minute<10 0\fi\number\minute}

\def\draftlabel#1{{\@bsphack\if@filesw {\let\thepage\relax
   \xdef\@gtempa{\write\@auxout{\string
      \newlabel{#1}{{\@currentlabel}{\thepage}}}}}\@gtempa
   \if@nobreak \ifvmode\nobreak\fi\fi\fi\@esphack}
        \gdef\@eqnlabel{#1}}
\def\@eqnlabel{}
\def\@vacuum{}
\def\draftmarginnote#1{\marginpar{\raggedright\scriptsize\tt#1}}
\def\draft{\oddsidemargin -.5truein
        \def\@oddfoot{\sl preliminary draft \hfil
        \rm\thepage\hfil\sl\today\quad\militarytime}
        \let\@evenfoot\@oddfoot \overfullrule 3pt
        \let\label=\draftlabel
        \let\marginnote=\draftmarginnote
   \def\@eqnnum{(\theequation)\rlap{\kern\marginparsep\tt\@eqnlabel}%
\global\let\@eqnlabel\@vacuum}  }

\def\numberbysection{\@addtoreset{equation}{section}
        \def\theequation{\thesection.\arabic{equation}}}

\def\underline#1{\relax\ifmmode\@@underline#1\else
        $\@@underline{\hbox{#1}}$\relax\fi}

\catcode`@=12
\relax

\numberbysection
\def\souligne#1{\underline{#1}}
\def\sigmab{\bar \sigma}
\def\cJ{{\cal J}}
\def\cJb{{\bar {\cal J}}}
\def\cC{{\cal C}^n}
\def\CP{{\cal CP}^n}
\def\CC{{\cal C}^{n+1}}
\def\EM{{\cal G}}
\def\EMO{{{\cal G}(0,0)}}
\def\Xb{{\bar X}}
\def\Yb{{\bar Y}}
\def\Ab{\bar A}
\def\Bb{\bar B}
\def\Ub{{\bar U}}
\def\zb{\bar z}
\def\ab{\bar a}

\def\eb{\bar e}
\def\fb{\bar f}
\def\ib{\bar \imath}
\def\jb{\bar \jmath}
\def\kb{\bar k}
\def\sb{\bar s}

\def\pb{\bar p}
\def\betab{\bar \beta}
\def\chib{\bar \chi}
\def\Lambdab{\bar \Lambda}
\def\gammab{\bar \gamma}

\def\1b{\bar 1}

\def\Cb{\bar C}
\def\xib{\bar \xi}
\def\mub{\bar \mu}

\def\bff{{\bf f}}
\def\bbff{\bar {\bf f}}
\def\bfe{{\bf e}}
\def\bbfe{\bar {\bf e}}
\def\bfX{{\bf X}}

\def\partialb{\bar \partial}

\def\Au{\underline A}
\def\Bu{\underline B}

\def\0b{\bar 0}
\def\lambdab{\bar \lambda}
\def\Yb{\bar Y}
\def\Jb{\bar J}
\def\bfv{{\bf v}}
\def\bbfv{\bar{\bf v}}
\def\rhob{\bar{\rho}}
\def\psis{\psi^+}
\def\bfu{{\bf u}}
\def\bbfu{\bar{\bf u}}
\def\vb{\bar v}
\def\bra#1{\langle {#1}|}
\def\ket#1{|{#1}\rangle}
\def\psif#1{\psi_{\displaystyle{f^{#1}}}}
\def\psifs#1{\psi^+_{\displaystyle{\fb^{#1}}}}
\def\psifz#1#2{\psi_{\displaystyle{f^{#1} }({#2})}}
\def\psifsz#1#2{\psi^+_{\displaystyle{\fb^{#1}}({#2})}}
\def\eJz{e^{\sum_0^n J_sz^{(s)}}}
\def\eJzb{e^{\sum_0^n \bar{J}_t\bar{z}^{(t)}}}
\def\ejz{e^{J_1z}}
\def\ejzb{e^{\bar{J}_1\bar{z}}}
\def\p{\Phi}
\def\bfet{\widetilde{\bf e}}
\def\bbfet{\widetilde{\bar{\bf e}}}
\def\bfvt{\widetilde \bfv}
\def\bfut{\widetilde \bfu}
\def\bbfvt{\widetilde{ \bbfv}}
\def\bbfut{\widetilde{ \bbfu}}
\def\gb{\bar g}
\def\LRA{\Leftrightarrow}
\def\ft{\tilde{f}}
\def\eH{e^{H(\zeta,[z])}}
\def\omegab{\bar \omega}
\def\gammab{\bar \gamma}
\def\ft{\tilde f}
\def\betab{\bar \beta}
\def\ftb{\tilde{\bar f}}
\def\pvphi{\partial \varphi}
\def\Ib{\bar I}
\def\Jb{\bar J}
\def\Kb{\bar K}
\def\Mb{\bar M}
\def\qb{\bar q}
\def\kb{\bar k}
\def\sb{\bar s}
\def\lambdab{\bar \lambda}
\def\omegab{\bar \omega}
\def\Lambdab{\bar \Lambda}
\def\gt{\tilde g}
\def\htd{\tilde h}
\def\ct{\tilde c}
\def\nablab{\bar \nabla}
\def\kappab{\bar \kappa}

\def\eb{\bar e}
\def\varphib{\bar \varphi}
\def\tb{\bar t}

\def\p{\Phi}
\def\bfet{\widetilde{\bf e}}
\def\bbfet{\widetilde{\bar{\bf e}}}
\def\gt{\widetilde g}
\def\Ft{\widetilde F}
\def\Deltat{\widetilde \Delta}
\def\phit{\widetilde \phi}
\def\Eq#1{ Eq.\ref{#1}}
\def\pzero{\Psi^{(0)}}
\def\pinfty{\Psi^{(\infty)}}
\def\pzeros{\Psi^{(0)*}}
\def\pinftys{\Psi^{(\infty) *}}
\def\Psit{\widetilde{\Psi}}
\def\Psist{\widetilde{\Psi}^*}
\def\sigmab{\bar \sigma}

\def\beq{\begin{equation}}
\def\eeq{\end{equation}}
\def\beqa{\begin{eqnarray}}
\def\eeqa{\end{eqnarray}}

\newtheorem{definition}{\bf Definition}
\newtheorem{theorem}{\bf Theorem}
\newtheorem{corollary}{\bf Corollary}
\newtheorem{proposition}{\bf Proposition}
\newtheorem{lemma}{\bf Lemma}
\newtheorem{situation}{\bf Situation}
\newtheorem{conjecture}{\bf Conjecture}
\def\qed{{\bf Q.E.D. }}
\def\proof{{\sl Proof:  }}
\begin{document}

\begin{titlepage}
\topmargin -1.5 true cm
\textheight 24.5 true cm
\textwidth 15 true cm
\oddsidemargin .5 true cm
\evensidemargin .5 true cm

\nopagebreak
\begin{flushright}

LPTENS--91/35,
NBI-HE--91-50\\
                January 1992
\end{flushright}

\vglue 1.5  true cm
\begin{center}
{\large\bf
CLASSICAL A$_{\hbox{\bf n}}$--W-GEOMETRY}

\vglue 1  true cm
{\bf Jean-Loup~GERVAIS}\\
{\footnotesize Laboratoire de Physique Th\'eorique de
l'\'Ecole Normale Sup\'erieure\footnote{Unit\'e Propre du
Centre National de la Recherche Scientifique,
associ\'ee \`a l'\'Ecole Normale Sup\'erieure et \`a l'Universit\'e
de Paris-Sud.},\\
24 rue Lhomond, 75231 Paris CEDEX 05, ~France.
\\
and
} \\
{\bf Yutaka~MATSUO}\footnote{
Present address: Dept. of Physics, Univ. of Tokyo,
Hongo 7-3-1, Bunkyo-ku, Tokyo, Japan.}\\
{\footnotesize Niels Bohr Institute,
University of Copenhagen, \\ Blegdamsvej 17,
DK-2100 Copenhagen $\emptyset$, Denmark.} \\
\medskip
{\footnotesize \LaTeX \ file available from
hepth@xxx.Lanl.GOV}
\end{center}

\vfill
\begin{abstract}
\baselineskip .4 true cm
\noindent
{\footnotesize
W-surfaces\cite{GM} in $\CP$ are studied by discussing 1) the extrinsic
geometries of chiral surfaces (Frenet-Serret and Gauss-Codazzi
equations) 2) the KP coordinates (W-parametrizations)  of the
target-manifold, and their fermionic (tau-function) description,
3) the intrinsic geometries  of the associated chiral surfaces
in the Grassmannians, and the associated higher
instanton-numbers of W-surfaces.
For regular points, the Frenet-Serret
equations for $\CP$--W-surfaces are shown  to give the
geometrical meaning of  the $A_n$-Toda
Lax pair,   and of the associated
conformally-reduced WZNW models, and Drinfeld-Sokolov equations.
KP coordinates are used to show that
W-transformations may be extended as
  particular  diffeomorphisms of the target-space.
The resulting W-parametrizations of $\CP$
provide  solutions of equations which come out as
 higher-dimensional generalizations of the WZNW and Drinfeld-Sokolov
equations. These are related
with the Zakharov-Shabat equations of the Toda hierarchy
by means of the Frenet-Serret formulae for KP coordinates.
Singular points are next studied from the
third  viewpoint mentionned above.
 Global Pl\"ucker formulae are derived  by combining the $A_n$-Toda
equations with  the Gauss-Bonnet theorem written for each
of the  associated surfaces. These  relations  involve
ramification-indices that are connected with
instanton-numbers, with singularities
of the Drinfeld-Sokolov equations, and  with
Polyakov's self-intersection-index.}

\end{abstract}
\vfill
\end{titlepage}
\section{Introduction  }
\markright{ 1. Introduction }

\label{1}

The intimate  relationship\cite{BG1} between
conformal Toda field theories\footnote{We only deal with
conformal Toda theories, in the present article.
They will   simply be called
Toda theories} and W-sym\-met\-ries\cite{Z} is well known by now.
In a recent letter\cite{GM},
we introduced  W-surfaces  as two-dimensional
surfaces
$\Sigma$ with chiral parametrizations embedded
into   K\"ahler manifolds. In the present article we develop
the arguments previously summarized
in \cite{GM}, for  the cases connected
with the Lie algebras  $A_n$
(It is well known that W symmetries are classified in parallel with
standard Lie groups).  We will show that the geometry of $W_{n+1}$
algebras is that of chiral surfaces embedded into   complex
projective spaces with $2n$ real dimensions, and describe its
 deep connection with the KP hierachy of integrable dynamics.
{}From the start, our geometrical viewpoint   seems  to be
different from the one of most of  the  interesting papers
which appeared recently about
W-algebras (see, for instance refs.\cite{SS}---
\cite{I}),
even if there is  some clear relationship with refs.\cite{SS}
\cite{OK}, and
with an earlier article on Toda theories \cite{S}. Yet we believe that
our work will provide a unifying framework for W-systems.

Since this article contains many differents points developed
one after the other, we present it as a succession of
theorems, propositions,
and so on, for clarity. Yet the language should be
 familiar to
physicists of the field.
The body of the paper is divided into
 three main sections after the present
introduction which is called section 1.

The \souligne{main section  2}  deals with the
regular points of  a W-surface $\Sigma$
 where the Taylor expansions of the
coordinates of the surface generate linearly independant vectors.
At first (\souligne{section 2.1}) the Gauss-Codazzi equations\cite{EH}
 are
written down, for chiral embeddings in $\cC$.
Since the tangents and normals are computed from
Taylor expansions, these equations may also be considered as
generalized Frenet-Serret relations. The expression  of this
moving frame
 involve  determinants of matrices whose entries are
 inner products of derivatives of the embedding functions.
The logarithms of these determinants are shown to be Toda-like
fields,  since the  Gauss-Codazzi-Frenet-Serret
 equations give  a  Toda Lax-pair\cite{LS}
 connected  with the algebra of the linear group $gl(n)= gl(1)\oplus
A_{n-1}$.
Next, in  \souligne{section 2.2}, we
 reformulate our scheme in such a way that
this additional $gl(1)$ is identified with the rescaling  used in order
to  describe
$\CP$ by homogeneous coordinates in $\CC$. This allows us
to derive the geometry of the $\CP$--W-surfaces
which are shown to be associated with the standard
$A_n$ Toda theories. The latter were recently related\cite{Dublin}
to the conformally reduced WZNW models. In \souligne{section 2.3},
we show that there is
a complete equivalence between these last models
and W-surfaces in $\CP$.
The generalized Frenet-Serret formula derived in section 2.2 is connected
with the Gauss decomposition of the solution of the WZNW equations. This
gives the geometrical meaning of this decomposition which plays a crucial
r\^ole in \cite{Dublin}.

The  \souligne{main section 3} deals with  the introduction
of higher KP coordinates.
First, the present geometrical interpretation makes
a frequent use of determinants, so that it is natural to
introduce fermionic
operators. Another motivation in the same direction is that
 the underlying dynamics is connected
with the higher KP systems where the free-fermion approach is remarkably
successful. \souligne{Section 3.1}  describes the basic points
of this approach\cite{M} where
all the expressions introduced so far are rewritten as matrix elements
between free-fermion states. Remarkably, the summations over  products
of such matrix elements which one needs to carry out may be handled
by means
of the  Hirota equation\cite{UT}, which we prove  in this section,
for completeness.
This equation becomes  easy to handle once it is re-interpreted
in the free-fermion language.
Next, in \souligne{section 3.2}, we use this free-fermion approach
to introduce
the additional coordinates of the KP hierarchy
in $\CP$--W-surfaces. They give particular
parametrizations of the target-manifold.
(We call them  W-parametrizations).
They are  such that the W-surface is
simply recovered by letting all additional
KP coordinates equal to zero. They allow us to extend
the W-transformations to the target-space, obtaining a special
class of diffeomorphisms.
 In  \souligne{section 3.3}, we next study the
W-parametrizations from the viewpoint of the Toda hierarchy.
The link between  Riemannian geometry and the latter is established
by showing that the integrability conditions for W-parametrizations
 coincide with the Zakharov-Shabat equations.
 The proof of this striking fact
is neatly carried out using the free-fermion technics described in
section 3.1.  This is a natural consequence of the duality
between the Zakharov-Shabat and the Hirota formulations
of the Toda hierarchy.       Next,
in \souligne{section 3.4}, it is noted that
 the W-parametrizations are  chiral
in the sense that the coordinate-functions only depend upon half of the
variables. Thus the metric-tensor is shown to
be factorized into chiral and anti-chiral components
in $2n$ variables, and  this gives a  higher dimensional analogue
of the WZNW equations for the Christoffel symbols, together with
a natural generalization of the Drinfeld-Sokolov
equations\cite{DS}.

Concerning the  \souligne{main section 4},  we first (in
\souligne{section 4.1}) reformulate our
approach in terms of the intrinsic geometry of the family of
associated surfaces in the Grassmannians $G_{n+1, k+1}$,
$k=1,\, \cdots,\, n$. This is needed to study  singular points
 and global aspects of W-surfaces following the
general scheme of our letter\cite{GM}. The aim is to establish the
generalization of the Gauss-Bonnet theorem to the W-surfaces
discussed above. In \souligne{section 4.2}, we recall the definition
of the instanton-number of a W-surface, given in Ref. \cite{GM}.
At singular points, there are obstructions to constructing
the moving frame.  In \souligne{section 4.3}, we define the corresponding
ramification-indices and study the behavior of  the tau-functions.
In \souligne{section 4.4}, the singularity structure of the
Drinfeld-Sokolov
operator\cite{DS} is related to these indices.
In \souligne{section 4.5}, we introduce a family of instanton-numbers
by applying  the definition of section 4.2 to each associated surface.
In this way, we associate $n$ higher topological invariants to a given
$\CP$ W-surface.
Applying  the Gauss-Bonnet theorem to
each associated surface,  we observe  that the infinitesimal
Pl\"ucker formula
coincides with the Toda equation, and derive the relationship
between the ramification-indices and higher instanton-numbers
(global Pl\"ucker formula for W-surfaces). Next, in
\souligne{section 4.6} these higher instanton-numbers  are
shown to generalize  the self-intersection numbers of ref.\cite{Pol}.

In the main \souligne{section 5}, an outlook of the present work is
presented.
The \souligne{appendix} involves four sections.
First, in \souligne{appendix A.1} we derive the Frenet-Serret
formula using the Fubini-Study metric\cite{GH}. This provides an example
where the target-space metric is not flat from the beginning. In
  \souligne{appendix A.2}, the bosonization of the KP free-fermions is
displayed for completeness.  Next we recall
that  the
free-fermion approach is very general and goes far beyond the $\CP$
case we have discussed. For example, one may get to a completely
different
situation by redefining the vacuum, that is by introducing a non-empty
Fermi sea. This is how free fermions appeared at $D=1$
in particular. Making a step in this direction in \souligne{appendix A.3},
we establish a ``dictionary'' between
the present language and the discrete formulation of the matrix-models.
Finally, in   \souligne{appendix A.4}, we take a different
viewpoint in order to
pave the way towards explicit generalizations of  the present work.
So far, we have
seen group theory ($A_n$-Toda dynamics) coming from the geometry.
We reverse the process and start from target-manifolds which
have a geometrical meaning since they are coset spaces. Looking at the
explicit solution of the Toda equations\cite{LS}, it is clear that the
most natural manifolds of this type are the group-orbits of the
fundamental weights and, for $A_n$, we obtain
the Grassmannians which were  the framework of the last main section.
Thus,  considering similar coset-spaces for other Lie groups,
should be helpful to generalize the present discussion.

Before beginning our journey through the  W-geometrical
landscape, we note
that it is quite attractive an idea that it comes out from the geometry
of embeddings. Indeed, 2D conformal systems are notoriously related to
string theories, and the present viewpoint goes in the same direction.
It is compatible with the fact that W-strings\cite{BG2}, if they exist,
 may come out spontaneously
when one looks for the true vacuum of the much-wanted
string-theory of
Nature.

%
%

\section{Local structure of the embedding at regular points}
\markright{ 2. Regular points  }
\label{2}
\subsection{The Gauss-Codazzi---Frenet-Serret equations}
\label{2.1}
In this section we only  deal with trivial target-manifolds.
This will be used later on to deal with  complex
projective spaces.

\begin{definition}{\bf $\cC$ target-manifolds. }
\label{target}
They will be taken to be   Riemannian
manifolds  with $2n$ real dimensions, noted $\cC$,
whose
points  represented by boldface letters $\bfX$, have components
 $X^{\Au}$, $\Au=1,\, \cdots,\, 2n$. It is assumed that

\noindent 1) there exists  a prefered class of coordinates
 $X^A$, $\Xb^{\Ab}$,
$1\leq A,\, \Ab \leq  n$,
 such that the line-element takes the form
$ds^2=2 \sum_{A}dX^A d\Xb^{A}$,

\noindent 2) there exists a conjugation-operation noted with a star
such that:
\beq
\label{2.1.2}
\left (X^{\Au}\right )^*=
\sum_{\Bu=1}^{2n} {\tilde C}^{\Au}_{\Bu} X^{\Bu},
\eeq
which leaves the line-element invariant.
\end{definition}

This is very close to the  standard definition of $\cC$,
but, contrary to the commmon pratice,
 we do not assume,  that
$X^A$ and $\Xb^{\Bb}$ are complex coordinates such that
$(X^A)^*=\Xb^{A}$. Our past knowledge of string theory
shows that one must be more flexible. For instance,
if we think of a string in a constant Minkowski-metric, the
components that  involve the time-direction, say  $X^0$
and $\Xb^{0}$, are real.
(more about this at  the end if this section).
We shall call chiral (resp. anti-chiral) components,  the
set $\{X^A, 1\leq A\leq n\}$
 (resp. $\{\Xb^{\Ab}, 1\leq \Ab\leq n\}$).

Our basic strategy is to study embeddings of two-dimensional surfaces
$\Sigma$ with chiral parametrizations. This chirality is defined
with respect to surface parameters noted $z$ and $\zb$.
 On the surface,
 we shall make use
of a  trival two-dimensional complex structure
 of the usual type. However,
taking $z$ to be a standard complex variable is not the only choice.
It corresponds  to a Euclidean parametrization where
 the real coordinates are $x_1=(z+\zb)/2$, and $x_2=(z-\zb)/2i$.
Another possibility is to
be   working
with  real surface-parameters. Then, as it is well known,
the square-root of $-1$ is represented by the
matrix $\left ({0\> -1\atop 1\> \> \> 0}\right )$ acting
on the two-component vector
$\left ({\zb \atop z}\right )$. The ``real'', and ``imaginary parts''
 are now $x_0=(z+\zb)/2$, $x_1=(z-\zb)/2$.
   This parametrization is of
the Minkowski type, where $x_0$ is a time-like parameter.

Next, a function is called chiral  if it  only depends upon one of the two
coordinates $z$ or $\zb$ (if $z$ is a complex variable this means
of course analytic or anti-analytic).
A basic object  of the present W-geometry is specified by the
\begin{definition}{\bf  $\cC$ W-surfaces. }
\label{W-surf}
A $\cC$ W-surface is a two-dimensional manifold $\Sigma$ with a
chiral embedding  into $\cC$ with an emphasis
on its extrinsic geometry.
 A chiral embedding
is defined by  equations of the form
\begin{equation}
\label{2.1.3}
X^A=f^A(z), \, A=1,\, \cdots,\, n, \quad
\Xb^{\Ab}=\fb^{\Ab}(\zb), \, \Ab=1,\, \cdots,\, n.
\end{equation}
\end{definition}
We do not assume any general link between the conjugation in $\cC$
and in the surface-parameter-space. Thus
 $f^A$ and
$\fb^{\Ab}$  are independent functions. For string applications this
is needed, basically, since the two chiral components may be
associated with the right- and left-moving modes which are
independent if the string is  closed. We shall give an example
of this fact, at the end of the section, by considering the
case of free bosonic strings.

Our first result is that  Toda field-equations naturally
arise
from the   Gauss-Codazzi
equations of the chiral  embedding of W-surfaces.
 These   equations  are
 integrability conditions for derivatives of
the tangents  and   of the normals
to  the surface. The latter are introduced
by extending
Frenet-Serret  formulae as follows. At each point of the surface, one
considers the Taylor expansion of $f^A$ and $\fb^{\Ab}$ up to
$n$th order,  and introduce
the corresponding matrix of inner products:
\begin{equation}
\label{2.1.4}
 g_{ \jb i }=g_{i \jb}\equiv \sum_{A\,\Bb} \delta_{\! A\Bb}\,
\>\>\partial^{(i)} \! f^A(z)\>
\partialb^{(\jb)}\!  \fb^{\Bb}(\zb),
\quad
1\leq i,\jb \leq n.
\end{equation}
$\partial $ and $\partialb$ are short hands for
$\partial/\partial z$ and $\partial /\partial \zb$ respectively.
$\partial^{(i)}$ stands for $(\partial)^i$. Later on we shall exhibit
a particular parametrization of $\cC$, called  W-parametrization,
  where the vectors
 $\partial^{(i)} \! f^A(z)$, $i>1$,  and
$\partialb^{(\jb)}\!  \fb^{\Bb}(\zb)$,  $\jb>1$ will become
tangent vectors, so that the covariance properties  of the present
discussion will become more transparent.
At this moment, we  are  concerned with generic regular
points of $\Sigma$ where the Taylor expansions
of $f^A$ and $\fb^{\Ab}$ give linearly independent
vectors.  Then  ${\bf f}^{(a)}$,
and $\bbff^{(a)}$, $a=1$, $\cdots $, $n$,
(upper indices in between parenthesis denote
derivatives) span the following

\begin{definition}{\bf Moving frame.}
\label{movingframe}
Consider the vectors $\bfe_a$, and $\bbfe_a$, $a=1, \cdots, n$,
with components
\begin{equation}
\label{2.1.5}
 e_a^A = {1\over \sqrt{\Delta_a \Delta_{a-1}}}
\left | \begin{array}{ccc}
g_{1 \1b} & \cdots  & g_{a \1b} \\
\vdots    &         & \vdots  \\
g_{1 {\overline {a-1}} } & \cdots  & g_{a {\overline {a-1}}} \\
f^{(1)A} & \cdots &  f^{(a)A}
\end{array}
\right |, \qquad  e_a^{\Ab}=0,
\end{equation}
\begin{equation}
\label{2.1.6}
\eb_a^A=0,\qquad
\eb_a^{\Ab} = {1\over \sqrt{\Delta_a \Delta_{a-1}}}
\left | \begin{array}{ccc}
g_{\1b 1} & \cdots  & g_{\ab 1} \\
\vdots    &         & \vdots  \\
g_{\1b a-1} & \cdots  & g_{{\overline {a}} a-1} \\
\fb^{(1)\Ab} & \cdots & \fb^{(a)\Ab}
\end{array} \right |,
\end{equation}
 $\Delta_a$ is  the determinant
\begin{equation}
\label{2.1.7}
\Delta_a\equiv \left | \begin{array}{ccc}
g_{1 \1b} & \cdots  & g_{a \1b} \\
\vdots    &         & \vdots  \\
g_{1 \ab} & \cdots  & g_{a \ab}
\end{array}
\right |.
\end{equation}
\end{definition}
Denote by  $({\bf x},{\bf y})$ the inner
product $\sum_A(x^A{\bar y}^A+
y^A{\bar x}^A)$. One has the

\begin{proposition}{~}
\label{ortho}
The moving frame defined above is orthonormal, that is
\begin{equation}
\label{2.1.8}
( \bfe_a, \bfe_b)=( \bbfe_a, \bbfe_b)=0,\quad
( \bfe_a, \bbfe_b) =\delta_{a,b}.
\end{equation}
\end{proposition}

\noindent {\sl Proof } These last relations
immediately follow from the fact that
\begin{equation}
\label{2.1.9}
(\bfe_a ,\bbff^{(b)})=0,\quad  \hbox{and}\>\>
(\bbfe_a ,\bff^{(b)})=0\quad  \hbox{for} \>
a > b
\end{equation}
 together with the definition Eq.\ref{2.1.7} of $\Delta_a$.
 {\bf Q.E.D.}

For the following it is important to note that, according to
Eq.\ref{2.1.9},  Eqs.\ref{2.1.5}, \ref{2.1.6} take the form
\begin{equation}
\label{2.1.10}
\bfe_a=\sum_{b\leq a} C_{ab}(z, \zb) \>
\sqrt{{\Delta_{a-1}\over \Delta_a}} \> \bff^{(b)}(z), \quad
\hbox{with}\> C_{aa}=1,
\end{equation}
\begin{equation}
\label{2.1.11}
\bbfe_a=\sum_{b\leq a} A_{ba}(z,\zb)\>
\sqrt{{\Delta_{a-1}\over \Delta_a}}\>  \bbff^{(b)}(\zb), \quad
\hbox{with}\> A_{aa}=1.
\end{equation}
This equation is also valid for $a=1$ if we define $\Delta_0$ to
be equal to one, as we shall do. The vectors
${\bfe}_1$ and  ${\bbfe}_1$ are tangents to the surface,
while the other vectors are clearly normals. Thus the Gauss-Codazzi
equations may be  derived  by studying their  derivatives
along $\Sigma$.
The main result of the present section is the
\begin{theorem}{\bf Generalized Frenet-Serret formulae.}

The derivative of the moving frame is given by
\begin{eqnarray}
\label{2.1.20}
\partial \bfe_a&=&{1\over 2}
\partial \ln \left ( {\Delta_{a}\over \Delta_{a-1}}\right )
\bfe_a+ \sqrt{{\Delta_{a-1}\Delta_{a+1}\over \Delta_a^2}}
\bfe_{a+1}, \quad a\leq n-1,\nonumber\\
\partial \bfe_n&=&{1\over 2}
\partial \ln \left ( {\Delta_{n}\over \Delta_{n-1}}\right )
\bfe_n,\nonumber\\
\partialb \bfe_a&=&-{1\over 2}
\partialb \ln \left ( {\Delta_{a}\over \Delta_{a-1}} \right )
\bfe_a- \sqrt{{\Delta_{a-2}\Delta_{a}\over \Delta_{a-1}^2}}
\bfe_{a-1}, \quad 2\leq a,\nonumber\\
\partialb \bfe_1&=&-{1\over 2}
\partialb \ln \left ( \Delta_{1}\right)
\bfe_1,
\end{eqnarray}
with similar equations for $\bbfe$.
\end{theorem}
\proof
It is easy to see that these derivatives  may be written as
\begin{equation}
\label{2.1.14}
\partial {\bfe}_a=\sum_b{\it R}_{a b} {\bfe}_b,\quad
\partialb {\bfe}_a=\sum_b{\it S}_{a b} {\bfe}_b;\quad
\partialb {\bbfe}_a=\sum_b{\overline {\it R}}_{a b} {\bbfe}_b,\quad
\partial {\bbfe}_a=\sum_b{\overline {\it S}}_{a b} {\bbfe}_b.
\end{equation}
According to  Eq.\ref{2.1.8}, one has
\begin{equation}
\label{2.1.15}
{\it R}_{a b}+{\overline {\it S}}_{b a}=
{\overline {\it R}}_{a b}+{\it S}_{b a}=0.
\end{equation}
Since $\partial \bbff^{(a)}=\partialb \bff^{(a)}=0$, it
follows from Eqs.\ref{2.1.9}, \ref{2.1.10}, \ref{2.1.11}
that ${\it S}_{a b}$ and
${\overline {\it S}}_{a b}$ vanish for $a<b$. Moreover, it is easy
to verify that $\partial {\bfe}_a$ and $\partialb {\bbfe}_a$ may be
respectively expanded in terms of $\bff^{(b)}$ and $\bbff^{(b)}$
with $b\leq a+1$ only. Thus ${\it R}_{a b}$ and
${\overline {\it R}}_{a b}$ vanish for $b > a+1$. Combining with
Eq.\ref{2.1.15},  one sees
that the only non-vanishing elements
are ${\it R}_{a a+1}\equiv \kappa_a$,
${\overline {\it R}}_{a a+1}\equiv \kappab_a$, for
$a\leq n-1$, and  ${\it R}_{a a}\equiv \sigma_a$,
${\overline {\it R}}_{a a}\equiv \sigmab_a$,
for $a\leq n$. Eq.\ref{2.1.14}
becomes
\begin{equation}
\label{2.1.16}
\partial {\bfe}_a
=\kappa_a {\bfe}_{a+1}+\sigma_a{\bfe}_{a},\quad
\partialb {\bfe}_a=
-\sigmab_a {\bfe}_a-\kappab_{a-1}  {\bfe}_{a-1}
\end{equation}
\begin{equation}
\label{2.1.17}
\partialb {\bbfe}_a
=\kappab_a {\bbfe}_{a+1}+\sigmab_a{\bbfe}_{a},\quad
\partial {\bbfe}_a=
-\sigma_a {\bbfe}_a-\kappa_{a-1}  {\bbfe}_{a-1}.
\end{equation}
Making use of Eqs.\ref{2.1.9}, \ref{2.1.10}, \ref{2.1.11}, one next
easily obtains
\begin{equation}
\label{2.1.18}
\sigma_a=-(\partial \bbfe_a,\, \bfe_a)={1\over 2}
\partial \ln \left ( {{\Delta_{a}}\over {\Delta_{a-1}}} \right ),
\quad
\sigmab_a=-(\partialb \bfe_a,\, \bbfe_a)=
{1\over 2}
\partialb \ln \left ( {{\Delta_a}\over {\Delta_{a-1}}} \right )
\end{equation}
\begin{equation}
\label{2.1.19}
\kappa_a= (\partial \bfe_a, \bbfe_{a+1} )=
\sqrt{{\Delta_{a-1}\Delta_{a+1}\over \Delta_a^2}},
\quad
\kappab_a= (\partialb \bbfe_a, \bfe_{a+1} )=
\kappa_a
\end{equation}
and the proof is completed by  using the fact that
the $\bfe$'s and
$\bbfe$'s form a complete basis. \qed

 This theorem generalizes,
for W-surfaces,    the Frenet-Serret form\-ulae
which are standard for curves.
Eqs.\ref{2.1.20} have a form
which is closely related to the $A_{n-1}$ Toda equations, if we define
the Toda-like fields by
\begin{equation}
\label{2.1.21}
\phi_a\equiv -\ln (\Delta_a), \quad \hbox{for}\>  a=1,\, \cdots n.
\end{equation}
This may be neatly expressed as follows. As is well known,
the  Lie algebra $A_{n-1}$
may be explicitly realized with  $n$  fermionic operators
$\flat_i$,  which satisfy
$\bigl [ \flat_i, \flat_j^+\bigr ]_+=\delta_{i,j}$.
One writes
\begin{equation}
\label{2.1.22}
h_i=\flat_i^+ \flat_i-\flat_{i+1}^+ \flat_{i+1},\quad
E_i=\flat_{i}^+ \flat_{i+1},\quad E_{-i}=E_i^+, \quad i=1,\cdots n-1,
\end{equation}
in the Chevalley basis where $h_i$ generate the Cartan subalgebra, and
$E_{\pm i}$ are associated with the standard set of primitive roots.
The basic difference in our case is that, contrary to the
 $A_{n-1}$--Toda case, $\phi_n$ is not zero, and we need another
``Cartan'' generator
\begin{equation}
\label{2.1.23}
h_n=\flat_n^+\flat  _n
\end{equation}
which is realized by the fermionic operators  but does not belong
to $A_{n-1}$. Altogether, the generators we have introduced
 satisfy the commutation relations
\begin{equation}
\label{2.1.24}
\bigl [h_i,\, E_{\pm j}\bigr ]= \pm K^{gl(n)}_{ji} \> E_{\pm j},
 \quad
\bigl [E_{j},\, E_{-k}\bigr ]= \delta_{j,k} \> h_{ j},
\end{equation}
where $i$ goes from $1$ to $n$, while $j$ and $k$ run from $1$ to
$n-1$. Concerning Lie algebras, we denote by $gl(n)$ the Lie algebra
of $n\times n$ matrices (Lie algebra of the linear group).
The  matrix
$K^{gl(n)}_{ij} $ may be regarded as the  $n\times n$  Cartan  matrix
of $gl(n) \sim A_{n-1}\oplus gl(1)$.
For $i$ and  $j$ between
$1$  and $n-1$, it
coincides with the Cartan matrix of $A_{n-1}$, and
\begin{equation}
\label{2.1.25}
K^{gl(n)}_{j\, n}=-\delta_{j,\,n-1}.
\end{equation}
The reality conditions are most simply discussed with  Minkowski
surface parameters ($z$ and $\zb$ real).  Then one is using the most
non-compact real form of
the Lie algebras we encounter.  In particular,
the Lie group associated with $gl(1)$ is the multiplication
by real positive numbers. This should be understood from now on.
Let us go back to  Eq.\ref{2.1.20}. Together with the
anti-chiral parts, it takes
 the form
\beq
\label{2.1.26}
\partial \bfe_a=\sum_b\omega_{z a}^{\>b} \bfe_b, \quad
\partialb \bfe_a=\sum_b\omega_{\zb a}^{\>b} \bfe_b,
\quad
\partialb \bbfe_a=\sum_b\omegab_{\zb a}^{\>b} \bbfe_b, \quad
\partial \bbfe_a=\sum_b\omegab_{z a}^{\>b} \bbfe_b.
\end{equation}
It follows trivially from Eq.\ref{2.1.8} that
$\omega_{za}^b+\omegab_{zb}^a =0$ and
$\omega_{\zb a}^b+\omegab_{\zb b}^a =0$.
The generators \ref{2.1.22}, \ref{2.1.23} commute with the number
operator $N=\sum_{i=1}^n \flat_i^+\flat_i$.
The subspace with $N=1$ has
dimension $n$. We identify it with the space span by the
$\bfe_a$ and write
\begin{equation}
\label{2.1.27}
\omega_{z a}^{\>b}  = <0|\flat_b \omega_{z} \flat_a^+ |0>,\quad
\omega_{\zb a}^{\>b}  = <0|\flat_b \omega_{\zb} \flat_a^+ |0>,
\end{equation}
where $|0>$ is the vacuum state of the oscillators $\flat_i$.
Using the formulae just given, one sees that Eq.\ref{2.1.20}
is equivalent to
\begin{eqnarray}
\label{2.1.28}
\omega_z =-{1\over 2} \sum_{i=1}^{n} h_i
\partial\phi_i
+\sum_{i=1}^{n-1} \exp \left ( \sum_{j=1}^{n}
K^{gl(n)}_{i j} \phi_j/2 \right )
E_{-i}\nonumber \\
\omega_{\zb} ={1\over 2} \sum_{i=1}^{n} h_i
\partialb\phi_i
-\sum_{i=1}^{n-1} \exp \left ( \sum_{j=1}^{n}
K^{gl(n)}_{i j} \phi_j/2 \right )
E_{i}.
\end{eqnarray}
Remarkably, one sees that the right member
is just the Toda Lax-pair\cite{LS}. Toda equations
are equivalent to the zero-curvature condition on $\omega$.
Let us remember that, in
the language of Riemannian
geometry\cite{EH},
the associated second fundamental form is given by
the projection of the  derivatives of the
tangent vectors onto the normals. Its non-vanishing components
are
\beq
\label{2.1.SF}
 (\partial^2\bff, \bbfe_a)  \equiv \Omega_{zz}^a=
 \delta_{a,\,2}\>\sqrt{\Delta_2\over \Delta_1},\quad
(\partialb^2\bbff, \bbfe_a)  \equiv {\bar \Omega}_{\zb \zb}^a=
 \delta_{a,\,2}\>\sqrt{\Delta_2\over \Delta_1}
\eeq
where $a\neq 1$.  Similarly, the third fundamental form
is given by the projection of the  derivatives of
the normals:
\beqa
\label{2.1.TF}
 (\partial\bfe_a, \bbfe_b) & = & \omega_{za}^b,
\quad
 (\partialb \bfe_a, \bbfe_b) = \omega_{\zb a}^b\nonumber\\
 (\partial \bbfe_a, \bfe_b)& =& \omegab_{z a}^b,
\quad
 (\partialb\bbfe_a, \bfe_b)  =  \omegab_{\zb a}^b,
\eeqa
where $a,\> b\geq 2$.
Going to the
second derivatives, we next derive the
\begin{theorem}{\bf Gauss-Codazzi equations.}

The integrability conditions of the Frenet-Serret equations
Eqs.\ref{2.1.20} coincide with the Toda equations associated
with $gl(n)$:
\begin{eqnarray}
\label{2.1.29}
0=\bigl [ \partial, \partialb \bigr ] \bfe_a & = &
\sum_bF_{z\zb a}^{\> \> b} \bfe_b
\equiv \sum_b<0|\flat_b  F_{z\zb }\flat_a^+ |0> \bfe_b
   \nonumber \\
F_{z\zb }&=&
\sum_{i=1}^{n} h_i \partial \partialb   \phi_i+
\sum_{i=1}^{n-1} h_i
\exp \left ( \sum_{j=1}^{n} K^{gl(n)}_{i j} \phi_j \right )
\end{eqnarray}
\end{theorem}
\proof Straghtforward computations using
Eqs.\ref{2.1.24}, \ref{2.1.25},
and  \ref{2.1.27}, \ref{2.1.28}. \qed

It is instructive to directly compare the above formulae with the
explicit solution of ref.\cite{BG1}.
 Eq.\ref{2.1.4} shows that
$g_{11}\equiv \exp(-\phi_1)$  has
 chiral components are $\chi^{A}\equiv f^{(1)\, A}$ and
$\chib^{A}\equiv \fb^{(1)\, A}$. A simple
calculation
starting from Eqs.\ref{2.1.7}, and  \ref{2.1.21}  shows that
\begin{equation}
\label{2.1.33}
e^{-\phi_k}=\sum_{i_1< \cdots <  i_k}
\left | \begin{array}{ccc}
\chi^{i_1}  & \cdots  & \chi^{ i_k} \\
\chi^{(1)\, i_1}  & \cdots  & \chi^{(1)\, i_k} \\
\vdots     & \cdots         & \vdots  \\
\chi^{(k-1)\, i_1}  & \cdots  & \chi^{(k-1)\, i_k} \\
\end{array}
\right |
\left | \begin{array}{ccc}
\chib^{ i_1}  & \cdots  & \chib^{ i_k} \\
\chib^{(1)\, i_1}  & \cdots  & \chib^{(1)\, i_k} \\
\vdots     & \cdots         & \vdots  \\
\chib^{(k-1)\, i_1}  & \cdots  &
\chib^{(k-1)\, i_k} \\
\end{array}
\right |.
\end{equation}
This exactly coincides with the explicit
form of the $A_{n-1}$-Toda
solutions of ref.\cite{BG1}. The only difference is that the
right member of the above is not equal to one for $k=n$, so that
$\phi_n$ does not vanish. However,
in the present case,  this right member
factorizes into the product of a
single  function of $z$ times another
function of $\zb$, so that $\phi_n$ is a solution of $\partial
\partialb \phi_n=0$.  These explicit formulae of course confirm
our previous calculations, that is,
  Eqs.\ref{2.1.29}.
The removal of the additional field,  might  simply be done
by imposing that the Wronskians of the functions $\chi^{ A}$ and
of the functions $\chib^{A}$ be equal
to one. At the present stage,
this would be an artificial condition without geometrical
 significance, since these functions are the first
derivatives of the embedding functions.  We will see
that this additional $gl(1)$ factor
will be removed naturally in the $\CP$ case.  This will be
the subject of the coming section. Beside
this $gl(1)$ factor, the
present situation has another unwanted feature.
The induced metric
on $\Sigma$ is $g_{11}=\exp(-\phi_1)$, while for the
Liouville theory, say, it is $\exp(2\phi_1)$ !
This disaster  will be repaired
in appendix A.1,  explicitly,  by deriving
proposition \ref{Liouville}.

\medskip
\noindent {\bf The example of free bosonic string. }

Let $Y^\alpha$, $\alpha=1,\, \cdots , \, 2n$, be
the space-time  coordinates of the string which are,
of course, real.
 As we shall see, it is
essential to work with the Minkowski signature, {\em without
performing the Wick rotation}. The target-space metric
${\widehat \eta}_{\alpha \beta}=\pm \delta_{\alpha,\,\beta}$
 is taken to be constant and diagonal.
For the sake of the coming argument, we shall
allow for several time-like directions (this possibility cannot
be ruled out a priori for W-strings\cite{BG2}). Thus we take
the target-space metric to be ${\widehat \eta}_{\alpha \beta}
=- \delta_{\alpha,\,\beta}$, for $\alpha=1,\, \cdots , \, s$,
${\widehat \eta}_{\alpha \beta}
= \delta_{\alpha,\,\beta}$, for $\alpha=s+1,\, \cdots , \, 2n$.
$s$ is the number of ``time'' axis.
This  metric will become off-diagonal, as required by the
definition  \ref{target}, if one defines
\beqa
\left.\begin{array}{cc}
X^\alpha&\equiv (Y^\alpha+ Y^{n+\alpha})/\sqrt{2} \\
\Xb^\alpha&\equiv (-Y^\alpha+ Y^{n+\alpha})/\sqrt{2}
\end{array}\right \},& \alpha=1,\, \cdots , \, s, \nonumber \\
\left.\begin{array}{cc}
X^\alpha\equiv (Y^\alpha+ i Y^{n+\alpha})/\sqrt{2}  \\
\Xb^\alpha\equiv (Y^\alpha- i Y^{n+\alpha})/\sqrt{2}
\end{array}\right \},& \alpha=s+1,\, \cdots , \, n.
\label{wp1}
\eeqa
 In this way the
inner product is indeed  $(\bfX_1,\bfX_2)=
\sum_A(X_1^A \Xb_2^A+ X_2^A \Xb_1^A)$.
Obviously, one has $(X^A)^* =X^A$, for $A\leq s$, and $(X^A)^* =\Xb^A$,
for $A>s$. This is an example of Eq.\ref{2.1.2}, which is not the usual
conjugation of $\cC$.
For a free string, the surface swept by the string-positions is given by

\beq
\label{wp2}
Y^\alpha=q^\alpha+p^\alpha\ln (z) + \tilde p^\alpha\ln (\zb)+
i \sum_{r\not=0} \left [ {a_r^\alpha\over r} z^{-r}+
{{\tilde a}_r^\alpha\over r} \zb^{-r}\right ],
\eeq
where $a_r^\alpha$ (resp. ${\tilde a}_r^\alpha$) are the right-moving
(resp. left-moving) oscillator-modes which satisfy $(a_r^\alpha)^*=
a_{-r}^\alpha$ (resp. $({\widetilde a}_r^\alpha)^*=
\widetilde a_{-r}^\alpha$). $q^\alpha$ (resp.
$(p^\alpha+\tilde p^\alpha)/2$) is the
 center-of-mass position (resp. total momentum) of the string which
must be real.
$(p^\alpha-\tilde p^\alpha)/2i$ is the winding-number
which is an integer.
Eq.\ref{wp2}  will  describe a W-surface,
 if the embedding  functions computed from
  Eqs.\ref{wp1} satisfy the Cauchy-Riemann relations
$\partial \fb^{\Ab}=\partialb f^A=0$. one gets
\beqa
\label{wp3}
p^\alpha&= p^{n+\alpha},\quad  \tilde p^\alpha= -\tilde p^{n+\alpha},
\quad
a_{r}^\alpha&=a_{r}^{n+\alpha}, \quad
\widetilde a_{r}^\alpha=-\widetilde a_{r}^{n+\alpha},
\qquad  \alpha\leq  s, \nonumber\\
p^\alpha&= ip^{n+\alpha},\quad  \tilde p^\alpha= -i\tilde p^{n+\alpha},
\quad
a_{r}^\alpha&=i a_{r}^{n+\alpha}, \quad
\widetilde a_{r}^\alpha=-i\widetilde a_{r}^{n+\alpha}, \qquad
 \alpha> s \nonumber \\
&
\eeqa
Clearly, the reality-condition forces us to take $f^A=\fb^{A}=0$,
for $A>s$. The number of components of the
W-string-surface is only equal to $2s$. This is why
the Minkowski metric was essential for the present example.
For $A,\Ab=1,\,\cdots\, s$, the embedding functions are
\beqa
\label{wp4}
f^A(z)&=\left \{(q^A+q^{A+n})/ 2+ p^A\ln (z)
+i \sum_{r\not=0} a_r^A z^{-r}/r\right\}\sqrt{2} \nonumber\\
\fb^{\Ab}(\zb)&= -\left \{(q^A-q^{A+n})/ 2  +
\tilde p^A\ln (\zb)+
i \sum_{r\not=0} \tilde a_r^{\Ab} \zb^{-r}/r\right\}\sqrt{2}.
\eeqa
For Euclidean world-sheet-parametrizations where $z$ is a complex
number,
they satisfy the conditions $\left (f^A(z)\right )^*=
f^A(\zb)$, and $\left (\fb^A(\zb)\right )^*=
f^A(z)$, in contrast with the  conditions of real
analyticity which are usually assumed (in particular for algebraic
curves). Physically, this reflects the fact that, for closed strings,
left and right modes are not correlated. On the contrary, if we
consider an open string with parameters running, say, in the
upper half-plane, the boundary condition is  that
$\left\{z\partial -\zb \partialb \right\}Y^\alpha=0$, for
$z=\zb$. As a result one has $\left (f^A(z)\right )^*=
\fb^A(\zb)$, and one recovers the standard mathematical situation
 of real analytic functions. A similar situation occurs for Liouville
theory with boundaries\cite{BG3},\cite{CG1}.

%
%

\subsection{$\CP$ $W$-surface and $A_n$ Toda Lax pair}
\label{2.2}

\begin{definition}{\bf $\CP$ target space.}
The complex projective space $\CP$ is defined to be the
quotient of the space $\CC$ of definition \ref{target}, by
the equivalence relation
\beq
\label{2.2.1}
\bfX\sim {\bf Y},
\qquad {\rm if}\quad X^A = Y^A\rho(Y), \quad
 {\rm and }
\quad \Xb^{\Ab} = \bar{Y}^{\Ab}\rhob(\Yb).
\eeq
where $\rho$ and $\rhob$ are arbitrary chiral functions.
\end{definition}
It will be convenient to denote the $n+1$
homogeneous coordinates by
$X^A$, $\Xb^{\Ab}$ with  $A,\Ab = 0,1,2,\cdots, n$.
In our  definition of $\CP$,
 $\rho $ and $\rhob$ are independent, since  we do not
impose any general reality condition on $X^A$, and $\Xb^{\Ab}$.
The metric which is invariant under the rescaling Eq.\ref{2.2.1} is
the Fubini-Study metric\cite{GH}
\beq\label{2.2.4}
G_{A\Ab} = \left (\delta_{A\Ab}\left (\sum_{B = 0}^n X^B\Xb^B
\right )
   - X^{\Ab}\Xb^{A}\right )\Bigl /(\sum_{B = 0}^n X^B\Xb^B)^2,
\eeq
whose K\"ahler potential is given by
${\cal K} = \ln\sum_{A = 0}^n X^A\Xb^{\Ab}$.
We note that  the variation Eq.\ref{2.2.1} shifts this potential
by  $\ln \rho+\ln \rhob$. The metric is  invariant if $\rho$
(resp.  $\rhob$)  is only function of $X^A$ (resp.  $\Xb^{\Ab}$),
as required by the above definition.
This will be called a local rescaling. At this point, there
are two ways
to parametrize $\CP$. On the one hand,  it
is customary to use the Fubini-Study metric and to impose the
condition $X^0=\Xb^0=1$. This procedure is developed in appendix A.1. It
gives an example of the treatment of the
 Gauss-Codazzi equations  with non-trivial target
metric. On the other hand, and for the
present purpose, it is more convenient to proceed as follows.
   First, instead of using the curved
Fubini-Study metric Eq.\ref{2.2.4},
we  use the {\it flat}
metric of $\CC$
\beq\label{2.2.15}
G_{A\Ab} = \delta_{A\Ab}, \qquad ( A, \Ab = 0,1,\cdots, n),
\eeq
We shall work with the $2(n+1)$ homogeneous coordinates,
without fixing the
local scale (its   choice
 is to be made only at
 the end). This is done by keeping the
$2(n+1)$  embedding-functions, and making our discussion covariant
under  the $gl(1)$ local-rescaling  symmetry,
\beq\label{2.2.16}
\bff^A(z)\rightarrow \rho(z)\> \bff^A(z),\quad
\bbff^{\Ab}(\zb) \rightarrow \rhob(\zb)\> \bbff^{\Ab}(\zb).
\eeq
For this, one  constructs the moving frame starting from
the {\it zero}-th order derivative of the embedding functions $\bff$
and $\bbff$.  The appropriate choice of the local
scale will turn out to depend upon the W-surface considered.

Except for these modifications, the construction of the moving
frame is completely parallel to the one of the  previous chapter.
\begin{definition}{\bf  Toda fields.}
\label{Todafields  }
Introduce the matrix of inner products
\beq
\label{2.2.9a}
\eta_{r\bar s}   = \sum_{A=0}^n f^{(r)\, A}(z) \fb^{(\sb)\, A}(\zb),
\quad 0\leq r,\sb \leq n.
\eeq
The Toda fields $\p_\ell$,  ($\ell=1,\cdots, n+1$)  are given by
\beq\label{2.2.9}
\p_\ell(z,\zb)=-\ln \tau_\ell(z,\zb), \qquad
\tau_\ell(z,\zb) \equiv
\left|
 \begin{array}{ccc}
  \eta_{0\0b} & \cdots & \eta_{\ell-1\0b}\\
  \vdots   &    ~   & \vdots                           \\
  \eta_{0 \overline{\ell-1}} & \cdots & \eta_{\ell-1{\overline{\ell -1}}}
 \end{array}\right|.
\eeq
Define also $\tau_0\equiv 1$ and $\p=_0\equiv 0$.
\end{definition}
\begin{definition}{\bf$\CP$ moving frame.}
\label{CPmoving}
The following vectors are orthonormal:
\beq  \label{2.2.19}
\bfet_\ell = \frac{1}{(\tau_\ell \tau_{\ell+1})^{1/2}} \bfvt_\ell,
\quad
\bbfet_\ell = \frac{1}{(\tau_\ell \tau_{\ell+1})^{1/2}} \bbfvt_\ell,
\eeq
\beq
\bfvt_\ell  =
\left|
 \begin{array}{ccc} \label{2.2.17}
  \eta_{0\0b} & \cdots & \eta_{\ell\0b}\\
  \vdots   &    ~   & \vdots                           \\
  \eta_{0 \overline{\ell-1}} & \cdots & \eta_{\ell{\overline{\ell -1}}}\\
  \bff  & \cdots & \bff^{(\ell)}
 \end{array}\right|,\quad
{\bbfvt}_\ell  =  \left|
 \begin{array}{ccc}
  \eta_{\0b 0} & \cdots & \eta_{\bar{\ell}0}\\
  \vdots   &    ~   & \vdots                           \\
  \eta_{\0b \ell-1} & \cdots & \eta_{\bar{\ell}{\ell -1}}\\
  \bbff  & \cdots & \bbff^{(\ell)}
 \end{array}\right|.
\eeq
\end{definition}
In the last equation, which is similar to Eqs.\ref{2.1.5}
and \ref{2.1.6}, the
determinants are to be computed for each components of the last lines,
and only non-vanishing components are written.
The vectors $\bfvt$ are introduced for later convenience.
They satisfy
\beq \label{2.2.18}
(\bfvt_\ell,\bbfvt_{\ell'}) = \tau_\ell\tau_{\ell+1}\delta_{
\ell \ell'}.
\eeq
\begin{proposition}{\bf Frenet-Serret formulae for $\CP$.}
\label{FSCP}

The above vectors satisfy ($\ell$ runs from $0$ to $n$, with
$\bfet_{-1}=\bfet_{n+1}\equiv 0$)
\beqa
\label{2.2.21}
\partial \bfet_\ell & = &
 \frac{1}{2}\partial (\p_\ell - \p_{\ell+1})\>\bfet_\ell
  + e^{(1/2)(2\p_{\ell+1}-\p_{\ell}
  -\p_{\ell+2})}\>\bfet_{\ell+1},
   \nonumber\\
\partialb \bfet_\ell & = &
 -\frac{1}{2}\partialb (\p_\ell - \p_{\ell+1})\>\bfet_\ell
  - e^{(1/2)(2\p_\ell-\p_{\ell-1}-\p_{\ell+1})}\>\bfet_{\ell-1}.
\eeqa
\end{proposition}
\proof Calculations similar to the ones of the previous section.
\qed

The next important point is the
\begin{theorem}{\bf Covariance under local rescaling.}
\label{covresc}

Under the transformation Eq.\ref{2.2.16}, the moving frame
transforms  covariantly:
\beqa
\tau_\ell  \rightarrow  \rho^\ell\rhob^\ell\tau_\ell,
&\quad& \p_\ell  \rightarrow \p_\ell -\ell \ln \rho-
\ell \ln \rhob,
\nonumber\\
\bfvt_\ell  \rightarrow   \rho^{\ell+1}\rhob^\ell \bfvt_\ell,&\quad&
\bbfvt_\ell \rightarrow  \rhob^{\ell+1}\rho^\ell \bbfvt_\ell,
\label{2.2.20}\\
\bfet_\ell  \rightarrow  (\rho/\rhob)^{1/2} \bfet_\ell,&\quad &
\bbfet_\ell  \rightarrow  (\rhob/\rho)^{1/2} \bbfet_\ell.\nonumber
\eeqa
\end{theorem}
\proof Trivial computations show that
$$
\bff^{(a)} \rightarrow \sum_{b=0}^a \left\{{a\choose b}
\partial^{(a-b)}\rho\right\}
\bff^{(b)}\>\equiv \sum_{b=0}^a \Lambda_{a b} \bff^{(b)},
\quad \Lambda_{a a}=\rho.
$$
\begin{equation}
\label{2.2.21l}
\bbff^{(a)} \rightarrow \sum_{b=0}^a \left\{{a\choose b}
\partialb^{(a-b)}\rhob\right\}
\bbff^{(b)}\>\equiv \sum_{b=0}^a \Lambdab_{b a} \bff^{(b)},
\quad \Lambdab_{a a}=\rhob.
\end{equation}

The transformation of the matrix  $\eta_{r\bar s}$
Eq.\ref{2.2.9a} is  $\eta_{r\bar s}\to \sum_{j \kb} \Lambda_{rj}
\eta_{j\bar k} \Lambdab_{\kb \sb}$. It follows that the tau-, $v$-,
and $\vb$-functions are multiplied by sub-determinants of
$ \Lambda$ and $ \Lambdab$.
Since $\Lambda$
(resp. $\Lambdab$) is lower (resp. upper) triangular, these
sub-determinants  are equal to the products of their
diagonal elements, and the result follows. \qed

\noindent Thus the above moving
frame may be called homogeneous.
\begin{corollary}{~}
\label{FSGI}
The Frenet-Serret formula Eq.\ref{2.2.21} are invariant under
local rescaling.
\end{corollary}
\begin{theorem}{\bf Toda equations.}
\label{Todaeq}
There exists a choice of
local rescaling such that
the Gauss-Codazzi equations coincide with the
$A_n$ Toda equations.
\end{theorem}
\proof
  The compatibility conditions of the Frenet-Serret equations give
\beqa \label{2.2.22}
\partial\partialb \p_\ell & = &
- \exp(2\p_\ell-\p_{\ell-1}-\p_{\ell+1})
\quad (\ell = 1,\cdots,n)\\
\label{2.2.23}
\partial\partialb \p_{n+1} & = & 0.
\eeqa
The first equation is precisely the $A_n$ Toda equation.  The second
equation
implies that $\tau_{n+1}$ is the product of two chiral functions.
For later use we introduce two functions $U_0(z)$, and $\Ub_0(\zb)$,
which are such that
\beq
\label{2.2.24}
\tau_{n+1} = U_0(z)\Ub_0(\zb).
\eeq
Given the $2(n+1)$ embedding functions, we apply the local rescaling
Eq.\ref{2.2.16} with
\beq\label{2.2.25}
\rho = (U_0(z))^{-1/(n+1)}, \qquad
\rhob = (\Ub_0(\zb))^{-1/(n+1)},
\eeq
which precisely puts $\tau_{n+1}=1$, killing the unwanted degree
of freedom. \qed

Finally we get the $A_n$ Toda equation {\it exactly}. By looking at
the explicit solution, similar to Eq.\ref{2.1.33}, one sees that
$U_0(z)$ (resp. $\Ub_0(\zb)$) is the Wronskian of the embedding
functions $f$ (resp. $\fb$).  These  do not
vanish at regular points, so
that this choice of parametrization is really possible.
We shall call it the Wronskian scale-choice. In the forthcoming it is
however, more convenient to work in a scale-invariant way, without
making this  particular choice.

 If one compares with the previous section, one
sees that,  in the present scheme,
the distinction between intrinsic and extrinsic geometries
is not so clear anymore,  however we shall soon show  that this
separation is not invariant by W transformations.

%
%

\subsection{Connection with the WZNW model}
\label{2.3}

\subsubsection{ \bf Preamble}
\label{preamble}
It has been shown\cite{Dublin} that there is a deep connection between
Toda equations and the so-called conformally reduced WZNW equations.
In this section we show how the latter, which contain more
degrees  of freedom than the former, are  directly related to
the present W-geometry. For completeness, we first recall the
\begin{definition}{\bf Conformally reduced
A$_{\hbox{\bf n}}$-WZNW model.}
\label{CWZNW}

Let $z$ and $\zb$ be Minowski surface-parameters,  let $\theta(z,\zb)$
be a $(n+1)\times (n+1)$ real matrix of determinant one, and
\begin{equation}
\label{2.3.0}
\cJ\equiv \theta^{-1}\partial\theta,\qquad
\cJb\equiv( \partialb \theta)\theta^{-1}.
\eeq
The conformally reduced WZNW equations are
\beq
\label{2.3.WZNW}
\partialb \cJ=\partial \cJb=0,
\eeq
\beq
\label{2.3.CR}
tr\left (\cJ E_{-\alpha}\right )=\mu_\alpha,\quad
tr\left (\cJb E_{\alpha}\right )={\bar \mu}_\alpha,
\eeq
 $\alpha$ runs over a set of positive roots. The parameters
$\mu_\alpha={\bar \mu}_\alpha=-1$ if $\alpha$ is primitive\footnote{
There may be arbitary constants for primitive roots, but the present
restricted choice is sufficient for our purpose.},
and  vanish otherwise.
\end{definition}
Of course there is a similar definition for complex $z$ obtained
by Wick's rotation.  We shall actually need the
following  generalization:
\begin{definition}{\bf Conformally reduced gl(n+1)-WZNW model.}
\label{CgWZNW}

Same as above, but the determinant of $\theta$ is arbitrary.
\end{definition}
This generalization incorporates
an  additional $gl(1)$ gauge degree of freedom.
As it is well known the general solution of Eqs.\ref{2.3.WZNW}
is $\theta=\theta_L(\zb)\theta_R(z)$ where $\theta_L$ and
$\theta_R$ are arbitrary chiral matrices. Then  conditions
Eqs.\ref{2.3.CR} lead to solutions of the
\begin{definition}{\bf Drinfeld-Sokolov equations.}
\label{DSE}
They are  of the form\cite{DS}
\beqa
\label{2.3.DS}
&\partial \Upsilon_r(z) -\sum_{s=0}^{n}{\cal D}_{rs}(z)\Upsilon_s(z)=0
\nonumber \\
&{\cal D}_{r s}=0, \>\hbox{\rm for}\> \> s-r >1, \quad
{\cal D}_{r r+1}=1
\eeqa
If $tr({\cal D})=0$ this (DS) equation is  associated
with $A_n$. Otherwise it is associated with $gl(n+1)$.
\end{definition}
It is easy to see that,  writing $\Upsilon_r^{(\ell)}=
\theta_R^{-1}(z)_{r \ell}$, and ${\bar \Upsilon}_r^{(\ell)}=
\theta_L^{-1}(\zb)_{\ell r}$, give
$2n$ solutions of the Drinfeld-Sokolov equations just
defined.

\subsubsection{ WZNW dynamics for $\CP$-- W-surfaces}
 Our starting point is
Eq.\ref{2.2.9a}:
\begin{equation}
\label{2.3.1}
\eta_{i\jb}   =   \sum_{A=0}^n f^{(i)\,A}(z) \,
\fb^{(\jb)\, A}(\zb),
\end{equation}
 It is quite clear
from the start that the above matrix is of the form
$\eta=\eta_R(z)\eta_L(\zb)$, and thus satisfies  equations of
the WZNW type. More precisely, we have the
\begin{theorem}{\bf Conformally reduced WZNW solutions from W surfaces.}
\label{WZNWfromW}
The matrix
$\theta \equiv \eta^{-1}$
is a solution of the conformally reduced $gl(n+1)$ WZNW equations
introduced by
the definition \ref{CgWZNW}.
\end{theorem}
\proof
The currents are given  by:
\begin{equation}
\label{2.3.2}
\partial \eta_{i\bar j}=-\cJ_{ik}\, \eta_{k\bar j},\quad
\partial \eta_{i\bar j}=-\eta_{i\bar k}\cJb_{\kb \jb}.
\end{equation}
They obviously satisfy
\begin{equation}
\label{2.3.3}
\partialb \cJ_{ik}= \partial \cJb_{\ib \kb}=0.
\end{equation}
Moreover, since by construction, $\partial f^{(i)A}=f^{(i+1)A}$, and
$\partialb \fb^{(i)A}=\fb^{(i+1)A}$,  one has, according
to Eq.\ref{2.3.1},
$\cJ_{i\, j}=-\delta_{j,\,i+1}$,
and $\cJb_{j\,  i}=-\delta_{j,\,i+1}$
for $i\leq n-1$, and this completes the derivation. \qed

  Concerning $\cJ_{n j}$,
it is well known that the $n+1$ functions $f^A$ are automatically
solutions of the differential equation ($A=0,\, \cdots,\, n$)
\begin{equation}
\label{2.3.4}
\left | \begin{array}{cccc}
f^0 & \cdots  & f^n & f^A \\
\vdots    &   & \vdots      & \vdots  \\
f^{(n+1)\, 0}  & \cdots  & f^{(n+1)\, n+1}  & f^{(n+1)\, A}
\end{array}
\right |
\equiv \left\{\sum_{k=0}^{n+1} U_{n+1-k}\partial^{(k)}\right \} f^A=0,
\end{equation}
which allows us to write
\begin{equation}
\label{2.3.5}
\partial^{(n+1)}f^A=\sum_{k=0}^n {U_{n-k}\over U_{0}}f^{(k)A}
\equiv \sum_{k=0}^n \lambda_k f^{(k)A},
\end{equation}
since $ U_{0}$, which is equal to the
Wronskian of the $n+1$ functions
$f^A$, does not vanish at regular generic points. Thus
one sees that $\cJ_{n j}=\lambda_j$ and $\cJb_{j n}=\lambdab_j$,
where $\lambdab_j$ is related to the differential equation satisfied
by the functions $\fb$. $\cJ$ and $\cJb$ are finally
given by
\begin{equation}
\label{2.3.6}
\cJb=-\Ib-\lambdab, \qquad \cJ=-I-\lambda,
\end{equation}
\begin{equation}
\label{2.3.7}
\Ib=\left (\begin{array}{ccccc}
0 & 0 & 0 &\cdots & 0 \\
1 & 0 & 0 &\cdots & 0 \\
0 & 1 & 0 &\cdots & 0 \\
\vdots & \, & \ddots &\ddots  & \vdots \\
0 & \,  & 0 &1 & 0 \end{array}
\right ), \>
I =\left (\begin{array}{ccccc}
0 & 1 & 0 &\cdots & 0 \\
0 & 0 & 1 &\cdots & 0 \\
\vdots & \,&\ddots & \ddots  & \vdots \\
0 & \cdots  & \, &0 & 1 \\
0 & \cdots  & \, &0 & 0\end{array}
\right )
\end{equation}
\begin{equation}
\label{2.3.8}
\lambdab=\left (\begin{array}{ccccc}
0 & 0 & 0 &\cdots & \lambdab_0 \\
0 & 0 & 0 &\cdots & \lambdab_1\\
\vdots & \, & \ddots &\,  & \vdots \\
0 & \, & \cdots &0  & \lambdab_{n-1}\\
0 & 0  & \cdots &0 &  \lambdab_n
 \end{array}
\right ),\>
\lambda=\left (\begin{array}{ccccc}
0 & 0 & 0 &\cdots & 0 \\
0 & 0 & 0 &\cdots & 0\\
\vdots & \, & \ddots &\,  & \vdots \\
0 & \, & \cdots &0  & 0\\
\lambda_0 & \lambda_1  & \cdots &\lambda_{n-1} &  \lambda_n
 \end{array}
\right ).
\end{equation}
The current $\cJ$ and $\cJb$ may be expressed in terms of the generators
introduced in section 2.1 (see Eqs.\ref{2.1.22}, \ref{2.1.23})
except that, here,
$i$ runs form zero to n, and we are dealing with $gl(n+1)$.
In this section we are always in  the sector $N\equiv \sum_{i=0}^n
\flat_i^+\flat_i=1$, that is, in the defining representation,  where
$h_i$ and $E_{\pm j}$ are $(n+1)\times (n+1)$ matrices. We  keep the
same  notation as in section \ref{2.1}  for the generators
since there may be no confusion.
\begin{corollary}{\bf Embedding functions as solutions
of DS equations.}
\label{DSfromW}

For  given $A$ and $\Ab$, the set of derivatives
of the embedding  functions
$\{f^{(j)\, A}, j=0, \cdots, n\}$ , and
$\{\fb^{(\jb)\, \Ab},  \jb=0, \cdots, n\}$, are solutions
of the $gl(n+1)$ DS  equations introduced in the definition \ref{DSE}.
\end{corollary}
\proof
It immediately  follows from Eqs.\ref{2.3.2}, and \ref{2.3.3}
that we have
\begin{equation}
\label{2.3.11DS}
\partial f^{(j)\, A}+\sum_k {\cal J}_{jk}f^{(k)\, A}=0,\quad
\partialb \fb^{(\jb)\, \Ab}+\sum_k {\bar{\cal J}}_{\kb \jb}f^{(\kb)\, \Ab}=0
\end{equation}
which coincides  with the Drinfeld-Sokolov equations  associated
with
 $gl(n+1)$ gauge introduced by the definition \ref{DSE}. \qed

One should note, however, that  the currents associated with
the embedding functions
are  of a more restricted type, since  there are many more vanishing
matrix elements in
Eqs.\ref{2.3.7}, and \ref{2.3.8} than
required  by the definition \ref{DSE}.
In this connection,
Eq.\ref{2.3.4}, and its anti-chiral counterpart
show that $U_\ell$ and
$\bar U_\ell$ should be regarded as W-potentials.
The fact that $U_0$ and $\bar U_0$ are not constant,
of course,  reflects the existence of the additional
$gl(1)$ degree of freedom. Accordingly, the current satisfy the
$gl(n+1)$ DS equations, and not the one related with $A_n$.

Next we give the geometrical interpretation of the degrees of
freedom that appear in the conformally reduced WZNW dynamics,
and not in the Toda equation. This results from the
\begin{theorem}{\bf Gauss decomposition from moving frame.}
\label{GFS}
The moving-frame equations
\begin{equation}
\label{2.3.11a}
\bfet_a=\sum_{b\leq a} C_{ab}(z, \zb) \>
\sqrt{{\tau_a\over \tau_{a+1}}} \> \bff^{(b)}(z), \quad
\hbox{with}\> C_{aa}=1,
\end{equation}
\begin{equation}
\label{2.3.11b}
\bbfet_a=\sum_{b\leq a} A_{ba}(z,\zb)\>
\sqrt{{\tau_a\over \tau_{a+1}}}\>  \bbff^{(b)}(\zb), \quad
\hbox{with}\> A_{aa}=1.
\end{equation}
are such that the matrix $\theta=\eta^{-1}$ has the Gauss decomposition
\beq
\label{2.3.14}
\theta_{rs}  =\sum_{a=0}^n \sum_{b=0}^n A_{ra} B_{ab} C_{bs}.
\end{equation}
Denote by ${\cal N}_{+}$ (resp. ${\cal N}_{-}$)
 the  sub-groups
generated by the step operators associated with
positive (resp. negative) roots,
and by ${\cal D}_0$ the group  generated by the
Cartan generators{ \bf including } $h_n$. Then
\beq
A\in {\cal N}_+,\quad  B\in {\cal D}_0, \quad
C\in {\cal N}_-.
\eeq
\end{theorem}
\proof  Eqs.\ref{2.3.11a}, and \ref{2.3.11b} are slight
modifications of Eqs.\ref{2.1.10} and \ref{2.1.11}.
Denote by $\varphi_{\pm}$ the set
of positive (resp. negative ) roots of $A_n$. The matrices $C_{ab}$
(resp. $A_{ab}$)  vanish unless $a\geq b$ (resp. $a\leq b$), and
their diagonal matrix elements are equal to $1$. Thus we may write
\begin{equation}
\label{2.3.12}
C=\exp
\left ( \sum_{\alpha \in \varphi_+} y^\alpha E_{-\alpha} \right ),
\quad
A=\exp
\left ( \sum_{\alpha \in \varphi_+} x^\alpha E_{\alpha}\right ).
\end{equation}
Eqs.\ref{2.3.11a},  \ref{2.3.11b} give
\begin{equation}
\label{2.3.13}
\bff^{(a)}(z) =\sum_{b} C^{-1}_{ab}(z, \zb) \>
\sqrt{{\tau_{b+1}\over \tau_b}} \> \bfet_b,
\quad
\bbff^{(a)}(\zb) =\sum_{b} A^{-1}_{ba}(z, \zb) \>
\sqrt{{\tau_{b+1}\over \tau_b}} \> \bbfet_b.
\end{equation}
Inserting this
into Eq.\ref{2.3.1}, one indeed verifies the decomposition
Eq.\ref{2.3.14}, if one lets
\begin{equation}
\label{2.3.15}
B_{ab}= {\tau_{a}\over \tau_{a+1}} \delta_{a,\, b}
=e^{\p_{a+1}-\p_{a}}  \delta_{a,\, b}.
\end{equation}
This completes the proof. \qed

In terms of the Cartan generators,  $B$ may be written as
\begin{equation}
\label{2.3.16}
B=\exp \left ( \sum_{i=0}^n \p_{i+1} h_i \right ).
\end{equation}
 In the Toda theory, the only
remaining degrees of freedom are the Toda fields $\p_i$. Thus we
see that  the geometrical interpretation of the matrices $A$ and $C$
is that they  specify the embedding.

\begin{corollary}{\bf Geometrical interpretation of
the WZNW equations.}
\label{GeomWZNW}
The WZNW equations for $A$, $B$, and $C$\cite{Dublin}, that is
\begin{equation}
\label{2.3.17}
(\partialb C ) C^{-1}=-B^{-1} \Ib B,\quad
A^{-1} \partial A= -B I B^{-1},
\end{equation}
are   direct consequences of the chirality conditions of the
W surface (Cauchy-Riemann equations): $\partial \fb^A=\partialb f^A=0$.
\end{corollary}
\proof Consider the last equation as an example. Eq.\ref{2.3.13}
gives
\begin{equation}
\label{2.3.18}
\sum_b \left ( (\partialb C^{-1}_{ab}) \bfe_b \sqrt{\tau_{b+1}
\over \tau_b}\right )+
\sum_b \left (  C^{-1}_{ab} \partialb \left (\bfe_b \sqrt{\tau_{b+1}
\over \tau_b}\right )\right )=0.
\end{equation}
It follows from Eqs.\ref{2.2.21} that
\begin{equation}
\label{2.3.19}
\partialb \left (\bfe_b \sqrt{\tau_{b+1}
\over \tau_b}\right )=e^{K_{b-1 j} \p_j}
\left (\bfe_{b-1} \sqrt{\tau_{b}
\over \tau_{b-1}} \right ),
\end{equation}
and  the previous equation gives $-(\partialb C ) C^{-1}
=\sum_{j=1}^n E_{-j}
\exp( \sum_iK^{gl(n+1)}_{j-1 i} \p_i)$ which is indeed equal
to $B^{-1} \Ib B$, according to Eqs.\ref{2.1.24}.
Treating similary, the equation for $\bbff$, we get
\begin{equation}
\label{2.3.17a}
(\partialb C ) C^{-1}=-B^{-1} \Ib B,\quad
A^{-1} \partial A= -B I B^{-1},
\end{equation}
which coincide\cite{Dublin} with the WZNW equations for $A$, $B$
and $C$. \qed

Our next topic is concerned with the additional field
$\p_{n+1}$, and the associated $gl(1)$ gauge invariance.
We shall prove the
\begin{proposition}{\bf gl(1)  invariance of the
gl(n+1) WZNW equations.}

\label{g1WZNW}
Given two arbitrary functions $\rho(z)$ and $\rhob(\zb)$, the
$gl(n+1)$ WZNW equations are invariant under the transformation
\beq
\label{2.3.gg1}
A\rightarrow \rhob \Lambdab^{-1} A, \quad
C\rightarrow \rho C \Lambda^{-1}, \quad
B\rightarrow B \bigl / \rho \rhob.
\eeq
where $\Lambda$ and $\Lambdab$ are given by Eq.\ref{2.2.21l}, that is
\beq
\label{2.3.gg2}
\Lambda_{a b}={a\choose b}
\partial^{(a-b)}\rho, \quad
\Lambdab_{b a}={a\choose b}
\partialb^{(a-b)}\rhob.
\eeq
\end{proposition}
\proof Clearly  $\left ( \partialb  C\right )  C^{-1}$ and
$A^{-1} \partial A$ are invariant. This is also trivially true
for the right-hand sides. \qed
 As a
 consequence we have the
\begin{proposition}{\bf WZNW Gauge equivalence.}
\label{ggeqv}

The solutions of the  $gl(n+1)$-WZNW equations are
gauge-equivalent to the  ones of the
$A_n$- WZNW equations.

\end{proposition}
\proof Since $\theta$ is a solution of the WZNW equations his
determinant is a product of chiral functions. Thus it may be set
equal to one by a $gl(1)$ transformation of the type introduced
by proposition \ref{g1WZNW}.
\qed

  Concerning the DS equations, the following is useful
\begin{lemma}{\bf Basic properties of the transformations $\Lambda$.}
\label{Lambda}

\noindent 1) For an  infinitesimal transformation
$\rho=1+\epsilon$,
$\Lambda=1+s$. One has
\begin{equation}
\label{2.3.28}
<b| \left ( \partial s +\left [ s,\, I \right ] \right ) | a >
= \left\{ \begin{array}{ccc}
0, & \hbox{if} & b<n \\
{n+1 \choose a} \partial^{n+1-a} \epsilon,  & \hbox{if} & b=n.
\end{array} \right.
\end{equation}
\noindent 2) The form (Eqs.\ref{2.3.6}, \ref{2.3.7}, \ref{2.3.8})
of the current $\cJ$ is gauge invariant.
\end{lemma}
\proof 1)  The derivation uses easy calculations based
on the standard recursion relation
for binomial coefficients:
${b+1\choose a}-{b\choose a-1}={b\choose a}$.

\noindent 2)
 It is sufficient to work with  infinitesimal transformations. Then
\begin{equation}
\label{2.3.30}
\cJ\rightarrow \cJ+\delta \cJ \equiv
\cJ+\partial s +\left [ s,\, I \right ] + \left [ s,\, \lambda \right ].
\end{equation}
In terms of the fermionic modes $\flat_j$, we may write
$\lambda =\sum_b \flat_n^+\flat_b \lambda_b$, and
\begin{equation}
\label{2.3.31}
\left [ s,\, \lambda \right ]= -
\sum_{a=0}^{n-1} \flat_n^+\flat_a \sum_{b\geq a} {b \choose a} \lambda_b
\partial^{b-a}\epsilon.
\end{equation}
\begin{equation}
\label{2.3.32}
\delta \cJ=
 \sum_{a=0}^{n-1} \flat_n^+\flat_a \left [
{n+1 \choose a} \partial^{n+1-a} \epsilon-
\sum_{b\geq a} {b \choose a} \lambda_b
\partial^{b-a}\epsilon\right ].
\end{equation}
Thus the form of $\cJ$ is indeed preserved with
\begin{equation}
\label{2.3.33}
\delta \lambda_a= {n+1 \choose a} \partial^{n+1-a} \epsilon-
\sum_{b\geq a} {b \choose a} \lambda_b
\partial^{b-a}\epsilon.
\end{equation}
This completes the proof. {\bf Q.E.D.}\newline
This lemma leads to the
\begin{proposition}{\bf DS gauge equivalence. }
\label{DSeqv}

The solutions of the $gl(n+1)$-DS equations are
gauge equivalent to those of the  $A_n$-DS  equations.
\end{proposition}
\proof This must be true since the corresponding WZNW are gauge
equivalent. Indeed it has been shown in ref.\cite{DS} that a
general DS current $\cal D$ of definition \ref{DSE} is
gauge equivalent to a current $\cJ$ of the form Eq.\ref{2.3.6},
\ref{2.3.7}, \ref{2.3.8}.  According to the
last lemma, one may thus perform  a $gl(1)$ transformation
such that $\lambda_n\rightarrow 0$, and  $h_n$ decouples.
\qed

The  $gl(1)$ invariance is
directly connected with the rescaling of $\CP$ W surfaces as shown by the
\begin{proposition}{\bf Rescaling. }
\label{rescaling}
The $gl(1)$ gauge transformation Eqs.\ref{2.3.gg1} corresponds to the
rescaling  Eq.\ref{2.2.16} of the embedding functions
\begin{equation}
\label{2.3.20}
f^A(z)\rightarrow \rho(z)f^A(z),\quad
\fb^{\Ab}(\zb) \rightarrow \rhob(\zb)\fb^{\Ab}(\zb).
\end{equation}
\end{proposition}
\proof Combining Eqs.\ref{2.2.20} with
Eqs.\ref{2.3.13}, one verifies Eqs.\ref{2.3.gg1} for $A$ and $C$.
Substituting Eqs.\ref{2.2.20} into Eq.\ref{2.3.16},
gives $B\to B\exp \left( -\ln (\rho \rhob)
\sum_{i=0}^n (i+1)h_i \right )$.
Making use of the explicit expressions Eqs.\ref{2.1.22}, \ref{2.1.23},
one verifies that $\sum_{i=0}^n (i+1)h_i $ is equal to the
identity operator, and this completes the proof.   \qed
 Of course the additional generator $h_n$ is
instrumental in the proof.
Finally we arrive at the

\begin{theorem}{\bf Equivalence between WZNW solutions
 and  $\CP$ W-surfaces.}

There exists a one-to-one correspondence  betweeen
 the solutions of the conformally reduced $A_n$ WZNW
 and  the W-surfaces in $\CP$.
\end{theorem}
\proof 1) First it is clear from the previous discussions that
there exists
a unique solution of the conformally reduced WZNW equations
associated with a given W-surface in $\CP$. Indeed
a W surface in $\CP$ is described by the homogeneous
formalism displayed in section \ref{2.2}, and there is a one-to-one
correspondence between the local rescaling of the homogeneous
description of $\CP$ and the $gl(1)$-gauge of the $gl(n+1)$
WZNW and DS equations.

\noindent 2) The proof of the converse goes as follows.
  Let $\theta$ be a
solution of the $A_n$--WZNW-model:
$\partial \theta=
\theta K$, $\partialb \theta = \Kb \theta$, such that,
for $\alpha\in \varphi_+$,
$-tr \left ( K E_{-\alpha} \right)$, and
$-tr \left ( \Kb E_{\alpha} \right)$ are equal to $1$,  if $\alpha$ is
primitive, and $0$ otherwise.  For the solution
\begin{equation}
\label{2.3.34}
\theta=\theta_{L}(\zb) \theta_R(z), \quad
K=\theta_R^{-1} \partial \theta_R, \quad
\Kb = \left ( \partialb \theta_L\right ) \theta_L^{-1}.
\end{equation}
the conditions     on $K$ and $\Kb$  are  left invariant by the gauge
transformations $\theta_L\to \alpha_L \theta_L$, and $ \theta_R \to
\theta_R \alpha_R$ such that $\alpha_L$ (resp. $\alpha_R$) belong to
 subgroups ${\cal N}_+$ (resp. ${\cal N}_-$).
One may verify \cite{DS} that there exist gauge tranformations
such that the gauge transformed  current $\cJ=\alpha_R^{-1}\partial
\alpha_R+ \alpha_R^{-1}K \alpha_R$, and
$\cJb =\left (\partialb \alpha_L\right ) \alpha_L^{-1}+
\alpha_L \Kb \alpha_L^{-1}$, take the form Eqs.\ref{2.3.6},
\ref{2.3.7},
\ref{2.3.8}. Of course, since $K$ and $\Kb$ belong to the Lie algebra
$A_n$, $\lambda_n$ and $\lambdab_n$ are found to vanish. Letting
$\eta=\theta^{-1}$, one sees that we have
$\eta=\eta_R(z) \eta_L(\zb)$ with
\begin{eqnarray}
\label{2.3.35}
\partial \eta_{R\, kl}&=\eta_{R\, k+1\, l},\> \hbox{for}\> k\leq n-1,\quad
\partial \eta_{R\, nl}&=\sum_{b=0}^{n-1} \lambda_b \eta_{R\, b l},
\nonumber \\
\partialb \eta_{R\, kl}&=\eta_{R\, k\, l+1},\> \hbox{for}\> l\leq n-1,\quad
\partialb \eta_{R\, kn}&=\sum_{b=0}^{n-1} \lambdab_b \eta_{R\, k b},
\end{eqnarray}
as consequences of the Drinfeld-Sokolov equations.
Thus we let
\begin{equation}
\label{2.3.36}
\eta_{R\, 0A}= {\widetilde f}^A, \qquad
\eta_{L\, A0}= {\widetilde {\fb}}^A.
\end{equation}
Eqs.\ref{2.3.35} are satisfied iff
\begin{equation}
\label{2.3.37}
\partial^{n+1}{\widetilde f}= \sum_{b=0}^{n-1} \lambda_b
\partial^{(b)} {\widetilde f}, \quad
\partialb^{n+1}{\widetilde {\fb}}= \sum_{b=0}^{n-1} \lambdab_b
\partial^{(b)} {\widetilde {\fb}},
\end{equation}
which means that the Wronskians of ${\widetilde f}$, and of
${\widetilde {\fb}}$ are constant. These conditions  are easily
removed by performing the rescalings
${\widetilde f}\to f=\rho {\widetilde f}$, ${\widetilde {\fb}}
\to \rhob \fb$, which are such that the new Wronskians are equal
to $\rho^{n+1}$, and $\rhob^{n+1}$, respectively. The $f$ and $\fb$
are coordinates of a W-surface with local rescaling  invariance.
Writing the Gauss decomposition $g=C^{-1}B^{-1} A^{-1}$ gives back the
Frenet-Serret formula, and this establishes the complete
correspondence between $\CP$--W-surfaces and the conformally reduced
$A_n$--WZNW-dynamics. {\bf Q.E.D.}

\section{ KP coordinates and W-geometry}
\markright{3. KP coordinates}
\label{3}
\subsection{Free-Fermion Description of Chiral Embedding}
\label{3.1}
In the construction of the moving frame, we have seen that
the determinants of the embedding functions play a
central r\^ole.  This fact leads us to suspect that
some kind of fermionic structure underlies the geometry of
 W-surfaces.  In the description of the Toda theory,
it is known\cite{M} that the free-fermions neatly describe their
solution-space as they do for the KP hierarchy.
In our situation, the embedding is connected to  the $A_n$
Toda theory,  and  the corresponding fermion theory
becomes non-relativistic in contrast with  the KP case.
The present free fermions are  identical to those which
appear in the matrix-model (more about this in appendix \ref{A.3}).
Although the present main section is devoted to the case of regular
points, this subsection
 deals  with the more general situation which we will encounter
in the next main section.

Let us summarize our free-fermion conventions
following\footnote{We actually make  a
slight modification by  interchanging  $\psi$ and $\psis$
to follow the usual convention.}  ref. \cite{M}.
\begin{eqnarray}
\left[ \psi_n, \psi_m \right]_+ & =
& \left[\psi^+_n, \psi^+_m\right]_+ = 0,\nonumber\\
\left[\psi_n, \psi^+_m\right]_+ & = & \delta_{n,m},
\qquad (~n,m~=~0,~1,~\cdots)
\label{3.1.1}\\
\psi_n\ket{\emptyset}  =  0, &\qquad&
\bra{\emptyset}\psis_n  =  0  \qquad \forall n.\label{3.1.2}
\end{eqnarray}
We use the semi-infinite indices $n=0,1,2,\cdots, \infty$ for
the fermion-operators.
The vacuum states $\ket{\emptyset}$ and $\bra{\emptyset}$
correspond to the
no-particle states.  The $n$--particle ground state is
 created from
them in the standard way:
\begin{equation}
\ket{n}  =  \psis_{n-1}\psis_{n-2}\cdots\psis_0\ket{\emptyset},
\quad
\bra{n}  =  \bra{\emptyset}
\psi_0\psi_1\cdots\psi_{n-1}.\label{3.1.3}
\end{equation}
The  current operators,
\begin{equation}
J_n = \sum_{s=0}^\infty \psis_{n+s}\psi_s,  \qquad
\bar{J}_n = \sum_{s=0}^\infty \psis_s \psi_{n+s}, \label{3.1.4}
\end{equation}
will be taken as  Hamiltonians as one  does  for  the
KP hierarchy\cite{DJKM}
(W-parameters) already mentionned in section \ref{2.1}.
The r\^ole   of these fermions may be understood as follows.
Take the case where $z$ is a complex variable. Then
the embedding functions $f^A$ are analytic, and  each  of them is
entirely determined
by its  Taylor expansion around a single point of its analyticity
domain.  Its belaviour at any other point of its  Riemann surface is
fixed by analytic continuation. The following  free-fermion formalism
realizes this continuation automatically. Consider the Taylor  expansions
at the point $z$:
\beq
f^A(z+x) = \sum_{s=0}^\infty
f^{(s)A}\! (z)\, {x^s\over s!},\quad
\fb^{\Ab}(\zb+{\bar x}) = \sum_{s=0}^\infty
\fb^{(s)A} \! (\zb)\,
{{\bar x}^s\over s!}.
\label{3.1.5}
\eeq
To these developements,  we associate the free-fermion operators,
\beq\label{3.1.6}
\psifz{A}{z} = \sum_{s=0}^\infty f^{(s)A}\! (z)\> \psi_s ,\quad
\psifsz{\Ab}{\zb} = \sum_{s=0}^\infty \fb^{(s)A}\!  (\zb)\>
\psis_s.
\eeq

The basic property of these  operators are
\begin{proposition}{\bf Fermionic representation of chiral
functions.}

\noindent 1) Any change of the Taylor-expansion point $z$, $\zb$
 can be absorbed by
the action of the Hamiltonians $J_1$, and $\Jb_1$.
In particular, one has
\beq\label{3.1.7a}
\psifz{A}{z}=e^{-J_1z}\>\psifz{A}{0}\> e^{J_1z}\quad
\psifsz{\Ab}{\zb}=e^{\Jb_1\zb}\>\psifsz{\Ab}{0}\> e^{-\Jb_1\zb}.
\eeq
2) The embedding functions are represented by the fermion
 expectation-values
\beq
\label{3.1.7}
f^A(z) = \bra{\emptyset}\psifz{A}{z_0} e^{J_1(z-z_0)}\ket{1},\qquad
\fb^{\Ab}(\zb) = \bra{1}e^{\Jb_1(\zb-\zb_0)} \psifsz{\Ab}{\zb_0}
\ket{\emptyset}.
\eeq
\end{proposition}
\proof
1) is a consequence of the identities,
$e^{-J_1z}\psi_s e^{J_1z}=\sum_{t=0}^s
z^t\psi_{s-t}/t!$ and of their  anti-chiral counterparts.

\noindent 2) comes from the relations
$\bra{\emptyset}\psi_s J_1^t\ket{\emptyset}
=\delta_{s,t}$.  \qed

\noindent Due to 1),  it is equivalent to work with $\psifz{A}{z_0}$, and
$\psifsz{\Ab}{z_0}$ at any fixed
$z_0$ and $\zb_0$.   Hence we put $z_0=\zb_0=0$ in the following
and write $\psifz{A}{0}$
and $\psifsz{\Ab}{0}$ as $\psif{A}$ and $\psifs{\Ab}$
for simplicity.  2) implies that we can translate the
chiral embedding into $\CP$ in the free fermion language.
The basic object of this approach is the
\begin{definition}{\bf Embedding operator.}
\label{Embop}
It is an operator in the fermionic Fock space defined by
\beq
\label{3.1.8}
\begin{array}{l}
\EM(z,\zb) = {\displaystyle\sum_{a=0}^{n+1}}
\EM(z,\zb)_a \\
\equiv
{\displaystyle\sum_{a=0}^{n+1}\sum_{0\leq A_1<\cdots<A_a \leq n}}
\psifsz{\Ab_1}{\zb}\cdots\psifsz{\Ab_a}{\zb}\ket{\emptyset}
\bra{\emptyset}\psifz{A_a}{z}\cdots\psifz{A_1}{z}.
\end{array}
\eeq
\end{definition}
It clearly follows from Eq.\ref{3.1.7a} that
\beq
\label{3.1.8a}
\EM(z,\zb)=\ejzb\EM(0,0) \ejz.
\eeq
$\EM$  is a sort of density matrix of the
embeddding functions.
It  has a natural restriction to
 the Fock space generated by the operators $\psif{A}$ and
$\psifs{\Ab}$ acting on the no-particle state, where it becomes
a finite matrix. Thus  $a$ may be regarded as specifying a
representation of $gl(n)$
  (note  that
$\EM_a=0$ if $a>n+1$).
This is the analogue of the $gl(\infty)$ matrix which appears in the
Toda hierarchy.  The main difference between these two is that
the rank of $\EM$ in Eq.\ref{3.1.8} is finite ($=n+1$),
i.e. it is degenerate.  If we take the limit $n\rightarrow\infty$,
it coincides with the matrix of the Toda hierarchy.
In the appendix A.3, we shall show that,
in this limit, the embedding problem is
equivalent to the orthogonal-polynomial method
of the multi-matrix model.

The readers might be curious about the relationship between
these ferm\-ions and the operators $\flat_j$ introduced
in section \ref{2.1}.
They obviously act on different indices of $f^{(s)A}$.
$\flat$-fermions
act on $A$ and $\psi$-fermions act on $s$.  There is a complicated
relation between the two, which is connected with the uniformization
of the Drinfeld-Sokolov equation.
We shall give explicit forms of these transformations in the
proof of the Hirota equation, for example,
see Eqs.\ref{3.1.18},\ref{3.1.19}.
This connection is simple in the limit
when $n\rightarrow \infty$ since, then,  we can choose the basis,
$f^{(s)A} \propto \delta_{A,s}$.

In order to get the relationship  between $\EM$
 and the embedding considered in section \ref{2},
we need the following
 \begin{theorem}{\bf Tau-functions.}
\label{taufunctions}
One has
\beq\label{3.1.9}
 \tau_a=\bra{a} \EM(z,\zb)\ket{a},
\eeq
that is,
the tau-functions  associated with  $\EM$
coincide with the functions $\tau_a$ defined by Eq.\ref{2.2.9}.
\end{theorem}
\proof It is easy to verify that, from Eq.\ref{3.1.7},
\beq
\eta_{j \kb}=\bra{\emptyset}\psi_{\kb} \EM(z,\zb)
\psis_j\ket{\emptyset}.
\end{equation}
The theorem follows by
computing the determinants of Eq.\ref{2.2.9},
by means of  Wick's theorem. \qed

\noindent These are non-chiral versions of the tau-functions of
the KP hierarchy.

Next  the basic tool
of the fermionic approach is the
\begin{theorem}{\bf Hirota equation\footnote{
The rest of this subsection is solely devoted to
the proof of this theorem.  Those who are familiar with
the Hirota equation, or do not bother about its proof  may
 simply skip it.}}
\label{Hirota}
The embedding operator $\EM$ satisfies
\beq\label{3.1.13}
\sum_{s=0}^\infty \psis_s\EM(z,\zb)\bigotimes \psi_s\EM(z,\zb)
= \sum_{s=0}^\infty\EM(z,\zb)\psis_s\bigotimes \EM(z,\zb)\psi_s.\eeq

\end{theorem}
\proof
First we remark that $z$ and $\zb$ can be
set equal to zero in order to
prove this theorem.  Indeed,  the explicit
form of the embedding operator $\EM$ Eqs.\ref{3.1.8},
and \ref{3.1.8a} implies that
the $z$ and $\zb$ dependence
can be eliminated by a suitable  re-definition of the
Taylor-expansion point.
Since our  proof is valid  for any
 $\psif{~}$ and $\psifs{~}$, and is carried out at the
level of formal series, it
automatically includes this modification, even if we
deal with $\EMO$ as we shall do.
The derivation is carried out  step by step.
The simplest situation
is studied first, before being gradually generalized.
\begin{situation}{}
\label{Situation1}
Assume that the embedding functions are given by ($ A, \Ab=0\cdots n$)
\beq
f^A(z) = {z^A}/{A!}
\quad
\fb^{\Ab}(\zb) = {\zb^{\Ab}}/{\Ab!}.
\label{3.1.14}
\eeq
\end{situation}
\proof
In this case, $\psif{A}  =  \psi_A$ and
$\psifs{\Ab}  =  \psis_{\Ab}$.
By a direct computation, we can easily confirm that
\beqa
\label{3.1.15}
\sum_{s=0}^\infty \psis_s\EMO_a\bigotimes \psi_s\EMO_b
&= \sum_{s=0}^\infty\EMO_{a+1}\psis_s\bigotimes \EMO_{b-1}\psi_s\noindent\\
&= \sum_{0\leq i \leq n}
      \psis_{i}\EMO_a
 \bigotimes \EMO_{b-1}\psi_{i}
\eeqa
\qed

We recall that,  in this simplest
case, the first $n+1$ fermion-operators
$\psif{A}$ and $\psifs{\Ab}$
coincide with the $\flat$ fermions, as already mentionned.

\begin{situation}{}
\label{Situation2:}
The case when the fermionic representation of
the embedding functions have the following form,
\beq
\psif{A}  =  \psi_{A} + \sum_{s=n+1}^\infty
f^{(s)A}\psi_s,\quad\psifs{\Ab}  =
\psis_{\Ab}+\sum_{s=n+1}^\infty \fb^{(s)\Ab}\psis_s.
\label{3.1.16}
\eeq
This is the canonical form of the embedding function at the regular
points.
\end{situation}
\proof
The idea is  to make a Bogoliubov transformation of the
free-fermion basis such that the problem reduces to Situation 1.
In doing so, we need to keep the orthonormality  properties of the
free-fermion basis.  We give the explicit form of such
transformation.  Introduce\footnote{Here, * does
not mean the complex
conjugation.}
\beqa
\pzero_\ell & =&  \left\{
      \begin{array}{ll}
       \psi_\ell + \sum_{s=n+1}^\infty
      f^{(s)\ell}\psi_s
       = \psif{\ell}& :\ell=0,1,\cdots,n\\
       \psi_\ell & :\ell>n
      \end{array}\right.
\nonumber\\
\pzeros_\ell & = & \left\{
      \begin{array}{ll}
       \psis_\ell & :\ell =0,1,\cdots,n\\
       \psis_\ell - \sum_{s=0}^n f^{(\ell)s}\psis_s
       & :\ell>n
      \end{array}\right.\label{3.1.17}\\
\pinftys_\ell &=&  \left\{
      \begin{array}{ll}
       \psis_\ell + \sum_{s=n+1}^\infty
      \fb^{(s)\ell}\psis_s= \psifs{\ell}
        & :\ell=0,1,\cdots,n\\
       \psis_\ell & :\ell>n
      \end{array}\right.
\nonumber\\
\pinfty_\ell &=&  \left\{
      \begin{array}{ll}
       \psi_\ell & :\ell=0,1,\cdots,n \\
       \psi_\ell - \sum_{s=0}^n \fb^{(\ell)s}\psi_s
        & :\ell>n
      \end{array}\right.\label{3.1.18}.
\eeqa
We remark that we need separate the Bogoliubov transformations for
the  chiral and anti-chiral embedding functions.  They are distinguished
by $(0)$ and $(\infty)$.
These new fermion-basis satisfy the standard anti-commutation relations,
\beqa
\left[\pzero_\ell, \pzeros_{\ell'}\right]_+ = \delta_{\ell\ell'},&
\left[\pzero_\ell, \pzero_{\ell'}\right]_+ = 0,&
\left[\pzeros_\ell, \pzeros_{\ell'}\right]_+ = 0,\nonumber\\
\left[\pinfty_\ell, \pinftys_{\ell'}\right]_+ = \delta_{\ell\ell'},&
\left[\pinfty_\ell, \pinfty_{\ell'}\right]_+ = 0,&
\left[\pinftys_\ell, \pinftys_{\ell'}\right]_+ = 0.
\label{3.1.19}
\eeqa
Furthermore, they keep the bilinear combinations
\beq\label{3.1.20}
\sum_{i=0}^\infty \psi_i\bigotimes\psis_i =
\sum_{i=0}^\infty \pzero_i\bigotimes\pzeros_i=
\sum_{i=0}^\infty \pinfty_i\bigotimes\pinftys_i.
\eeq
Due to these identities, the LHS and RHS of the Hirota equation
can be rewritten as
\beqa
\sum_{i=0}^\infty\psis_i\EMO_a\bigotimes\psi_i\EMO_b
&=& \sum_{i=0}^\infty\pinftys_i\EMO_a\bigotimes\pinfty_i\EMO_b,
\nonumber\\
 \sum_{i=0}^\infty\EMO_{a+1}\psis_i\bigotimes\EMO_{b-1}\psi_i
&=& \sum_{i=0}^\infty\EMO_{a+1}\pzeros_i\bigotimes\EMO_{b-1}\pinfty_i.
\nonumber \\ &
\label{3.1.21}
\eeqa
In terms of the new basis, the $\EMO$ matrix of \Eq{3.1.8}
becomes
\beq\label{3.1.22}
\EMO_a =\sum_{0\leq A_1<\cdots<A_a \leq n}
\pinftys_{\Ab_a}\cdots\pinftys_{\Ab_1}\ket{\emptyset}
\bra{\emptyset}\pzero_{A_1}\cdots\pzero_{A_a}.
\eeq
It is now clear that the problem reduces to
Situation 1.
\qed

\begin{situation}{}
\label{Situation 3}
Define the following sets of non-negative integers,
\beqa
\Xi_0 &=& \{ 0, 1+\beta_1, 2+\beta_1+\beta_2,
\cdots, n+\beta_1+\cdots+\beta_n\},\nonumber\\
\Xi_\infty &=& \{ 0, 1+\betab_1, 2+\betab_1+\betab_2,
\cdots, n+\betab_1+\cdots+\betab_n\},
\label{3.1.23}
\eeqa
where $\beta_i$ and $\betab_i$ ($i=1,2,\cdots, n$)
are two sets of non-negative integers. Define also
$\Xi_0^{(-)}$ and $\Xi_\infty^{(-)}$ as the set of non-negative
integers that do not  belong to
$\Xi_0$ and $\Xi_\infty$, respectively.
We also introduce two mappings $\sigma$ and
$\sigmab$ from the set $\{ 0,1,2,\cdots, n\}$
to $\Xi_0$ and $\Xi_\infty$,  respectively:
\beq\label{3.1.24}
\sigma(0) = \sigmab(0) =0,\quad
\sigma(\ell) = \ell+\sum_{j=1}^\ell\beta_j,\quad
\sigmab(\ell) = \ell+\sum_{j=1}^\ell\betab_j,
\eeq
for $i=1,\cdots, n$.
With these notations, we define the embedding
functions\footnote{
We remark that,  in this case,  the embedding becomes
 singular at $z=0$.  The geometrical property of such embeddings
will be the main topics of section 4.
The integers $\beta$, $\betab$
are called {\it ramification indices}.}
in this situation by,
\beq
f^A(z) = \frac{z^{\sigma(A)}}{\sigma(A)!}+
\sum_{s\in \Xi_0^{(-)}} \frac{f^{(s)A}}{s!}z^{s},
\quad
\fb^{\Ab}(\zb) = \frac{\zb^{\sigmab(\Ab)}}{{\sigmab(\Ab)}!}+
\sum_{s\in \Xi_\infty^{(-)} } \frac{\fb^{(s)\Ab}}{s!}\zb^{s}.
\label{3.1.25}
\eeq
\end{situation}
\proof
This problem reduces to situation 2 by
the mappings $\sigma$ and $\sigmab$.
More explicitly, we modify the definition of the
Bogoliubov-transformed fermions by,
\beqa
 \pzero_\ell & =&  \left\{
      \begin{array}{ll}
       \psi_\ell + \sum_{s\in\Xi_0^{(-)}}
      f^{(s)\sigma^{-1}(\ell)}\psi_s
       = \psif{\sigma^{-1}(\ell)}& \ell \in \Xi_0\\
       \psi_\ell & \ell \in \Xi_0^{(-)}
      \end{array}\right.
\nonumber\\
\pzeros_\ell & = & \left\{
      \begin{array}{ll}
       \psis_\ell & \ell \in\Xi_0\\
       \psis_\ell - \sum_{s=0}^n f^{(\ell)s}\psis_{\sigma(s)}
       & \ell \in\Xi_0^{(-)}
      \end{array}\right.\label{3.1.26}\\
\pinftys_\ell &=&  \left\{
      \begin{array}{ll}
       \psis_\ell + \sum_{s\in\Xi_0^{(-)}}
      \fb^{(s)\sigmab^{-1}(\ell)}
        \psis_s= \psifs{\sigmab^{-1}(\ell)}
        & \ell\in \Xi_\infty\\
       \psis_\ell & \ell \in \Xi_\infty^{(-)}
      \end{array}\right.
\nonumber\\
\pinfty_\ell &=&  \left\{
      \begin{array}{ll}
       \psi_\ell & \ell \in \Xi_\infty\\
       \psi_\ell - \sum_{s=\in \Xi_\infty^{(-)}}
\fb^{(\ell)s}\psi_{\sigmab(s)}
        & \ell \in \Xi_\infty^{(-)}
      \end{array}\right.\label{3.1.26a}.
\eeqa
These fermions are  orthonormal,
 and the embedding operator is given by
\beq
\label{3.1.27}
\EMO_a =\sum_{0\leq A_1<\cdots<A_a \leq n}
\pinftys_{\sigmab(\Ab_a)}\cdots
\pinftys_{\sigmab(\Ab_1)}\ket{\emptyset}
\bra{\emptyset}\pzero_{\sigma(A_1)}\cdots\pzero_{\sigma(A_a)}.
\eeq
In terms of this basis, both sides  of the Hirota equation produce
the following term,
\beq\label{3.1.28}
\sum_{0\leq i \leq n}
      \pinftys_{\sigmab(i)}\EMO_a
 \bigotimes \EMO_{b-1}\pzero_{\sigma(i)}.
\eeq
\qed

\begin{situation}{}
\label{Situation4: }
The proof of the Theorem, i.e. the general situation.
\end{situation}
\proof
By means of $(n+1)\times (n+1)$ constant matrices,
$S^{(0)}$ and $S^{(\infty)}$
$\ft^A = (S^{(0)-1})^A_B f^B$ and $
\ftb^{\Ab} = (S^{(\infty)-1})^{\Ab}_{\Bb}
 \fb^{\Bb}$, we can return to the previous normal form
Eqs.\ref{3.1.25} (more about this in section \ref{4.3}).
Denote by $\widetilde{\Psi}$
and $\widetilde{\Psi}^*$ the fermion basis
in terms of the normal form.  Introduce ${\Psi}$
and ${\Psi^*}$ which correspond to the original embedding functions
by\footnote{We omit the superscripts
$(0)$ and $(\infty)$ because
they are discussed in complete parallel. }
\beq\label{3.1.29}
\Psi_\ell = \sum_{\ell'=0}^n S_{\ell\ell'}\Psit_{\ell'},\quad
\Psi^*_\ell = \sum_{\ell'=0}^n \Psit_{\ell'}(S^{-1})_{\ell'\ell}.\eeq
for $\ell \in \Xi$ and $\Psi_\ell = \Psit_\ell$ ,
$\Psi^*_\ell = \Psist_\ell$ for $\ell \in \Xi^{(-)}$.
It is easy to check that the new $\Psi$- and $\Psi^*$-basis have
exactly the same properties as  the normal basis which we introduced
in  the
previous situation.
\qed

%
%

\subsection{W-parametrization from KP coordinates}
\label{3.2}

\begin{definition}{\bf W-transformations.}
\label{Wtrans}

A general infinitesimal W-transformation is a change of
embedding functions which takes the form
\beq
\label{3.2.1}
\delta_W f^A(z)=\sum_{j=0}^n w^j(z) \partial^{(j)} f^A(z),
\quad \delta_W \fb^{\Ab}(\zb)=\sum_{j=0}^n
\bar w^j(\zb) \partial^{(j)} \fb^{\Ab}(\zb),
\eeq
where $w^j(z)$, and $\bar w^j(\zb)$ are arbitrary functions
of one variable.
\end{definition}
This is the standard definition.
The purpose of this section is to introduce a special
class of parametrizations for  the target-manifold which is
such that these W-transformations of $\Sigma$ may be extended
as special types of diffeomorphisms of $\CP$.
\begin{definition}{\bf W-parametrizations of $\CC$. }
\label{Wparam}

Given a W-surface embedded into $\CC$,
the associated W-parameters of
the target space are
$n+1$ variables $z^{(0)}$, $z^{(1)}=z$, $z^{(2)}$, $\cdots$,
$z^{(n)}$,
noted  $[z]$, and $n+1$ variables $\zb^{(0)}$, $\zb^{(1)}=\zb$,
$ \zb^{(2)}$, $\cdots$, $\zb^{(n)}$,
noted  $[\zb]$.  The change of coordinates from $X^A$, $\Xb^{\Ab}$
to $[z]$,  $[\zb]$ is defined by
\beq
\label{3.2.2}
X^A=f^{A}([z]),\qquad \Xb^{\Ab}=\fb^{\Ab}([\zb])
\eeq
where
 $f^{A}([z])$, and
 $\fb^{\Ab}([\zb])$, are  the solutions of the
equations
\beq
\label{3.2.3}
\frac{\partial}{\partial z^{(\ell)}} f^A([z])
= \frac{\partial^\ell}{\partial z^\ell} f^A([z]),
\quad
\frac{\partialb}{\partial \zb^{(\ell)}} \fb^{\Ab}([\zb])
= \frac{\partialb^\ell}{\partial \zb^\ell} \fb^{\Ab}([\zb])
\eeq
with the initial conditions $f^A([z])= f^A(z)$
for  $z^{(0)},\ z^{(2)}, \cdots, z^{(n)}=0$, and
$\fb^{\Ab}([\zb])= \fb^{\Ab}(\zb)$
for  $\zb^{(0)},\ \zb^{(2)}, \cdots, \zb^{(n)}=0$.
\end{definition}
These coordinates  coincide with  the higher variables
of the KP hierarchy. Indeed,  their definition  is
most natural in  the free-fermion language,
where it is easy to see that
\beq
\label{3.2.4}
f^A([z]) = \bra{\emptyset}\psif{A}\eJz\ket{1},\quad
\fb^{\Ab}([\zb]) = \bra{1}\eJzb\psifs{\Ab}\ket{\emptyset}.
\eeq
The dependence in $[z]$ and $[\zb]$ is
dictated by the action
of the higher currents $J$, $\bar{J}$,
defined by Eq.\ref{3.1.4},that is,
$J_1z\to \sum_{i=0}^n J_iz^{(i)}$,
$\Jb_1\zb\to \sum_{i=0}^n \Jb_i\zb^{(i)}$ in Eq.\ref{2.2.9}.
Thus we shall call them \souligne{KP coordinates}.
Accordingly, the embedding
operator, and tau-functions  are  re-defined by modification
of the Hamiltonian in Eqs.\ref{3.1.7}, and \ref{3.1.8a}, that is:
\begin{definition}{\bf Generalized tau-functions,
and embedding operator.}
\beq
\label{3.3.calG}
\EM([z],[\zb]) \equiv
\eJzb \EMO \eJz
\eeq
\beq \label{3.2.tau}
\tau_\ell([z],[\zb])= \bra{\ell}\EM([z],[\zb])\ket{\ell}.
\eeq
\end{definition}
{}From the physicist's viewpoint,
it is illuminating to realize that these tau-functions
play the r\^ole of partition functions since the relevant
fermionic matrix elements are obtained by taking derivatives
of them with respect to $[z]$ and $[\zb]$. This is contained in
the so-called bosonization rules derived in  Appendix B, using
 a method which is an adaptation of the proof of the
relativistic
fermion \cite{DJKM} to the present non-relativistic ones:
\begin{theorem}{\bf Bosonization rules.}
\label{Bosrules}
\beqa
 \bra{\ell+1}\EM([z][\zb])
\psis_{\ell+s}\ket{\ell} & = & \chi^{Sch}_s\! ([\partial])
 \> \tau_{\ell+1}([z][\zb]),\nonumber\\
 \bra{\ell}\EM([z][\zb])
\psi_{\ell-s}\ket{\ell+1} & = & \chi^{Sch}_s\! (-[\partial])
  \>\tau_{\ell+1}([z][\zb])\nonumber\\
 \bra{\ell+1}\psis_{\ell-s}\EM([z][\zb])\ket{\ell}
& = & \chi^{Sch}_s\! ([\partialb])
  \>\tau_{\ell}([z][\zb]),\nonumber\\
 \bra{\ell}\psi_{\ell+s}\EM([z][\zb])
\ket{\ell+1}&  =&  \chi^{Sch}_s\! (-[\partialb])
 \>  \tau_{\ell}([z][\zb]),\label{3.3.17}
\eeqa
where  the differential operators are given by
Schur's polynomials,
\beq
\label{3.3.18}
\chi^{Sch}_s\! ([\partial])  =  \sum_{i_1 + 2i_2 + \cdots + si_s = s}
   \left(\prod_{\alpha =1}^s\frac{1}{i_\alpha!}\left(
    \frac{1}{\alpha}\frac{\partial}{\partial z^{(\alpha)}}
\right)^{i_\alpha}
        \right).
\eeq
\end{theorem}
For example, one has
\beq
\chi^{Sch}_0\! ([\partial]) = 1,\quad
\chi^{Sch}_1\! ([\partial]) = \frac{\partial}{\partial z^{(1)}},\quad
\chi^{Sch}_2\! ([\partial]) = \frac{1}{2}\left(
\frac{\partial}{\partial z^{(2)}}
  +\left(\frac{\partial}{\partial z^{(1)}}\right)^2\right).
  \label{3.3.19}
\eeq
Clearly the relationship with matrix-models is
striking. We shall come back to this point in appendix \ref{A.3}.

Going back to our main line, we note that
Eqs.\ref{3.2.4} give the extension to $\CC$ of
the $\eta$  matrix \ref{2.3.1}:
\beq
\label{3.2.5}
\eta_{i\jb}([z],[\zb]) =
\sum_{A=0}^n\partial_if^A([z])\, \partialb_j\fb^{A}([z]),
\quad \partial_i\equiv \frac{\partial}{\partial z^{(i)}},
\quad\partialb_i\equiv \frac{\partialb}{\partialb \zb^{(i)}}.
\eeq
Now, only first-order  derivatives appear.
As a matter of fact this
expression
coincides with the
 {\it true } Riemannian metric with respect to the KP
coordinates.  We call the corresponding frame,
span by  $\partial_s \bff$,
$\partialb_s \bbff$, the  \souligne{W-frame}.
In terms of the W-frame and modified $\eta$,
the moving frame\footnote{
In this context the name ``moving frame" is somehow
misleading.  They are actually {\bf the local Lorenz frame} of the
target space.} is given by,
\begin{definition}{\bf Moving frame with
the KP coordinates.}
\label{mvKP}
\newline
The following set of vectors  is orthonormal
($\eta$ is given by Eq.\ref{3.2.5})
\beqa
\label{3.2.v}
\bfet_\ell([z],[\zb])&=&\frac{1}{\sqrt {\tau_\ell([z],[\zb])
\, \tau_{\ell+1}([z],[\zb])}}
\bfvt([z],[\zb]),\nonumber \\
\bfvt_\ell([z],[\zb]) &=& \left|
 \begin{array}{ccc}
  \eta_{0\0b} & \cdots & \eta_{\ell\0b}\\
  \vdots   &    ~   & \vdots   \\
  \eta_{0 \overline{\ell-1}} & \cdots &
\eta_{\ell{\overline{\ell -1}}}\\
\partial_0\bff([z])  & \cdots & \partial_\ell\bff([z])
 \end{array}\right|\nonumber \\
\bbfet_\ell([z],[\zb])&=&
\frac{1}{\sqrt{\tau_\ell([z],[\zb])
\, \tau_{\ell+1}([z],[\zb])}}
\bbfvt([z],[\zb]),\nonumber\\
\bbfvt_\ell([z],[\zb])& =& \left|
 \begin{array}{ccc}
  \eta_{\0b 0} & \cdots & \eta_{\bar\ell 0}\\
  \vdots   &    ~   & \vdots                           \\
  \eta_{\ell-1 \0b} & \cdots & \eta_{\bar\ell\ell -1}\\
  \partialb_0\bbff([\zb])  & \cdots & \partialb_\ell\bbff([\zb])
 \end{array}\right|.
\eeqa
\end{definition}

Each W-parametrization  depends upon the W-surface considered.
The latter is obviously  recovered  by letting $z^{(k)}=
\zb^{(k)}=0,$ for $k=2,\cdots,n$.
This link between the target-space parametrization and
the W-surface allows us to relate its intrinsic and extrinsic
geometries.  This is a key step in this whole scheme.

Next we show that the W-transformations  may
be extended to $\CC$ by requiring that
the differential equations Eq.\ref{3.2.3}
be left invariant.   This will
give a special class of diffeomorphisms
 of $\CC$.
 We shall
only consider the holomorphic sector explicitly.
The calculation in the anti-holomorphic
sector is completely analogous.
First we consider the limit $n\rightarrow\infty$.
It is known that the W-transformations become linear
(so-called $w_{1+\infty}$ transformations) and the coordinate
transformations are simplified. We work at the level of formal series,
without considering convergence problems.
The desired results follow from the
\begin{lemma}{\bf Invariance of the differential equations.}
\label{equadiff}

Given an arbitrary function $\epsilon(z)$, define the functions
$\epsilon^{(s)}([z])$, $s=0$, $\cdots$, $\infty$, from the
generating function
\beq
\label{3.2.6}
e^{-H(\zeta,[z])}
\epsilon(\partial_\zeta)\eH =  \sum_{s=0}^\infty
\zeta^s\epsilon^{(s)}([z]),\quad
H(\zeta,[z])\equiv \sum_{s=0}^\infty \zeta^s z^{(s)}.
\eeq
1) To first order, and for any given positive integer $\ell$,
 the differential equations \ref{3.2.3} are left invariant by the
 change of W-parameters
\beq
\label{3.2.7}
\delta^{(\ell)}_\epsilon z^{(r)}= \epsilon^{(r-\ell)}([z]),
\> r\geq \ell,\qquad \delta^{(\ell)}_\epsilon z^{(r)}=0,
\>  r<\ell.
\eeq
2) Conversely, any first-order
reparametrisation of $[z]$ that leaves
the differential equations \ref{3.2.3}
invariant is a linear combination
of the above.
\end{lemma}

\noindent \proof 1) One makes use of the inverse Laplace
transform.  Write
\beq
\label{3.2.8}
f^A(z) = \int_{a-i\infty}^{a+i\infty}
d\zeta e^{\zeta z}\ft^A(\zeta),
\eeq
where $a$ is to the right of the singularities of $\ft^A$. It
follows from the definition \ref{Wparam} that ($H(\zeta, [z])$
is
defined in Eq.\ref{3.2.6})
\beq
\label{3.2.9}
f^A([z]) = \int d\zeta \eH \ft^A(\zeta).
\eeq
Making use of Eqs.\ref{3.2.6}, \ref{3.2.8}, one may rewrite
the variation of $f^A([z])$
under the form
\beq
\label{3.2.10}
\delta^{(\ell)}_\epsilon f(z)=
\int_{a-i\infty}^{a+i\infty} d\zeta
\eH\epsilon(-\partial_\zeta)(\zeta^\ell \ft(\zeta)),
\eeq
which is indeed a solution of the
differential equations \ref{3.2.3}.

\noindent 2) Conversely, consider a variation
$\delta z^{(r)}= \rho^{(r)}([z])$.
If the variation of $f^A$ is a solution,
one should be able to write
$$
 \delta  \eH\equiv  \sum_{r=0}^\infty \rho^{(r)}([z])\zeta^r \eH
=\sum_{q=0}^\infty \zeta^q P_q(\partial_\zeta) \eH,
$$
where $P_q(\partial_\zeta)$  is a differential operator with
constant coefficients. To each $P_q$, and making use of the
generation function Eq.\ref{3.2.6},
we associate a family of functions $P_q^{(s)}([z])$
by the equations
\beq
\label{3.2.11}
e^{-H(\zeta,[z])}
P_q(\partial_\zeta)\eH =  \sum_{s=0}^\infty
\zeta^sP_q^{(s)}[z]),
\eeq
and we obtain
$$
 \delta  \eH=\sum_q\zeta^q\sum_s \zeta^s P_q^{(s)}([z])
$$
\beq
\label{3.2.12}
\delta z^{(r)}=\sum_q\delta_{P_q}^{(q)} (z)=
\sum_{q\leq r} P_q^{(r-q)}([z]).
\eeq
This completes the proof. {\bf Q.E.D.}
\medskip

Next, it follows from Eq.\ref{3.2.6} that Eq.\ref{3.2.7} gives
\beq
\label{3.2.13}
\delta^{(\ell)}_\epsilon f^A([z])=
\sum_{s=0}^\infty \epsilon^{(s)}([z])
{\partial\over \partial z^{(\ell+s)}}
f^A([z])=\sum_{s=0}^\infty \epsilon^{(s)}([z])
 \partial^{(\ell+s)}
f^A([z]).
\eeq
It is easy to see that, on the W surface
 (that is for $z^{(1)}=z$, $z^{(0)}=z^{(2)}=\cdots$ $=
z^{(n)}=0$), Eq.\ref{3.2.6}  gives $\epsilon^{(s)}=0$, for
$s\not= 0$, and $\epsilon^{(0)}(z)=\epsilon (z)$. Thus the function
$\epsilon (z)$ specifies the variation of the embedding functions
themselves:
\beq
\label{3.2.14}
\delta^{(\ell)}_\epsilon f^A(z)= \epsilon(z) \partial^{(\ell)}
f^A(z).
\eeq
Clearly the W-transformations introduced by the definition
\ref{Wtrans} are linear transformations of such variations.
The lemma  \ref{equadiff}  thus leads to the
\begin{theorem}{\bf W-diffeomorphisms. }

Each W transformation has a unique
local extension to $\CC$ that leaves
the differential equations Eqs.\ref{3.2.3} invariant.
\end{theorem}

The  above discussion moreover shows that we have
\beq
\label{3.2.15}
\delta_W f^A([z])=\sum_r W^r \partial_r f^A([z]),\quad
W^r=\sum_{s=0}^r w^{r-s\, (s)}.
\eeq
The functions $W^r$ should be regarded as the components of the
 tangent vector associated with the W-transformation considered.

The Lie algebra of W-transformations coincides  with the bracket
algebra of the associated tangent vectors. The corresponding change of
coordinates is $\delta_Wz^{(r)}=W^r$, and one sees that
{ \em the W-surface is moved by the W-transformations}.
For a W-surface, there is no covariant separation
between intrinsic and extrinsic geometries.

Our next topic is the local rescaling  introduced
in Eq.\ref{2.2.16}, that is,  the transformation
$f^A(z)\to \rho(z) f^A(z)$. Infinitesimal transformations
of this type  may be regarded as special
cases of the definition \ref{Wtrans}, where $w^j=0$, for $j\not=0$.
Moreover, one may check that the covariance properties
theorem \ref{covresc} of the moving frame are extended away
from the W-surface by the definition of the W-parametrzation.
The above discussion thus gives  the
\begin{corollary}{\bf Local rescaling. }
\label{u1gauge}

The infinitesimal rescaling $\delta f^A(z)
=\sigma(z) f^A(z)$ is equivalent to the following change
of W-parameters
\beqa
\label{3.2.17}
\delta z^{(s)}=\sigma^{(s)}([z]),\quad
e^{-H(\zeta,[z])}
\sigma(\partial_\zeta)\eH = \sum_{s=0}^\infty
\zeta^s\sigma^{(s)}([z]).
\eeqa
The generalized moving frame  introduced by definition \ref{mvKP}
is covariant under these transformations.
\end{corollary}

The fact that the rescaling  is equivalent to
a change of W-coordinates is  understood by noting that
$\eH$ is of the form   $\exp (z^{(0)})$ times a factor
that does not depend upon this variable. Thus $z^{(0)}$ is
really the scaling factor of the W-coordinates.

Next we   deal with the case where
the dimension of the target-space is finite and equal to $n$.
We prove that the modification is given by
\begin{theorem}{\bf W-diffeomorphism for finite n. }
\label{Wfiniten}

For finite $n$, the W-reparametrizations
Eq.\ref{3.2.15} become
\beq
\label{3.2.22}
\delta_W f^A([z])=\sum_{r=0}^n \left ( W^r +
\sum_{s>0} W^{n+s}\lambda_r^{(s)}([z])\right )
\partial_r f^A([z])
\eeq
Similarly, the reparametrisation
Eq.\ref{3.2.7} is to be rewritten
as
\beq
\label{3.2.23}
\delta_\epsilon^{(\ell)} z^{(s)} = \epsilon^{(s-\ell)}([z])+
\sum_{t=1}^\infty \epsilon^{(n-\ell +t)}([z])
\lambda^{(t)}_s([z]).
\qquad (s=0,\cdots,n),
\eeq
where, according to eq.\ref{3.2.7}, $\epsilon^{(r)}$ is defined to
be
zero for $r<0$. The notations of this theorem are explained in
the proof.
\end{theorem}

\proof
In this case there are relations between the embedding functions.
They are derived  from the equations
($A=0,\, \cdots,\, n$)
\beqa
\label{3.2.18}
&\left | \begin{array}{cccc}
f^0(z) & \cdots  & f^n(z) & f^A(z) \\
\vdots    &   & \vdots      & \vdots  \\
f^{(n)\, 0}(z)  & \cdots  & f^{(n)\, n}(z)  & f^{(n)A}(z)\\
f^{(n+s)\, 0}(z)  & \cdots  & f^{(n+s)\, n}(z)  & f^{(n+s)A}(z)
\end{array}
\right| \nonumber \\
&\equiv
\left\{ U_0(z) \partial^{(n+s)}-\sum_{t=0}^n U^{(s)}_t(z)
\partial^{(t)}\right\} f^A(z)=0.
\eeqa
$U_0$ which is the Wronskian of the functions $f^0,\, \cdots,\,
f^n$, does not vanish at regular points of $\Sigma$. Thus we may
eliminate the higher derivatives by the relation
\beq
\label{3.2.19}
\partial^{(n+s)} f^A(z) =
\sum_{t=0}^n \lambda^{(s)}_t(z)\partial^{(t)} f^A(z),
\qquad
\lambda^{(s)}_t(z) = U^{(s)}_t(z)\bigl /U_0(z).
\eeq
When the W-parametrizations are defined according to the
definition \ref{Wparam}, it is easy to see that this last condition
is  extended by construction. Indeed, one  has
\beqa
\label{3.2.20}
&\left | \begin{array}{cccc}
\partial_0 f^0([z]) & \cdots  &
\partial_0 f^n([z]) & f^A([z]) \\
\vdots    &   & \vdots      & \vdots  \\
\partial_nf^{0}([z])  & \cdots  & \partial_nf^{n}([z])
& \partial_nf^{A}([z])\\
\partial_{n+s}f^{0}([z])  & \cdots  & \partial_{n+s}f^{n}([z])
& \partial_{n+s}f^{A}([z])
\end{array}
\right|\nonumber \\
&\equiv
\left\{ U_0([z])\partial_{n+s}-\sum_{t=0}^n U^{(s)}_t([z])
\partial_{t} \right\}f^A([z])=0
\eeqa
and the W-parameters automatically satisfy the conditions
\beq
\label{3.2.21a}
\partial_{n+s} f^A([z]) =
\sum_{t=0}^n \lambda^{(s)}_t([z])\partial_{t} f^A([z]),
\quad
\lambda^{(s)}_t([z]) = U^{(s)}_t([z])\bigl /U_0([z]).
\eeq
By this, we can eliminate all dependence in the higher coordinates
$z^{(k)}$, with $k >n$.
\qed

The reparametrization is an explicit function of $f$.
This reflects the fact that the W parametrization
depends upon the W-surface considered.
This is why the embedding  functions
$f^A$ transform nonlinearly.

Finally, it is clear that, since we have worked in a way which is covariant
under the local rescaling defined by the corollary \ref{u1gauge},
we have the
\begin{corollary}{\bf W-parametrization of $\CP$.}
\label{WparamCP}
The W-parameters of the definition \ref{Wparam}
give a parametrization of $\CP$ if one identifies any two
points which are connected by  a   transformation  of
the form Eq.\ref{3.2.17} introduced
in the corollary \ref{u1gauge}.
\end{corollary}

%
%

\subsection{Extended Frenet-Serret Formula and Toda Hierarchy}
\label{3.3}

In this section, we study the generalization of the
Frenet-Serret  equations Eqs.\ref{2.1.20}, that include
the
KP coordinates introduced above to define W-parametrizations.
As  expected, such study leads us to the Lax pair
for the Toda {\it hierarchy}.  We shall deal with
the  $\CP$ W-surface, by working in $\CC$ with the
generalized homogeneous coordinates introduced in the
previous section (corollary \ref{WparamCP}).
Our  result is expressed by the\footnote{
{}From now on, we omit the arguments $[z],[\zb]$ unless they are
explicitly needed}
\begin{theorem}{\bf Frenet-Serret formulae for KP coordinates.}
\label{FSKP}

Consider the chiral vectors
\beq
 \label{3.3.1}
  \bfut_\ell = \frac{1}{\tau_\ell}\bfvt_\ell,
  \quad
  \bbfut_\ell = \frac{1}{\tau_{\ell+1}}\bbfvt_\ell,
  \qquad
  (\bbfut_\ell, \bfut_{\ell'}) = \delta_{\ell\ell'}
\eeq
where $\bfvt_\ell$ and $\bbfvt_\ell$ are given by Eqs.\ref{3.2.v}.
It follows from the differential equations Eqs.\ref{3.2.3} that
they obey the differential equations
\beqa
 \label{3.3.2}
  \partial_p \bfut_\ell & = & \sum_{\ell' = \ell}^{\ell+p}
   \tau_\ell^{-1}(F^p)_{\ell \ell'}\tau_{\ell'} \bfut_{\ell'}\\
 \label{3.3.3}
  \partialb_p \bfut_\ell & = & -\sum_{\ell' = \ell-p}^{\ell-1}
   \tau_{\ell+1}(G^p)_{\ell \ell'}
    \tau_{\ell'+1}^{-1}\bfut_{\ell'}\\
 \label{3.3.4}
   \partialb_p \bbfut_\ell & = & \sum_{\ell' = \ell+1}^{\ell+p}
    \bbfut_{\ell'}\tau_{\ell'+1}(G^p)_{\ell' \ell}\tau_{\ell+1}^{-1}\\
 \label{3.3.5}
  \partial_p \bbfut_\ell & = & -\sum_{\ell' = \ell-p}^{\ell}
   \bbfut_{\ell'}\tau_{\ell'}^{-1}(F^p)_{\ell' \ell}
    \tau_{\ell}.
\eeqa
where $F^p$ and $G^p$, which  are the p-th power of
matrices $F^1\equiv F$ and $G^1\equiv G$, respectively,
are given by
\beqa
 \label{3.3.FG}
(F^p)_{\ell \ell'}=
\frac{1}{\tau_{\ell'}\tau_{\ell'+1}}
      \sum_{s=0}^\infty
       \bra{\ell'+1}\EM \psis_{s+p}\ket{\ell'}
        \bra{\ell}\EM\psi_{s} \ket{\ell+1}\nonumber \\
(G^p)_{\ell'\ell} =
       \frac{1}{\tau_{\ell'}\tau_{\ell'+1}}
      \sum_{s=0}^\infty
         \bra{\ell+1}\psis_s\EM\ket{\ell}
           \bra{\ell'}\psi_{s+p}\EM \ket{\ell'+1}.
\eeqa
\end{theorem}
\medskip

Using a matrix-notation which is self-explanatory,
one may view the
general structure of the above  equations as follows,
\beqa
\partial_p\bfut  =   H_1^{-1}(F^p)^+H_1\bfut, &\quad&
\partialb_p\bfut =  - H_2 (G^p)^-H_2^{-1}\bfut,\nonumber\\
\partialb_p\bbfut  =   \bbfut H_2(G^p)^-H_2^{-1}, &\quad&
\partial_p\bbfut =  - \bbfut H_1^{-1} (F^p)^+H_1,
\label{3.3.6}\\
(H_1)_{\ell\ell'} = \tau_\ell\delta_{\ell\ell'},&\quad&
(H_2)_{\ell\ell'} = \tau_{\ell+1}\delta_{\ell\ell'}.\nonumber
\eeqa
The Lax operators which appear in Eqs.\ref{3.3.2}--\ref{3.3.5}
are exactly those of the $A_n$-type Toda hierarchy.
The integrability conditions for them, therefore,  quite naturally
give the famous Zakharov-Shabat equations. Thus we have the
\begin{corollary}{\bf Solutions of Zakharov-Shabat equations.}
The integrability conditions of the generalized Frenet-Serret
formulae of theorem\ref{FSKP} are
\beqa
\label{3.3.ZS}
\left[\partial_p - H_1^{-1}(F^p)^+H_1,
\quad\partialb_q + H_2(G^q)^-H_2^{-1}\right]
 & = & 0,\nonumber\\
\left[\partial_p - H_1^{-1}(F^p)^+H_1,
\quad\partial_q - H_1^{-1}(F^q)^+H_1\right]
             & = & 0,
            \nonumber\\
\left[\partialb_p + H_2(G^p)^-H_2^{-1},
\quad \partialb_q + H_2(G^q)^-H_2^{-1}\right]
 & = & 0,
\eeqa
which coincide with the Zakharov-Shabat equations.
\end{corollary}
This is the most explicit proof of the relation between the
extrinsic geometry of the W-surfaces and the Toda hierarchy.

According to the general scheme, of section \ref{3.1},
$F$ and $G$  are given by derivatives of
tau-functions. Eqs.\ref{3.3.17} -- \ref{3.3.19} give
\begin{proposition}{}
\beqa
(F^p)_{\ell \ell'} & = &
 \frac{1}{\tau_{\ell'}\tau_{\ell'+1}}
\sum_{s=0}^{\ell-\ell'+p}
\left(\chi^{Sch}_s\! ([\partial])\>\tau_{\ell'+1}\right)
\left(\chi^{Sch}_{\ell-\ell'+p-s}\! (-[\partial])\>\tau_{\ell}\right)
\nonumber\\
(G^p)_{\ell' \ell} & = &
\frac{1}{\tau_{\ell'}\tau_{\ell'+1}}
  \sum_{s=0}^{\ell-\ell'+p}
\left(\chi^{Sch}_s\!([\partialb])\>\tau_{\ell'+1}\right)
\left(\chi^{Sch}_{\ell-\ell'+p-s}\! (-[\partialb])\>\tau_{\ell}\right).
\label{3.3.20}
\eeqa
\end{proposition}
The first few terms are very simple,
\beq\label{3.2.21}
F_{\ell \ell+1}   =   \frac{\tau_\ell}{\tau_{\ell+1}},\quad
F_{\ell\ell}   =   \partial_1\ln(\tau_{\ell+1}/\tau_\ell),
\quad
F_{\ell\ell-1}   =   \frac{\tau_\ell}{\tau_{\ell+1}}\partial_1^2
    \ln\tau_\ell\quad
   \cdots
\eeq

The rest of this section is devoted to the detailed proof of
theorem \ref{FSKP}.
In subsection \ref{3.3}.1, and
since we need to treat the higher KP coordinates
systematically, we first translate the moving-frame equations
into the free-fermion language. In subsection \ref{3.3}.2,
explicit formulae
for the Lax operators $F$,$G$ are given.
In subsection \ref{3.3}.3, we  finally spell out the actual
derivation of the generalized Frenet-Serret equations.

\subsubsection{Free-fermion representation of the moving frame}
In order to automatically solve the differential equations
Eqs.\ref{3.2.3}, we re-write all expressions of
 the moving-frame equations,
in  the free-fermion operator-formalism.
The basic point is that the Hirato equation \ref{3.1.13} remains
valid when the
higher coordinates are included. Indeed, using the same argument as
in section \ref{3.1},  we can again reduce the derivation to the point
$z^{(k)}=\zb^{(k)}=0, \> \forall k$. Thus the
full power of the fermionic method is still on.
{}From Eqs.\ref{3.2.v}, it is straightforward to derive the
following  neat expressions for
$\bfvt_\ell$ and $\bbfvt_\ell $,
\beqa
\bfvt_\ell &=&
  \sum_{s=0}^\ell \bff^{(s)}\> \bra{\ell} \EM \psi_s
   \ket{\ell+1}\nonumber\\
\bbfvt_\ell &=&
  \sum_{s=0}^\ell \bbff^{(s)}\> \bra{\ell+1}\psis_s \EM([z],[\zb])
   \ket{\ell},\label{3.3.7}
\eeqa

Instead of working with the vectors $\bfvt_\ell$ and $\bbfvt_\ell$,
it is more useful to introduce the  ket-  and bra-fermionic-states
which correspond to them:
\beqa
\bfvt_\ell & \leftrightarrow & \sum_{s=0}^\ell
 \eJz \psis_s\ket{\emptyset}\bra{\ell}\EM \psi_s\ket{\ell+1}
  \equiv \ket{v_\ell},
   \nonumber\\
\bbfvt_\ell & \leftrightarrow & \sum_{s=0}^\ell
 \bra{\ell+1}\psis_s\EM\ket{\ell}\bra{\emptyset}\psi_s\eJzb
  \equiv \bra{\vb_\ell}.\label{3.3.10}
\eeqa
These states  satisfy
\beq\label{3.3.11}
\bra{\vb_\ell}\EM([0],[0])\ket{v_{\ell'}}
= \bbfvt_\ell^{\Ab}\bfvt_{\ell'}^A\delta_{A\Ab}
= \tau_{\ell}\tau_{\ell+1}\delta_{\ell\ell'}.
\eeq
It is easy to derive  the equations of $\bfvt$, $\bbfvt$ from
those of $\bra{\vb_\ell}$ and $\ket{v_\ell}$.

\subsubsection{Definition of the Lax operator}
Using  the free-fermion representation of the moving frame,
we first justify the introduction  of the Lax operators $F$, $G$.
Since one has $\partial_p\eJz=J_p\eJz$, we first study the action
of the $J_p$'s on the states $\bra{\vb_\ell}$ and $\ket{v_\ell}$.
Define $F$ and $G$ by
\beq
J_1\ket{v_\ell}  \equiv  \sum_{\ell'=0}^{\ell+1}
 F_{\ell \ell'}\ket{v_{\ell'}}, \qquad
\bra{\vb_\ell}\bar{J}_1  \equiv  \sum_{\ell'=0}^{\ell+1}
 \bra{\vb_{\ell'}}G_{\ell' \ell}\label{3.3.12}
\eeq
Since $\ket{v_\ell}$ and $\bra{\vb_\ell}$ are one-particle states,
we can easily derive the following lemma,
\beq
J_p\ket{v_\ell}  \equiv  \sum_{\ell'=0}^{\ell+p}
 (F^p)_{\ell \ell'}\ket{v_{\ell'}},\quad
\bra{v_\ell}\bar{J}_p  \equiv  \sum_{\ell'=0}^{\ell+p}
 \bra{\vb_{\ell'}}(G^p)_{\ell' \ell}\label{3.3.13}
\eeq
We can calculate the explicit formulae for $F$ and $G$ by using
the generalized Hirota equation (theorem \ref{Hirota}).
For example,
\beqa
F_{\ell \ell'} & = & \frac{1}{\tau_{\ell'}\tau_{\ell'+1}}
  \bra{\vb_{\ell'}}\EM J_1\ket{v_\ell}\nonumber\\
 & = & \frac{1}{\tau_{\ell'}\tau_{\ell'+1}}
      \sum_{s=0}^\infty\sum_{s'=0}^\infty
       \bra{\ell'+1}\psis_{s'}\EM \ket{\ell'}
        \bra{\ell}\EM\psi_{s} \ket{\ell+1}
        \bra{\emptyset}\psi_{s'}\EM\psis_s \ket{\emptyset}
          \nonumber\\
 & = & \frac{1}{\tau_{\ell'}\tau_{\ell'+1}}
      \sum_{s=0}^\infty
       \bra{\ell'+1}\EM \psis_{s+1}\ket{\ell'}
        \bra{\ell}\EM\psi_{s} \ket{\ell+1}.
\label{3.3.14}
\eeqa
Similarly,
\beq\label{3.3.15}
G_{\ell'\ell} =
       \frac{1}{\tau_{\ell'}\tau_{\ell'+1}}
      \sum_{s=0}^\infty
         \bra{\ell+1}\psis_s\EM\ket{\ell}
           \bra{\ell'}\psi_{s+1}\EM \ket{\ell'+1}.
\eeq
The summations in Eqs.\ref{3.3.14}--\ref{3.3.15} can be
taken from $\ell'-1$ to $\ell$ since the other terms vanish.
  Due to the lemma (\ref{3.3.13}),
the powers of $F$, $G$ have similar forms,
\beqa
(F^p)_{\ell \ell'} & = &
  \frac{1}{\tau_{\ell'}\tau_{\ell'+1}}
      \sum_{s=0}^\infty
       \bra{\ell'+1}\EM \psis_{s+p}\ket{\ell'}
        \bra{\ell}\EM\psi_{s} \ket{\ell+1},
   \nonumber\\
(G^p)_{\ell'\ell} & = &
       \frac{1}{\tau_{\ell'}\tau_{\ell'+1}}
      \sum_{s=0}^\infty
         \bra{\ell+1}\psis_s\EM\ket{\ell}
           \bra{\ell'}\psi_{s+p}\EM \ket{\ell'+1}.
  \label{3.3.16}
\eeqa

\subsubsection{Proof of the theorem }

We give a proof of \ref{3.3.2}--\ref{3.3.FG} by direct computations.
Before that, it is useful to observe that,
by differentiation of Eq.\ref{3.3.11},
\beq
 \label{3.3.22}
  \partial_p (\tau_\ell\tau_{\ell+1})\delta_{\ell\ell'}
  = (\partial_p\bra{\vb_\ell})\EM([0],[0])\ket{v_{\ell'}}
    +\bra{\vb_\ell}\EM([0],[0])(\partial_p\ket{v_{\ell'}}).
\eeq
The first term on the RHS vanishes if $\ell'>\ell$.
Hence the second term should also vanish in this case.
By the orthogonality of $\bra{\vb}$,
and $\ket{v}$, we can conclude that
$\partial_p\ket{v_{\ell'}}$ is the linear combination of
$\ket{v_{\ell'+p}}$, $\ket{v_{\ell'+p-1}}$,
$\cdots$ , $\ket{v_{\ell'}}$.
On the other hand, a direct computation gives
\beqa
  \partial_p\ket{v_{\ell}}& = &J_p\ket{v_\ell} +
    \sum_{s=0}^\infty
    \eJz \psis_s\ket{\emptyset}\bra{\ell}\EM J_n\psi_s\ket{\ell+1}
      \nonumber\\
    & = &\sum_{\ell'=0}^{\ell+p}(F^p)_{\ell\ell'}\ket{v_{\ell'}} +
    \sum_{s=0}^\infty
    \eJz \psis_s\ket{\emptyset}\bra{\ell}\EM J_n\psi_s\ket{\ell+1}.
  \label{3.3.23}
\eeqa
We remark that the second term is a linear combination of
$\ket{v_0},\cdots,\ket{v_\ell}$.
{}From these two arguments, we can conclude that the second term
cancells with $(F^p)_{\ell\ell'}\ket{v_{\ell'}}$ for
$\ell' = 0,1,\cdots,\ell-1$.  This fact can be confirmed
from the following explicit calculations.
By using the Hirota equation (theorem \ref{Hirota}), one obtains
\beq   \begin{array}{l}
 \sum_{s=0}^\infty \bra{\vb_m}\EM
 \psis_s\ket{\emptyset}\bra{\ell}\EM J_p\psi_s\ket{\ell+1}\\
 =  \left\{\begin{array}{lc}
   0 & (\ell<m)\\
  \label{3.3.24}
    \tau_{\ell+1}\partial_p\tau_\ell & (\ell=m)\\
   -\sum_{q=0}^\infty \bra{m+1}\EM \psis_{q+p}\ket{m}
       \bra{\ell} \EM \psi_q\ket{\ell+1} & (\ell>m).
   \end{array}\right. \end{array}
 \label{m2.24}
\eeq
The $\ell=m$ term can be alternatively written as
\beq\label{3.3.25}
-\sum_{q=0}^\infty \bra{\ell+1}\EM \psis_{q+p}\ket{\ell}
       \bra{\ell} \EM \psi_q\ket{\ell+1} +
       \tau_\ell\partial_p\tau_{\ell+1}.
\eeq
By combining, Eqs.\ref{3.3.16} with Eqs.\ref{3.3.23}--\ref{3.3.25}, we get the
desired result,
\beqa
\partial_p \ket{v_\ell} & = &
  \sum_{\ell'=\ell+1}^{\ell+p}(F^p)_{\ell\ell'}\ket{v_{\ell'}}
   +(\partial_p\tau_{\ell+1})\ket{v_\ell}\nonumber\\
    & = &
\sum_{\ell'=\ell}^{\ell+p}(F^p)_{\ell\ell'}\ket{v_{\ell'}}
   +(\partial_p\tau_{\ell})\ket{v_\ell}.\label{3.3.26}
\eeqa
If we combine this with Eq.\ref{3.3.22}, we get another formula,
\beqa
\partial_p\bra{\vb_\ell} & = &
 -\sum_{\ell'=\ell-p}^{\ell-1}
 \bra{\vb_{\ell'-1}}(\tau_{\ell'}\tau_{\ell'-1})^{-1}
 (F^p)_{\ell'\ell}\tau_{\ell}\tau_{\ell-1}
 +(\partial_p\ln\tau_\ell)\bra{\vb_\ell}\nonumber\\
                         & = &
 -\sum_{\ell'=\ell-p}^{\ell}
 \bra{\vb_{\ell'-1}}(\tau_{\ell'}\tau_{\ell'-1})^{-1}
 (F^p)_{\ell'\ell}\tau_{\ell}\tau_{\ell-1}
 +(\partial_p\ln\tau_{\ell+1})\bra{\vb_\ell}.\label{3.3.27}
\eeqa
Equations \ref{3.3.2} and \ref{3.3.5} can be obtained from these
formulae by scaling the moving frame appropriately s.t.
we remove the diagonal terms.
The proof of Eqs.\ref{3.3.3},\ref{3.3.4} is exactly parallel.
This completes our proof of the Frenet-Serret formula including
the higher coordinates. {\bf Q.E.D.}

%
%

\subsection{Generalised WZNW equations and Riemannian Geometry}
\label{3.4}
    In the present section we show that the W-parametrization
of the target-spaces  $\CC$, and $\CP$,
which were  defined in section \ref{3.2}
(definition \ref{Wparam}),
 are  such that the correspondence between
W-surfaces and conformally reduced
WZNW has a natural multi-dimensional
extension away from the W-surface
$\Sigma$. This will also lead to
an extension  of the Drinfeld-Sokolov
equations.   These extensions show
the intimate connection between
these equations and the Riemannian
geometry of the target-space. First we have the
\begin{theorem}{\bf  Christoffel connection for
W-parametrization.}\label{chLC}

With the W-parametrization, the Christoffel
symbols are  chiral and given by
\beq
\label{3.4.1}
\left\{ \begin{array}{cc}
(\gamma_p([z]))_j^k=\delta_{p+j,\, k}, \> {\rm if}\>p+j\leq n,
 \nonumber\\
    = \lambda^{(p+j-n)}_j([z]) \> {\rm if}\>p+j>n
\end{array} \right. ; \quad
\left\{ \begin{array}{cc}
(\gammab_{\pb}([\zb]))_{\jb}^{\kb}
=\delta_{\pb+\jb,\, \kb}, \> {\rm if}\>\pb+\jb\leq n,
 \nonumber\\
    = \lambdab^{(\pb+\jb-n)}_j([\zb]) \> {\rm if}\>\pb+\jb> n
\end{array} \right.,
\eeq
where $\lambda$ $\lambdab$ are defined in Eqs.\ref{3.2.18}
and \ref{3.2.19}.
\end{theorem}
\proof With the W-parameters, the metric tensors
is  given by Equation \ref{3.2.5}, that is,
 $\eta_{i\jb} = \sum_{A,\Bb}\delta_{A\Bb}
\partial_if^A([z])\partialb_{\jb}\fb^{\Bb}([z])$.
The Christoffel symbols are such
that its  covariant derivatives
  vanish.
 Since this metric tensor
is factorized into a  product of two
chiral matrices, we immediately get
\beq
\label{3.4.2}
\partial_p \eta_{i\jb}=\sum_j (\gamma_p([z]))_i^j
\eta_{j\jb},\quad
\partialb_{\pb} \eta_{i\jb}=\sum_{\kb}
(\gammab_{\pb}(\zb))_{\jb}^{\kb} \eta_{i\kb},
\eeq
where the Christoffel symbols $\gamma$, and $\gammab$ are such
that
\beq
\label{3.4.3}
\partial_p (\partial_\ell \bff)  =
(\gamma_p)_\ell^{\ell'}(\partial_{\ell'} \bff), \quad
\partialb_p (\partialb_\ell \bbff)  =
(\partialb_{\ell'} \bbff)(\gammab_{\bar p})^{\ell'}_\ell
\eeq
Making use of Eqs.\ref{3.2.13} together with its anti-chiral
counterpart, one easily deduces the explict expressions
 Eqs.\ref{3.4.1}.
Clearly the  Christoffel connection satisfies
\beq
\label{3.4.4}
\partialb_{\qb}  (\gamma_p)_i^j =\partial_q (\gammab_{\pb})_{\jb}^{\kb}=0.
\eeq
It is thus  chiral, and
this completes the proof.
\qed

Next it is clear that Eqs.\ref{3.4.3} may  be considered as
 multi-variable
extensions  of the Drinfeld-Sokolov equation \ref{2.3.11DS}.
Thus we introduce the
\begin{definition}{\bf Generalized Drinfeld-Sokolov equations.
}
\label{GDS}

 They are defined as a set
of n partial-differential equations of the form
\beq
\label{3.4.5}
{\cal L}_{p}\Upsilon([z])=0, \qquad
{\cal L}_{p}\equiv \partial_p -{\cal V}^{(p)}([z]),
\eeq
where  $\Upsilon$ is  a column-vector $\{\Upsilon_\ell,
0\leq \ell \leq n\}$, and where
  the $(n+1)\times (n+1)$ matrices  ${\cal V}^{(p)} $ are such
that
\beq
\label{3.4.6}
\left( {\cal V}^{(p)}\right )^j_k=
0,\> {\rm if} \> p+j> k,\quad
\left( {\cal V}^{(p)}\right )^j_k=
1,\> {\rm if} \> p+j= k.
\eeq
\end{definition}

Next it seems appropriate to generalize the
link between WZNW and Drinfeld-Sokolov equations as well.
Thus we introduce the
\begin{definition}{\bf Generalized WZNW equations. }
\label{GenWZNW}

They are
partial-differ\-ential equations of the form
\beq
\label{3.4.9}
\partialb_{\pb}\left (\theta^{-1}([z],[\zb])\>
\partial_q \theta([z],[\zb])\right )=
\partial_q\left (\left(\partialb_{\pb} \theta([z],[\zb])\right)
\>
\theta^{-1}([z],[\zb])\right )=0
\eeq
where $\theta$ is a  $(n+1)\times (n+1)$ matrix
which is real for Minkowski $z$ and $\zb$ coordinates.
\end{definition}
Of course, in the same  way as in section \ref{2.3}, the present
definition includes the $gl(1)$ factor, so that we do not assume
that the determinant of $\theta$ is equal to one.
The
W-parametrizations of $\CC$  automatically give solutions of
these equations, and one easily verifies the
\begin{theorem}{\bf WZNW from Christoffel connection.}

The matrix $\theta=\eta^{-1}$, where $\eta$ is the metric tensor of
the W-coordinates, is a solution of the generalized WZNW equations,
\label{3.4.9a}
satisfying the following constraints:
\beqa
\label{3.4.10}
Tr\left (\theta^{-1}\partial_p \theta E_{j+p,j}\right )&=
-1, \> 0\leq j\leq n-p,
\nonumber \\
Tr\left ((\partialb_{\pb} \theta) \theta^{-1} E_{j,j+\pb}\right )&=
-1, \> 0\leq j\leq n-\pb,
\eeqa
and $Tr\left (\theta^{-1}\partial_p \theta
E_{-\alpha}\right )=0$ (resp.
$Tr\left ((\partialb_{\pb} \theta) \theta^{-1}
E_{\alpha}\right )=0$) for all other
positive roots (the step operator $E_{jk}$ is defined by
$\left( E_{jk}\right )_a^b=\delta_{a,\,j}\delta_{k,\, b}$).
\end{theorem}
It seems appropriate to call  this
last   system the conformally
reduced WZNW equations, since they are the direct generalizations
of the
standard notion.
Finally it is tempted to give the following
\begin{conjecture}{\bf Equivalences.}
There is a one-to-one correspondence between the W-param\-etrizations
of $\CC$ (resp. $\CP$) and the generalized WZNW and Drinfeld-Sokolov
equations (definitions \ref{GenWZNW}, and \ref{GDS})
for $gl(n+1)$ (resp. for $A_n$).
\end{conjecture}

 As we have observed in section
\ref{2.3}, the Frenet-Serret formulae
give the geometrical interpretation of
the Gauss decomposition of the
metric
$\eta_{j\kb}(z,\zb)= C^{-1}_{js}B^{-1}_{ss'}A^{-1}_{s'\kb}$ on
the
W-surface.
The triangular matrices $C$ and $A$ give the relation between the
vectors ${\tilde {\bfe}}_a$,
${\tilde {\bbfe}}_a$, and the W-frame $\bff^{(a)}$,
and $\bbff^{(a)}$ (see
Eqs.\ref{2.3.11a}, and \ref{2.3.11b}).
Making use of the method developed in
section \ref{3.2}, one immediately sees that the argument may be
extended to the target-space, where
the Gauss decomposition of
the
matrix  $\theta([z],[\zb])$ gives  the general relationship
between the moving frame (= vielbein) span by the vectors
${\tilde {\bfe}}^a([z],[\zb])$, ${\tilde {\bbfe}}^a([z],[\zb])$,
and the W-frame span by $\bff^{(a)}([z])$,
and $\bbff^{(a_)}([\zb])$.
In terms of the   vectors $\bfut$, and
$\bbfut$ defined by Eqs.\ref{3.3.1}, and Eqs.\ref{3.2.v}
 one
has
\beqa
\label{3.4.11}
\bff^{(a)}([z])& =\sum_{b} C^{-1}_{ab}([z], [\zb]) \>
\bfut_b([z], [\zb]), \nonumber\\
\bbff^{(a)}([\zb])& =\sum_{b} \left (AB\right )^{-1}_{ab}([z],
[\zb]) \>
  \bbfut_b([z], [\zb]).
\eeqa
  In the previous  section, we have  derived
 the generalized Frenet-Serret equations,
\beq\label{3.4.12}
\begin{array}{ccc}
\partial_p \bfut_\ell & = &
(\omega_p)_{\ell\ell'} \bfut_{\ell'},\quad
\partialb_{\pb} \bfut_\ell  =
(\omega_{\bar p})_{\ell\ell'} \bfut_{\ell'},\\
\partialb_{\pb} \bbfut_\ell & = &
\bbfut_{\ell'}(\omegab_{\pb})_{\ell\ell'},\quad
\partial_p \bbfut_\ell  =
\bbfut_{\ell'}(\omegab_{ p})_{\ell'\ell}.
\end{array}
\eeq
The matrices    $\omega,\ \omegab$ take the form
\beq\label{3.4.13}
\begin{array}{ccc}
\omega_p & = & (H_1^{-1}F^pH_1)^+,\quad
\omega_{\bar p}  =  -(H_2G^{\pb}H_2^{-1})^-,\\
\omegab_{\bar p} & = & (H_2G^{\pb}H_2^{-1})^-,\quad
\omegab_p  =  -(H_1^{-1}F^pH_1)^+.
\end{array}
\eeq
where $H_i,\ \ (i=1,2)$ is the diagonal
matrices $F_{\ell\ell'}$, and
$ G_{\ell'\ell}$ vanish if $ \ell>\ell'+1$.

Conversely, if a  local lorenz frame of any kind satisfies
this type of equations,
their integrability condition
is equivalent to the Zakharov-Shabat equation.
The general argument of the Toda theory\cite{UT}
tells us that there exists a tau-function such that the
coefficients of the Zakharov-Shabat equation
are given in the form \Eq{3.3.6}.
Since the tau-function is defined by the embedding operator
$\EM$, this argument shows
 that the local lorentz frame can be identified with the
moving frame of a W-surface and
their coordinates can be identified
with the higher coordinates.
 Actually, what we are doing in this paper
is a reversal procedure of
the whole scheme of the Toda equation,
 i.e. we start from the
geometry of the explicit solutions
(W-surface) to get the equation of motion
(Zakharov-Shabat equation).
Thus we reach an important conclusion that
\begin{theorem}{\bf Characterization of W-parametrizations.}

 The reparametrization
of the KP coordinates can be identified with those  coming from
the W-transformations if and only if they
do not change the form of Frenet-Serret
equation.
\end{theorem}

Combining the above formulae, one sees that  $C^{-1}$ and
$(AB)^{-1}$ which appear in  the
Gauss decomposition give the transformation  between
the Lax operator of the Toda
hierarchy Eqs.\ref{3.4.12}, \ref{3.4.9},
  and the generalized
Drinfeld-Sokolov equations
Eqs.\ref{3.4.4}, \ref{3.4.5}. Indeed one finds that
\beq
\label{3.4.14}
\begin{array}{ccc}
 (\partial_p -\gamma_p)C^{-1} & = &
C^{-1}(\partial_p -\omega_p)\\
 \partialb_{\pb} C^{-1} & = &
C^{-1}(\partialb_{\pb}- \omega_{\bar p})\\
(AB)^{-1}\partial_p & = &(\partial_p -\omegab_p) (AB)^{-1}\\
(AB)^{-1}(\partialb_{\pb}-\gammab_{\pb}) & = &
(\partialb_{\pb} -\omegab_{\bar p}) (AB)^{-1}.\end{array}
\eeq
This equation without the higher
coordinates was discussed previously
in ref.\cite{Dublin}.
Clearly the integrability conditions of
the Frenet-Serret  and
 Drinfeld-\-Sokolov equations
are equivalent.
\section{Global structure of the embedding}
\markright{4. Global structure}

\subsection{Associated mappings}
\label{4.1}
In this last part we deal with singular
points of W-surfaces. We shall
 mostly be interested in  the global  aspects.
They  will be described
by $n$ topological numbers which will be related to the ramification
indices of the singularities by a global Pl\"ucker equation
that generalises the Gauss-Bonnet formula. Our guideline is the
beautiful discussion of ref. \cite{GH}.
For these purposes, we need to change the viewpoint which we took
until now.  So far, we have discussed the extrinsic geometry
of the W-surfaces using the moving frame.  We derived the
Toda field equations from the Frenet-Serret formulae
and Gauss-Codazzi equations.
In defining the global indices, this method is not so convenient
at present, since we do not yet know how to make use of
higher topological invariants of the target space.

The way to replace the extrinsic geometry by the intrinsic one
is to  introduce the
\begin{definition}{\bf  Associated mappings.}

Consider the family of osculating hyperplanes with contact
of order $k$ denoted ${\cal O}_k$
($ k=1,\cdots,n$) to the original W-surface.
With $\CP$ as the target space, this family defines an
embedding into the Grassmannian $G_{n+1,k+1}$,
 which we  call  the
$k$th associated mappings to  the original W-surface.
\end{definition}
This  formulation
looks different, but  is equivalent to the construction
of the moving frame and  only uses the intrinsic geometries
of the induced metrics for $k=1,\cdots,n$.
In practice, what this means is that,  instead of
forming moving-frame vectors $\bfe_k$ out of
$\bff,\cdots, \bff^{(k)}$ ( $k=1,\cdots, n$), we consider the nested
osculating planes ${\cal O}_1\subset{\cal O}_2\subset\cdots
\subset{\cal O}_n$.  It is obvious that those two have the same
information.
The rest of this section is devoted to the
explicit form of these mappings.

For pedagogical purpose, we recall the well-known
Grassmannian aspect of  the hyperplanes in $\CP$.
The Grassmannian manifold $G_{n+1,\, k}$ is the set of
$(n+1)\times k$ matrices ${\cal F}_k$ with the equivalence
relation ${\cal F}_k\sim a{\cal F}_k$, where $a$ is a $k\times  k$ matrix.
In our case, it is natural to consider another set of
$(n+1)\times k$ matrices ${\bar {\cal F}}_k$ simultaneously,
in order to deal with
each chiral component independently.
For arbitrarily
given ${\cal F}_k$ and ${\bar {\cal F}}_k$, we can uniquely define
hyperplanes in $\CP$
by equations    of the form
$X^A(t)=\sum_j f^{j,\, A}t_j$,
$\Xb^{\Ab}(\tb)=\sum_j\fb^{j,\, \Ab}\tb_j$, where
$t_j$, and $\tb_j$ are arbitrary parameters. $f^{j,\, A}$ and
$\fb^{j,\, \Ab}$ are the matrix elements of ${\cal F}_k$ and
${\bar {\cal F}}_k$.
The equivalence
relation ${\cal F}\sim a {\cal F}$ and
${\bar {\cal F}} \sim \ab {\bar {\cal F}}$ simply expresses  the fact
that the geometrical hyperplane does not change if we replace
$t_j$ and $\tb_j$ by linear combinations. Thus the Grassmannian
$G_{n+1, \, k}$ is the space of $k$-dimensional hyperplanes in
$\CP$.  Following ref.\cite{KN}, it is natural to base
its K\"ahler structure  on the potential
${\cal K}_k\equiv \ln (\det {\cal F}_k{\bar{\cal F}}_k^T)$ which coincides with
the K\"ahler potential of $\CP$  for $k=1$.

Consider an embedding of
2D surface into   $G_{n+1,\, k}$ with chiral parametrization
 --- this time, however, we do not introduce its extrinsic geometry.
  It is  defined by its chiral components
${\cal F}_k(z)$ and ${\bar {\cal F}}_k(\zb)$ which are
respectively given by
\begin{equation}
\label{4.1.3}
\left ( \begin{array}{ccc}
f^{0,\, 0}(z) & \cdots  & f^{0,\, n}(z) \\
\vdots    &         & \vdots  \\
f^{k,\, 0}(z) & \cdots & f^{k,\, n}(z)
\end{array}
\right ), \quad
\left ( \begin{array}{ccc}
\fb^{0,\, 0}(\zb) & \cdots  & \fb^{0,\, n}(\zb) \\
\vdots    &         & \vdots  \\
\fb^{k,\, 0}(\zb) & \cdots & \fb^{k,\, n}(\zb)
\end{array}
\right ).
\end{equation}
The difference with the usual situation is
that, for us, $\fb^{i,j}$ is
not assumed to be the complex conjugate
of $f^{ij}$. It is immediate that
the metric induced on this surface is derivable from  the potential
${\cal K}^{(k)}$ which is such that
\beq
\label{4.1.4}
\begin{array}{l}
e^{{\cal K}^{(k)}}=\\
\displaystyle{\sum_{0\leq i_1<\cdots <i_k\leq n}}
\left |  \begin{array}{ccc}
f^{0,\, i_1}(z) & \cdots  & f^{0,\,i_k }(z) \\
\vdots    &         & \vdots  \\
f^{k-1,\, i_1}(z) & \cdots & f^{k-1,\, i_k}(z)
\end{array}
\right |
\left | \begin{array}{ccc}
\fb^{0,\,i_1 }(\zb) & \cdots  & \fb^{0,\,i_k }(\zb) \\
\vdots    &         & \vdots  \\
\fb^{k-1,\, i_1}(\zb) & \cdots & \fb^{k-1,\, i_k}(\zb)
\end{array}
\right |.
\end{array}
\eeq
The construction of ${\cal O}_{k-1}$
at each point $z$, $\zb$, goes as follows.
One lets
\begin{equation}
\label{4.1.5}
\left \{
\begin{array}{cc}
f^{s,l}&=
\partial^{(s)} f^l(z)\\
\fb^{s,l}&=
\partialb^{(s)} \fb^l(\zb)\\
\end{array} \right.,\>
\hbox{ for}\> s=0,\,  \cdots, \, k-1,\>
l=0, \, \cdots, \, n.
\end{equation}
For a fixed $k$, the $k$-planes
generate a surface in $G_{n+1, \, k+1}$ called the $k$th associated surface.
In technical terms: from the embedding  $\Sigma\to \CP$,
we have  canonically  constructed
the kth associated embedding $\Sigma\to
G_{n+1,\, k+1} \hookrightarrow
P(\Lambda^{k+1} \cC)$.  Next,
according to Eq.\ref{4.1.4},      the induced metric on the
$k$th associated surface
 in $G_{n+1,\, k+1}$ is simply,
\begin{equation}
\label{4.1.6}
g_{z\, \zb}^{(k)}=\partial \partialb \ln \tau_{k+1}(z,\zb),\quad
g_{z\, z}^{(k)}=g_{\zb\, \zb}^{(k)}=0,
\end{equation}
so that the Toda field $\p_{k+1}\equiv
-\ln (\tau_{k+1})$ appears naturally.
Thus  $-\p_{k+1}$ is equal to the  K\"ahler potential
of the $k$th associated surface.
At this point, it is very clear that by considering the associated
surfaces, we can restrict ourselves to  intrinsic geometries.

In the discussion of the main section 2, the Toda equation
came out from the Gauss-Codazzi equation.
Here, it is equivalent to the local Pl\"ucker formula
as we shall see.

\subsection{The instanton-numbers of a W-surface}
\label{4.2}

A key point in the coming discussion is to use topological
quantities that are instanton-numbers. As a preparation,
 we  recall the fact, pointed out in \cite{GM},
that W-surfaces are instantons of the associated
non-linear $\sigma$-model. The general situation is as follows.
W-surfaces are characterized by their chiral parametrizations
which   thus satisfy
the Cauchy-Riemann relations. These are self-duality equations
so that the coordinates of a W-surface  define fields that
are solutions of the associated non-linear $\sigma$ model,
with an  action  equal to
the topological instanton-number.  For a general K\"ahler
manifold $M$ with coordinates $\xi^\mu$ and $\xib^{\mub}$, and
metric $h_{\mu \mub}$, the action associated with 2D manifolds
of $M$ with equations $\xi^\mu=\varphi^\mu(z,\,\zb)$,
and $\xib^{\mub}=\varphib^{\mub}(z,\, \zb)$ is
\begin{equation}
\label{4.2.1}
S={1\over 2} \int d^2 x\>  h_{\mu \mub} \> \partial_j  \varphi^\mu
\partial_j \varphib ^{\mub}.
\end{equation}
In this short digression  we let $z=x_1+i x_2$,
and $\partial_j=\partial/\partial x_j$.
The instanton-number is defined by
\begin{equation}
\label{4.2.2}
Q={i\over 2\pi}\int d^2 x \> \epsilon_{jk}\>
h_{\mu \mub} \> \partial_j  \varphi^\mu
\partial_k \varphib ^{\mub}.
\end{equation}
For W-surfaces and their associated surfaces,
 $\partialb \varphi^{\mu}=\partial \varphib^{\mub}=0$, and
one has $S=\pi Q$. $Q$ is proportional to the integral
of the determinant of the induced metric.
Applying the last formula
to the $k$th associated surface, we get
\begin{definition}{\bf Higher instanton-numbers of the W-surface.}

The
$k$th instanton number of the W-surface
 $Q_{k+1}$ is defined by,
\begin{equation}
\label{4.1.7}
Q_{k+1}\equiv {i\over 2\pi} \int _{\Sigma}
dzd\zb g_{z\, \zb}^{(k)}, k=1,\cdots, n-1.
\end{equation}
\end{definition}
Its topological nature is obvious from Eq.\ref{4.1.6} which shows
that the integrand is indeed a total derivative. The collection
of the ($k$-th) instanton-numbers together with the original one
$Q\equiv Q_1$
gives a set of topological quantities
which characterize the global properties of the original W-surface.

%
%

\subsection{ Singular points of embeddings}
\label{4.3}

In the main section 2, we have constructed the moving frames
at the point where the tau-functions
are regular.
 When those
functions become irregular,  we meet an obstruction
to derive the moving frames. In
the WZNW language, this  signals that
there appears a global obstruction to the Gauss decomposition.
Toda equations  should be modified at these points.
In the following, we study the structure of
such singularities and the behavior of the tau-functions.

Let us discuss the $\CP$ case, where the structure of
such singularity was already studied
in detail in mathematics\cite{GH}.
As always, we use the notation $f^A(z)$ (resp.
$\fb^{\Ab}(\zb)$) to describe  the chiral (resp. anti-chiral)
part of the embedding.  In the following, we only discuss
the chiral components explicitly. Consider a  point
$z_0$ which is a singular point of the embedding.
We assume, as one does in
mathematics\cite{GH},  that we may reduce the problem to
the case where there are no branch point around $z_0$ (If there was,
for instance, a non-trivial
monodromy-matrix acting on  the $f^A$'s around $z_0$
to begin with,
one would assume that this matrix  is
diagonalizable and that its eigenvalues are rational. In this way,
by taking a finite covering,  one would be reduced to the case
we are discussing).
Now we remark first that
 if some  $f^A$'s  blow up  at $z_0$,
 we can remove that singularity by applying a
local rescaling
$f^A(z)\rightarrow \rho(z)f^A(z)$, with $\rho(z_0)=0$.
The idea is that one divides by the most singular behavior so that
every $f^A(z)$ has a finite limit  at $z=z_0$.   Now the
study of the singularity structure is replaced by
the study of the zeros of $f$, and of its derivatives at $z_0$.
By appropriate reshuffling,
\beq \label{4.3.2}
f^A(z)\rightarrow \tilde{f}^A(z)= \sum_BS^A_Bf^B(z),
\eeq
with a suitable constant matrix $S$, we can get the following
{\em normal form} for $f$ at $z=z_0$,
\beqa
\ft^{0}(z)& = & O(1),\quad
\ft^{1}(z) =  O\left((z-z_0)^{1+\beta_1(z_0)}\right),\nonumber\\
\ft^{2}(z)& = & O\left((z-z_0)^{2+\beta_1(z_0)+\beta_2(z_0)}\right)
,\cdots
\nonumber\\
\ft^{n}(z) &=&
  O\left((z-z_0)^{n+\beta_1(z_0)+\cdots+\beta_n(z_0)}\right)
             \label{4.3.3}
\eeqa
We define
\begin{definition}{\bf Ramification indices.}

The non-negative integers $\beta_\ell(z_0)$
($\ell = 1,\cdots,n$) which appear in Eq.\ref{4.3.3}
describe the local behavior
of the embedding function at $z=z_0$.  We call these numbers
{\bf ramification indices} following  the terminology
of  the mathematical literature\cite{GH}.
We introduce similar indices  $\betab$ to describe the
behavior of the anti-chiral embedding functions.
We define here also  the {\bf total ramification index}
$\beta_\ell$, $\betab_\ell$ as follows,
\beq
\label{4.3.8}
\beta_\ell = \sum_{z}\beta_\ell(z),\qquad
\betab_\ell = \sum_{\zb}\betab_\ell(\zb),
\eeq
where $z$ and $\zb$ run over all the singular points of $\Sigma$.
\end{definition}
The $\beta$'s are integer since we assumed that there were
no branch point.  Regular points of the embedding are
characterized by the vanishing of all ramification indices.

{}From this explicit form of the  local behavior of the embedding
functions, it is easy to
calculate the behavior of the tau-functions
at $z=z_0$, which is explicitly obtained by,
\begin{theorem}{\bf Behavior  of the tau-functions.     }
\beq\label{4.3.5}
\tau_\ell = O\left(
(z-z_0)^{\beta_{\ell-1}+2\beta_{\ell-2}+\cdots+(\ell-1)\beta_1}
    (\zb-\zb_0)^{\betab_{\ell-1}+2
      \betab_{\ell-2}+\cdots+(\ell-1)\betab_1}
      \right).
\eeq
\end{theorem}
  \proof
The explicit computations of the first few ones look as follows,
\beqa
\tau_1 & \sim & \ft_0\ftb_0+\
{\rm higher\ order\ terms}\nonumber\\
       & \sim & O(1)\nonumber\\
\tau_2 & \sim &
   \left|
   \begin{array}{cc}
    1 & (z-z_0)^{1+\beta_1}\\
    0 & (*)(z-z_0)^{\beta_1}
   \end{array}\right|
   \left|
   \begin{array}{cc}
    1 & (\zb-\zb_0)^{1+\betab_1}\\
    0 & (*)(\zb-\zb_0)^{\betab_1}
   \end{array}\right|
  + \cdots\label{4.3.4}\\
      & \sim & O((z-z_0)^{\beta_1}
(\zb-\zb_0)^{\betab_1})\nonumber\\
\tau_3 & \sim &
   \left|
   \begin{array}{ccc}
    1 & (z-z_0)^{1+\beta_1}& (z-z_0)^{2+\beta_1+\beta_2}\\
    0 & (*)(z-z_0)^{\beta_1} &
      (*)(z-z_0)^{1+\beta_1+\beta_2}\\
    0 & (*)(z-z_0)^{\beta_1-1}
      &  (*)(z-z_0)^{\beta_1+\beta_2}
   \end{array}\right|
   \nonumber\\
   & &   \times\left|
    \begin{array}{ccc}
    1 & (\zb-\zb_0)^{1+\betab_1}&
(\zb-\zb_0)^{2+\betab_1+\betab_2}\\
    0 & (*)(\zb-\zb_0)^{\betab_1} &
      (*)(\zb-\zb_0)^{1+\betab_1+\betab_2}\\
    0 & (*)(\zb-\zb_0)^{\betab_1-1}
      &  (*)
      (\zb-\zb_0)^{\betab_1+\betab_2}
   \end{array}\right|+\cdots\nonumber\\
      & \sim & O((z-z_0)^{2\beta_1+\beta_2}
            (\zb-\zb_0)^{2\betab_1+\betab_2}).\nonumber
\eeqa
Here we omit $(z_0)$ in the $\beta$'s and $(*)$'s are
some non-vanishing numerical constants.
Obviously, this type of computation
can be performed for every tau-functions.
\qed

Special combinations of $\tau$-functions
appear in the Toda equations.
They behave as
\beq\label{4.3.7}
\frac{\tau_{\ell+1}\tau_{\ell-1}}{\tau_\ell^2}
= O\left((z-z_0)^{\beta_\ell(z_0)}
(\zb-\zb_0)^{\betab_\ell(\zb_0)}\right).
\eeq

In Eqs.\ref{2.2.19},  we need to divide vectors by
tau-functions.  Unless all ramification
index vanish, we get divergence in those formulae.
In the WZW language, it shows that the
Gauss decomposed matrices $A$ and $C$ become singular at
these points.
In terms of the Toda equation, it leads to
\begin{theorem}{\bf  Modified Toda equation}

At the singular point, the Toda equations   have an  extra
$\delta$-function source-term given by
\beq\label{4.3.9}
\partial\partialb \p_\ell
+ e^{\sum_{s}K_{\ell s} \p_s}
+ \pi\sum_{z_0,\zb_0\in \Sigma}\sum_{t=1}^{\ell-1}
(\ell-t)(\beta_t(z_0)+\betab_t(\zb_0))\delta^{(2)}(z-z_0)
=0.
\end{equation}
\end{theorem}
\proof
The behavior at the singularity can be evaluated directly
through the behavior Eq.\ref{4.3.5} of the tau-functions, and
the well-known formula
\beq\label{4.3.10}
\partial\partialb\ln(z-z_0) =\partial\partialb\ln(\zb-\zb_0)
= \pi \delta^{(2)}(z-z_0).
\eeq
\qed

%
%

\subsection{Ramification Indices  and DS Operator}
\label{4.4}

In the following, we study the relation between the
ramification-indices  and the singularity of the
Drinfeld-Sokolov operator (DS operator).
This  will be important since
it is known that the phase-space of the classical W-algebra
is described by the gauge equivalence class of the
DS operator. For regular points, this indeed results from the discussion
of section \ref{2.3}. Ultimately, we
should clarify the relation between
this phase-space and the space of W-surfaces.
A related motivation  is to  give a  direct bridge with
the generalization to  the quantum situation.
In the previous section (4.3), we saw that the  local
rescaling  symmetry is essential to eliminate the singularity
so that it should not be fixed globally.  In the same way,
we have to start from the DS operator of the section 2.3
with additional $h_n$ generator.
Let us recall it from the section 2.3,
\beq\label{4.4.1}
{\cal L} = \partial - I - \lambda.
\eeq
where  $I$ is given in Eq.\ref{2.3.7} and
$\lambda$ is the lower triangle matrix, (Borel
subalgebra).
The gauge symmetry of ${\cal L}$ is generated by
the strictly lower-triangular  matrices.
It follows from section 2.3 that the chiral component
of the metric $\eta_R$ is a solution of
the Drinfeld Sokolov equation (DS equation),
\beq\label{4.4.2}
{\cal L} \eta_R = 0,
\eeq
with $\lambda$ given in Eq. \ref{2.3.8}.
For our  purpose, it will be better to switch to
the diagonal gauge where only nonvanishing elements of
$\lambda$ are the diagonal elements. We call $\Upsilon$
the gauge transformed of $\eta_R$.

Inspired by the instanton-solutions of
${\cal CP}^1$ non-linear $\sigma$-model,
we shall study the solution of the DS equation where the singular part of
the DS potential $\lambda$ has the following form,
\beqa\label{4.4.5}
\lambda & = &
\left (\begin{array}{ccccc}
\pvphi_0 & 0 & 0 &\cdots & 0 \\
0 & \pvphi_1 & 0 &\cdots & 0\\
\vdots & \, & \ddots &\,  & \vdots \\
0 & \, & \cdots &\pvphi_{n-1}  & 0\\
0 & 0  & \cdots & 0 &  \pvphi_{n}
 \end{array}
\right ),\\
\pvphi_i& =& \frac{\alpha_i-\alpha_{i+1}+\alpha_S}{z} +O(1),
\qquad \alpha_0=\alpha_{n+1}\equiv 0.
\label{4.4.6}
\eeqa
This $\alpha_S$ is the reflection of the fact that
$gl(1)$ local-scale   is not yet fixed.
We will calculate the singular behavior of the solution of
DS equation ${\cal L} \Upsilon=0$
at $z=0$ and find the relation between $\alpha$
and the ramification index.

Let us start from the simplest situation, i.e. the $n=2$ case.
Then, the singular part of the DS equation is simply,
\beq\label{4.4.7}
\partial
      \left (\begin{array}{c}
      f_0\\ f_1\end{array}\right) =
     \left(\begin{array}{cc}
      -\alpha_1/z & 1\\
      0 & \alpha_1/z\end{array}
      \right)
      \left(\begin{array}{c}
      f_0\\ f_1\end{array}\right).
\eeq
where $(f_k)_i =  z^{-\alpha_S}\Upsilon_{ki}$
In the following, we suppress the index $k$.
This equation reduces to the second order
scalor differential equation,
\beq\label{4.4.8}
(\partial^2+ u(z))f_0=0,\qquad u(z)=
\frac{1}{z^2}(\alpha_1^2+\alpha_1)+\
O\left(\frac{1}{z}\right).
\eeq
This  is a well known example of the differential
equation with regular holonomy.
The  behavior of its solution $\bff = \left ({f_0\atop  f_1}\right)$
at $z=0$ is determined by
\beqa\label{4.4.9}
f_0\sim z^{-\alpha_1}&\quad f_1
\sim z^{\alpha_1+1} &(\alpha_1\neq
      -\frac{1}{2})\\
f_0\sim z^{1/2}&\quad f_1\sim z^{1/2}
\ln z & (\alpha_1=-\frac{1}{2}).
\label{4.4.10}
\eeqa
Converting to $\Upsilon$'s, we can choose the overall scaling factor
$\alpha_S$ in order to remove the singularities.
If we use this solution as the embedding into ${\cal CP}^1$,
we can get the relation between $\alpha_1$ and ramification index
$\beta_1$.  We find that
\beq
\beta_1 = 2\alpha_1 \quad(\alpha_1\geq 0),\quad
\beta_1 = 2|\alpha_1+1|\quad (\alpha_1\leq -1)\label{4.4.11}
\eeq
Some of its features should be noticed,
\begin{enumerate}
\item For each ramification index, there are two possible values for
      $\alpha_1$.
\item When $\alpha_1 = -1$, although the DS operator has apparent
      singularity at $z=0$, corresponding solution is perfectly regular.
\item When $\alpha_1=-1/2$,  the monodromy matrix is not diagonalizable,
      and this case is not covered by the present analysis. This is
      the so-called parabolic case.
\end{enumerate}
As for the first point, we have met this situation elsewhere.
In the free-field approach to CFT, we represent  primary fields
by means of  vertex operators $\exp(\alpha_1 \varphi)$.  Its conformal
dimension  is then given by $\alpha_1(\alpha_1+1)$.
Hence for each conformal dimension, we have two vertex operators
which have the same dimension with $\alpha_1$ and $-1-\alpha_1$.
Since our $\varphi$ fields will be replaced by free boson operator,
the two situations have actually same origin.
The second point of our remark can be understood similarly.
In CFT, we have non-trivial  operators with vanishing weight apart from
 the trivial operator 1. Our DS operator
with false singularity is apparently similar.
In any case, we can show that these two situations are
actually gauge equivalent, which is clear from the fact that
they have same potential $u(z)$ in \ref{4.4.8}.
In the quantum case, rational theories will indeed lead to rational
values for  $\alpha_1$. In the classical case
these cuts  may be
removed by switching to the covering space  whenever we meet cuts.
This is why we
 only consider the situations  where the $\alpha$'s are integers.

Let us generalize our discussion to  the $gl(n+1)$ DS operator.
In this case, the singular part of the
differential equation which generalizes
Eq.\ref{4.4.7} is given by,
\beq\label{4.4.12}
(\partial- \frac{\alpha_n}{z})
(\partial+ \frac{\alpha_n-\alpha_{n-1}}{z})
\cdots (\partial+\frac{\alpha_2-\alpha_1}{z})
(\partial+ \frac{\alpha_1}{z})f_0=0.
\eeq
The behavior of the solutions of this equation is given by $
\bff_0\sim z^\alpha$ where $\alpha$ satisfies following
characteristic  equation,
\beq\label{4.4.13}
(\alpha +\alpha_1)(\alpha-1 +\alpha_2-\alpha_1)
\cdots
(\alpha-n+1 +\alpha_n-\alpha_{n-1})(\alpha-n -\alpha_n)=0.
\eeq
If any two solution of this equation are equal, we meet the
logarithmic singularities, which  do not fit in the present analysis.
Otherwise, the behavior
of the embedding function  is found to be
\beqa
f_{0}  =  O(z^{-\alpha_1}),  &\quad&
f_{1}  =  O(z^{1+\alpha_1-\alpha_2})\nonumber\\
f_{2}  =  O(z^{2+\alpha_2-\alpha_3}),\label{4.4.14}&\cdots,&
f_{n}  =  O(z^{n+\alpha_n}).
\eeqa
There are $n!$ diferent set of $\alpha$ which produce
the same ramification index.  Since they are all gauge equivalent,
we can restrict our analysis to one of them.
To specify the choice, we can postulate that $f_{i}/f_{i+1}$
is regular at $z=0$ for every $i$.
The ramification index in this situation is simply given by
\beq\label{4.4.15}
\beta_i = \sum_{j=1}^n K^{A_n}_{ij}\alpha_i,
\eeq
where $K^{A_n}_{ij}$ is the Cartan matrix of $A_n$.
In this way, we can get a clear group-theoretical correspondence between
the ramification-indices  and the singularity-index of
the DS operator.  This result should be considered as the
local version of the Pl\"ucker formula we will encounter in
the coming section.

%
%

\subsection{Pl\"ucker formulae}
\label{4.5}
The method will be to exhibit  relations  between the curvatures
 $R_{z\, \zb}^{(k)} $  and metric tensors
of the associated embeddings.  There are
two versions of these formulae.  First,
as a direct consequence of the Toda equation, we obtain the
\begin{theorem}{\bf Infinitesimal Pl\"ucker formulae.}
\label{Plucker1}
At the regular points of the embedding one has
\begin{equation}
\label{4.5.9}
R_{z\, \zb}^{(k)} \sqrt{g_{z\, \zb}^{(k)}}
=-g_{z\, \zb}^{(k+1)}+2g_{z\, \zb}^{(k)}
-g_{z\, \zb}^{(k-1)}.
\end{equation}
\end{theorem}
\proof
This is derived by
computing  the curvature
\begin{equation}
\label{4.5.8}
R_{z\, \zb}^{(k)}\sqrt{g_{z\, \zb}^{(k)}}
\equiv -\partial \partialb \ln  g_{z\, \zb}^{(k)}
=-\partial \partialb \ln \left ( {\tau_{k+2} \tau_{k}\over \tau_{k+1}^2}
\right ).
\end{equation}
The first equality is a simple consequence of the general form
Eq. \ref{4.1.4} of the K\"ahler potential.
The second is a consequence of the
specific mapping of $\Sigma$ in $G_{n+1, k+1}$ (Eq.\ref{4.1.5}) which is
such that the Toda equation is verified automatically.
Making use of Eq.\ref{4.1.6} one easily completes the proof.
\qed

These  infinitesimal Pl\"ucker formulae give us the
global following relations
\begin{theorem}{\bf Global Pl\"ucker formulae}
\label{Plucker2}
The genus $g$ of a W-surface is related to its  instanton-numbers
and ramification-indices by the relations
\begin{equation}
\label{4.5.13}
2-2g+\beta_k=2Q_k-Q_{k+1}-Q_{k-1}, \quad \left \vert
\begin{array}{cc} k&=1,\cdots, n\nonumber \\
Q_{n+1}&\equiv 0, \>
Q_{0}\equiv 0 \end{array}\right.
\end{equation}
\end{theorem}
\proof
First we apply the  Gauss-Bonnet theorem for
each of the $k$th associated surfaces by computing
$\int_{\Sigma_\epsilon} R_{z\, \zb}^{(k)}
\sqrt{g_{z\, \zb}^{(k)}}$. The  integral is first computed over
$\Sigma_\epsilon$ where  small neighborhoods of singularities
are removed.
The ramification
indices at singularity was previously defined so that at a singular
point the induced metric of the $k$th associated surface behaves as
\begin{equation}
\label{4.5.10}
g_{z\, \zb}^{(k)}\sim (z-z_0)^{\beta_k(z_0)}
 (\zb-\zb_0)^{\betab_k(\zb_0)}
\tilde g_{z\, \zb}^{(k)},
\end{equation}
where $\tilde g_{z\, \zb}^{(k)}$ is regular at $z_0$, $\zb_0$.
 Since we do not
assume that $\overline {f(z)}=\fb(\zb)$,  $\beta_k(z_0)$ and
$\betab_k(\zb_0)$ may be different. By letting $\epsilon\to 0$,
one sees, that the contribution of the singularities
to the Gauss-Bonnet formula is proportional
to the k-th ramification index
\begin{equation}
\label{4.5.11}
\beta_k\equiv {1\over 2}
\sum_{ (z_0, \zb_0)\in \Sigma}\left ( \beta_k(z_0)+
\betab_k(\zb_0)\right ).
\end{equation}
The contribution of the regular part does not depend upon k, since
changing k there, is equivalent to
using a different complex structure, while the
result is equal to the Euler characteristic
that does not depend upon it.
The Gauss-Bonnet theorem for the
$k$th associated surface  finally gives
\begin{equation}
\label{4.5.12}
{i\over 2\pi} \int _{\Sigma} dzd\zb R_{z\, \zb}^{(k)}
=2-2g+\beta_k.
\end{equation}
Combining these last relations with Eqs.\ref{4.1.7},
completes the derivation.   \qed

Using these formulae, we find that
there are n independent topological numbers
($Q_1$, $\cdots$, $Q_{n}$), which characterize
the global topology of
W-surfaces. A direct consequence of this
observation is that    W$_{n+1}$-string
  have n coupling constants which play the same r\^ole
as  the genus for  the usual
string theories. Eq. \ref{2.1.10} may be understood as the index theorem
for W-surfaces.

%
%

 \subsection{Relation with self-intersection numbers}
\label{4.6}

A few  years ago, Polyakov\cite{Pol}  introduced    a modified
Goto-Nambu action, with a topological term
involving the extrinsic geometry of the string-manifold.
In this section, we connect his discussion with the  one
carried out in the present article.

The Goto-Nambu action, is proportional to
$\int dz d\zb \sqrt{-det(\widehat g)}$.  $\widehat g$ is the
induced metric which takes the K\"ahler form.  This gives
\beq
\label{4.6.5}
S_{GN}\propto i \int dz d\zb \widehat g_{z\zb}=
i \int dz d\zb \sum_A\partial f^A \partialb \fb^A,
\eeq
which  is indeed a tological integral, since $\widehat g_{z\zb}=
\partial \partialb \sum_Af^A \fb^A$. As expected it coincides
with the instanton-number $Q$ of Eq.\ref{4.2.2} associated
with $\widehat g$. Since the target space is
topologically trivial, $Q$ actually
vanishes for any $\cC$-W-surfaces,
contrary to the situation of $\CP$-W-surface.

In ref.\cite{Pol},  an  additional topological term was
defined from the
second fundamental form. Let denote by $D\equiv \partial-
\partial(\ln \Delta_1)$ and $\bar D\equiv
\partialb-\partialb (\ln \Delta_1)$, the covariant derivatives
on the W-surface (with the notations
of the previous sections  $\widehat g_{z\zb}=\Delta_1$).
 It follows from Eq.\ref{2.1.SF}  that
\beq
\label{4.6.6}
D \left(\partial\bff\right) =
\sum_{a=2}^n\Omega^a_{zz}\bfe_a,
\qquad
\bar D \left(\partialb\bbff\right) =
\sum_{a=2}^n{\bar \Omega}^a_{\zb\zb}\bbfe_a,
\eeq
where, according to Eqs.\ref{2.1.5}, and \ref{2.1.6},
\beq
\label{4.6.7}
\partial\bff=\sqrt{\Delta_1}\bfe_1
\qquad
\partialb\bbff=\sqrt{\Delta_1}\bbfe_1.
\eeq
The self-intersection number of ref.\cite{Pol} is
\beq
\label{4.6.8}
\nu_1 =
\frac{i}{2\pi}\int dz d\zb
\Omega_{zz}^a {\bar \Omega}_{\zb \zb}^a (\widehat g^{z\zb})^2
\sqrt{-det(\widehat g)}=\frac{i}{2\pi}
\int dz d\zb {\Delta_2\over \Delta_1^2}.
\eeq
It was originally defined in the case when the target space is
${\cal C}^2$, where a W-surface
intersects with itself at several isolated points.
The index $\nu_1$ counts the number of these isolated points with
suitable signs determined by  their mutual orientations
at intersection points.
This explains the original terminology, ``self-intersection number".

{}From our viewpoint, it is clear that there is a close
analogy between
$\nu_1$ and $Q_1$ in Eq.\ref{4.1.7}.  The only difference between them
is  that we need to replace $\Delta_a$ by the $\tau_a$.
This analogy
 suggest a  re-interpretation of $\nu_1$  as   the  generalized
instanton-number of a certain associated mapping.
It enables us to obtain a generalization of that index
 for the W-surface in higher $\cC$ target spaces
where there seems to be no interpretation of ``self-intersection".

This time, for each W-surface in $\cC$,
we define  the $k$th associated mapping
as being  from  $\Sigma$ into $G_{n,k}$,
($k=1,\cdots,n-1$).  Each point $z\in\Sigma$,
is mapped into an osculating frame spanned by
$\bff^{(1)},\cdots, \bff^{(k)}$ at $z$
which defines a point in $G_{n,k}$.
Although the original $\cC$-W-surface has only vanishing
instanton number, these associated mapping give nontrivial indices,
which  obviously are analogous to those of the
 of the $\CP$ W-surfaces and of their 1 associated surfaces.
The only difference is that in $\CP$ case we constructed an osculating
frame out of $\bff^{(0)}=\bff,\cdots, \bff^{(k-1)}$.
Since $\Delta_a$ can be obtained from $\tau_a$ by replacing
$\bff^{(a-1)}$ by $\bff^{(a)}$, we obtain,
\begin{definition}{\bf Generalized `intersection numbers'.}

They are defined by the instanton number $\nu_k$ of
the $k$-th associated mapping of $\cC$-W-surface,
\beq\label{4.6.9}
\nu_k=\frac{i}{2\pi}\int dzd\zb g^{(k)}_{z\zb}
=\frac{i}{2\pi}\int dzd\zb\partial\partialb \Delta_k=
\frac{i}{2\pi}\int dzd\zb\frac{\Delta_{k+1}\Delta_{k-1}}{
\Delta_k^2}.
\eeq
\end{definition}
These $n-1$ integrals have exactly
same topological meanings as those
of higher instanton numbers in $CP^{n-1}$-W-surface.  Polyakov's
index corresponds to the {\it first} index, i.e. the instanton
number
of the associated mapping into $CP^{n-1}$.

It is easy to express  our new indices out of the third fundamental
forms.  Rewrite Eq.\ref{2.1.TF}, with $a\not= 1$ as
\beqa
\label{4.6.10}
\left[D-{1\over 2}\sum_{l=1}^{a-1} \partial g_{z\, \zb}^{(\ell)}\right ]
 \left(\sqrt{\Delta_1}\bfe_a\right)& =
L_a  \sqrt{\Delta_1}\bfe_{a+1},
\nonumber \\
\left [ \bar D-{1\over 2}\sum_{l=1}^{a-1}
\partialb g_{z\, \zb}^{(\ell)}\right ]
 \left(\sqrt{\Delta_1}\bbfe_a\right)& =
L_a  \sqrt{\Delta_1}\bbfe_{a+1}.
\eeqa
Since one has $L_a=\sqrt{g_{z\, \zb}^{(a)}}$, it follows that the
numbers $\nu_a$ are such that
\beq
\label{4.6.11}
\nu_a =\frac{i}{2\pi}  \int dz d\zb L_a^2.
\eeq

\section{Outlook}
\markright{5. Outlook}

 This article has gone quite a  way towards describing W-geometries.
Yet, many  problems remain untouched, many more
aspects deserve our attention.
Let us mention a few of them.

Concerning
the $A_n$ case itself, it will be interesting to consider the
Poisson-bracket structure and its relation with the Lie-brackets
of tangent vectors to $\CP$. This should be straightforward. Another
 point is to derive the light-cone formulation of W-gravity in the
present frame-work. It should correspond to a particular parametrization
of $\CP$.

Clearly the next problem is to derive the
other Toda dynamics and WZNW theories from W-geometries. The coset
viewpoint discussed in the appendix is a first step in that direction.

A much more difficult task is of course to consider quantum W-geometries.
It is our expectation that the quantum group structure already
exhibited\cite{G} \cite{CG2}
for W gravity will emerge. Indeed, in
the present classical  discussion, the algebra $gl(n+1)$ plays a
crucial r\^ole. In the same way as for Toda theories\cite{CG2},
 it is likely that
quantum effects will lead to its  associated
mathematical ``quantum'' deformation.

The similarity between $\CP$ W-geometry and matrix-models, shows
that the method just discussed is very general and may be
a unifying framework for all the problems related to conformal
theories and  strings. In particular it may be convenient in the  search
for the true string-ground-state.

We may forsee interesting progress in the future.

\bigskip
\noindent{\Large \bf Acknowledgements:}
\smallskip

\noindent This research  was supported
in part by the Twinning Program
of the E.C. Community Stimulation Action. One of us (J.-L. G.)
is grateful to the Niels Bohr Institute for the warm hospitality
extended to him during his visit. The other author benefited a
great deal from the highly stimulating atmosphere
of this Institute
where he was a postdoctoral fellow when a large fraction  of
this work was
carried out. This research  was initiated while
Y. M.  was a postdoctoral fellow
at the \'Ecole Normale Sup\'erieure, and he is grateful
to the members of this group for the very efficient time
he spent there.
\appendix
\section{Appendix}
\markright{Appendix}
\subsection{ Frenet-Serret equations for
$\CP$}
\label{A.1}
In section 2.2, we wrote down the Frenet-Serret
formulae for $\CP$ through the approach which is covariant
by local rescaling.
We have seen that the additional  Toda field in $\CC$
W-surface can be cancelled using this covariance.
However, in order to accomplish it, we had to include
the zero-th derivative term of the embedding functions
to define the moving frame.  The correspondence with the
second and third fundamental forms became not so direct
due to this strategy. Moreover, this method seems
to be very specific to the $A_n$ Toda theories, and
it is useful to apply the general strategy we put forward
in our letter\cite{GM}.

The standard description of $\CP$ makes  use the
Fubini-Study metric Eq.\ref{2.2.4}, denoted $G_{A\Bb}$,
together  with  the so-called
inhomogeneous coodinate system, which satisfy
\beq
X^0 = \Xb^0 = 1.
\label{A.A.1}
\eeq
Thus  the embedding functions  are supposed
obey the conditions  $f^0(z)=1$, and $\fb^0(\zb)=1$.
This is clearly
not always  possible
since,  starting from any parametrization,
one goes to the present one by dividing the coordinates by
$X^0$,
or $\Xb^0$. This scale  choice may be made only if $f^0$ and
$\fb^0$
have no zero. Let us assume that this is true,  in the
present
section, in order to proceed.
We define an analogue of the metric tensor Eq.\ref{2.1.4}
\beq\label{A.A.2}
\gt_{i\jb}=\sum_{A,\Bb} G_{A\Bb} \>\partial^{(i)} f^A(z)
\> \partialb^{(\jb)}\ft^{\Bb}(\zb).
\eeq
The apparent drawback to use the strategy of section 2.1
for the curved space is that the higher order derivatives
of the embedding functions do not transform covariantly
under the target space reparametrizations.
Hence  formulae  like Eq.\ref{A.A.2} make sense only
when we work with  a particular  coordinate system similar
to the W-parametrization of section \ref{3.2}.
The interesting point is that
every argument in section 2.1 is valid with only minor
modifications to this situation, which is, to be treated following
our earlier general scheme\cite{GM}. In the present case, the moving
frame
case is given by Eqs.\ref{2.1.5}, \ref{2.1.6} after  replacing
$g_{i\jb}$ by $\gt_{i\jb}$.  We define $\tilde{\Delta}$ and
$\phit$ as in Eqs.\ref{2.1.7} and \ref{2.1.21} by using the
same replacement.  A minor modification is needed in
definition of the derivatives $\partial \bfe$ and $\partialb
\bbfe$.  They should be consistently replaced by the
covariant derivative,
\beq\label{A.A.3}
(\nabla {\bfe}_a)^A \equiv
\sum_Bf^{(1) B}\left ({\partial {e}_a^A\over \partial X^B} +
\Gamma_{BC}^A {e}_a^C\right ),\quad
(\nablab {\bbfe}_a)^{\Ab} \equiv
\sum_{\Bb}\fb^{(1) {\Bb}}\left (
{\partial {\eb}_a^{\Ab}\over \partial \Xb^{\Bb}} +
\Gamma_{\Bb\Cb}^{\Ab} {\eb}_a^{\Cb}\right ),
\eeq
throughout the discussion.  This modification is needed in order
to
keep the condition Eq.\ref{2.1.15} where the metricity is used.
The Christoffel symbols take simple forms for the Fubini-Study
metric,
\beq\label{A.A.4}
\Gamma_{BC}^A= -\frac{\delta_{AC}\Xb^B+\delta_{AB}\Xb^C}{
\sum_{D=0}^{n}X^D\Xb^D}
,\qquad \Gamma_{\Bb\Cb}^{\Ab}=
-\frac{\delta_{\Ab\Cb}X^{\Bb}+\delta_{\Ab\Bb}X^{\Cb}}{
\sum_{D=0}^{n}X^D\Xb^D}.
\eeq
Using  of covariant derivatives in Eqs.\ref{A.A.3}   ensures
the validity of the discussion that leads to Eqs.\ref{2.1.16}.
The Gauss-Codazzi equations  become, in agreement with ref.\cite{GM}
\beqa\label{A.A.5}
\bigl [ \nabla, \partialb \bigr ] \bfe_a&=&
\sum_b\Ft_{z\zb a }^b \bfe_b,\\
\left (\bigl [ \nabla, \partialb \bigr ] \bfe_a\right )^A
&={\displaystyle \sum} & f^{(1)B} \fb^{(1)\Bb}
\left (\bigl [\nabla_{\Bb},
\partialb_{\Bb} \bigr ] \bfe_a\right )^A
=\sum f^{(1)B} \fb^{(1)\Bb} {\cal R}_{B \Bb\, a}^{\> \> \, b}
e_b^A.\label{A.A.6}
\eeqa
with
\beq\label{A.A.7}
\Ft_{z\zb }=
\sum_{i=1}^{n} h_i \partial \partialb   \phit_i+
\sum_{i=1}^{n-1} h_i
\exp \left ( \sum_{j=1}^{n} K^{gl(n+1)}_{i j} \phit_j \right ).
\eeq
As we see here, due to the contribution of the
 target-space curvature, we do not directly  get
the exact form of the Toda equations.
However,  $\CP$
is known to possess  a  constant
sectional-curvature, that is to say,
the curvature-tensor satisfies\footnote{
The notation {$\bf{b,d\ (\bar{a},\bar{c})}$} represents
tangent
vectors which only have  chiral (anti-chiral)
non-vanishing components.}
\beq\label{A.A.8}
({\bf \bar a}, {\cal R}({\bf b, \bar c}){\bf d})
= ({\bf \bar a, d})({\bf \bar c, b})
- ({\bf \bar a, b})({\bf \bar c, d}).
\eeq
Writing,
\beqa
\Ft_i & = & \partial\partialb\phit_i +
\exp\left(\sum_{j=1}^nK^{gl(n+1)}_{ij}
\phit_j\right),\quad (i=1,\cdots,n-1)\nonumber\\
\Ft_n & = & \partial\partialb\phit_n.
\label{A.A.9}
\eeqa
Eq.\ref{A.A.6}
is more explicitly rewritten as
\beqa\label{A.A.10}
\Ft_{\ell} - \Ft_{\ell-1} & = & (\bbfe_\ell, {\cal R}
(\partial \bff, \partialb\bbff)\bfe_\ell) = \Deltat_1\qquad
\ell = 2,\cdots, n\nonumber\\
\Ft_{1} & = & \Deltat_1.
\eeqa
Solving these equations, we get
\beq\label{A.A.11}
\Ft_{\ell} = \ell \Deltat_1.
\eeq

The relation between the present  discussion and the one of  section
2.2 is clarified by establishing the following
\begin{proposition}{}
\label{taudelta}
The relation between the tau-functions and
$\Deltat$-functions is
given by
\beq\label{A.A.12}
\Deltat_\ell = \tau_{\ell+1}/\tau_1^{\ell+1}.
\eeq
\end{proposition}
\proof
First, we observe that the inner product in Eq.\ref{A.A.2}
is given by the free-fermion inner product (here $\cal G$
stands for ${\cal G}(z,\zb)$)
\beqa
  f^{(i)\, A}  \fb^{(\jb)\, \Ab} G_{A\Ab}
       &=&
       (\tau_1\partial^{(i)}\partialb^{(\jb)}\tau_1
        -\partial^{(i)}\tau_1\partialb^{(\jb)}\tau_1)\bigl /
            \tau_1^2
      \nonumber\\
      & = &
 {(\bra{\emptyset}\psi_0\EM\psis_0\ket{\emptyset}
       \bra{\emptyset}\psi_{\jb}\EM\psis_i\ket{\emptyset}
      - \bra{\emptyset}\psi_{\jb}\EM\psis_0\ket{\emptyset}
       \bra{\emptyset}\psi_0\EM\psis_i\ket{\emptyset})\over
          \tau_1^2}\nonumber\\
       &=&
       \frac{\bra{1}\psi_{\jb}\EM\psis_i\ket{1}}{\tau_1^2}.
 \label{A.A.14}
\eeqa
In the last line, we applied  Wick's theorem.
Further use of this theorem gives the proof of the
proposition,
\beq
\Delta_a \equiv \tau_1^{-2a}{\rm det}_{i,j=1\cdots a}
               \bra{1}\psi_{\jb}\EM\psis_i\ket{1}
      = \tau_1^{-a-1}
      \bra{a+1}\EM\ket{a+1}
       =  \tau_{a+1}\tau_1^{-a-1}.\label{A.A.15}
\eeq
\qed

This leads us to replace
the Toda-like field $\phit_{\ell}$ by  the true Toda field
$\p_\ell=-\ln \tau_\ell$. It is  such that
\beq\label{A.A.13}
\phit_\ell = \p_{\ell+1} - (\ell +1)\p_1.
\eeq
In terms of this Toda field,  Eqs.\ref{A.A.10} and \ref{A.A.11}
become,
\beqa
\partial\partialb\p_{n+1}& = &
(n+1)(\partial\partialb \p_1 +\exp(2\p_2-\p_1))\nonumber\\
\partial\partialb \p_\ell +\exp(2\p_\ell-\p_{\ell+1}
-\p_{\ell-1})& = & \ell(\partial\partialb \p_1 +\exp(2\p_2-\p_1)),
\label{A.A.16}\\
 & \  & (\ell = 2\cdots ,n).\nonumber
\eeqa
This result becomes closer to the well-known $A_n$ Toda equation.
However, the coincidence is still not exact.
Clearly the above equations are consequences of  Toda equations,
 that is
 $\partial\partialb \p_1 +\exp(2\p_2-\p_1)=0$,
$\partial\partialb \p_\ell +\exp(2\p_\ell-\p_{\ell+1}
-\p_{\ell-1})=0$, $2\leq \ell\leq n$,
$\partial\partialb\p_{n+1}=0$. However
we missed the first and the last. On the other hand, proposition
\ref{taudelta} can be used to complete the  derivation.  Indeed,
it was shown in ref.\cite{BG1} that, apart from the last one,
 the above Toda equations  are automatically true
if the $\ell$th Toda field is  equal to $\tau_\ell$.
Combining this last observation with Eqs.\ref{A.A.16}, one
gets the desired result. Finally, we make use of
proposition \ref{taudelta} to derive the
\begin{proposition}{\bf Liouville solutions.}
\label{Liouville}

\noindent The induced metric of
W-surfaces embedded in ${\cal CP}^1$ satisfies  Liouville's
equation.
\end{proposition}
\proof The induced metric is  $\gt_{11}$ (see Eq.\ref{A.A.2}).
According to proposition \ref{taudelta}, and Eq.\ref{2.2.24},
one may write
\beq
\label{A.A.17}
\gt_{11}={\tau_2\over \tau_1^2}={U_0(z)\Ub_0(\zb)\over \tau_1^2}.
\eeq
With the present choice of coordinates (Eq.\ref{A.A.1}),
$f^0(z)=\fb^0(\zb)=1$, and the Wronskians
$U_0(z)$, and $\Ub_0(\zb)$ are respectively  equal to
$f^{(1)1}(z)$, and $\fb^{(1)1}(\zb)$. One gets
\beq
\label{A.A.18}
\gt_{11}=f^{(1)1}(z)\fb^{(1)1}(\zb)\Bigl /
\left (1+f^{1}(z)\fb^{1}(\zb)\right )^2,
\eeq
which takes the form of Liouville's general solution. \qed

\noindent This solves the mystery pointed out in section \ref{2.1}.

Although the approach of  this appendix has  drawbacks,
it  certainly has the  merit of  being  applicable
to arbitrary  K\"ahler target-manifolds. Moreover,
 in this approach the relationship  with
extrinsic geometry is
manifest.

\subsection{ Bosonization rule for non-relativistic
fermion}
\label{A.2}

In this article, we have used  non-relativistic fermions to describe
 W-surfaces.  Bosoni\-zation of these fermions\footnote{ The
bosonization of nonrelativistic fermions in  the context of the $c=1$
matrix model was discussed in \cite{DJ}, using the method
developed by Jevicki-Sakita\cite{JS}. The relation between
our bosonization rule and theirs is  still not  clear.}
  gives the explicit
form of the Frenet-Serret formula in terms of the tau-functions
and their derivatives with respect to the KP coordinates.
Special care is needed because we discuss fermion states
with finite particle-numbers.  The modifications of the
ordinary bosonization
rules \cite{DJKM} for this case  were investigated
in refs.\cite{IM}, and \cite{M}.
In the following, we give a more direct proof.

Let us first describe the main statement of our bosonization rule.
Writing,
\beq\label{A.B.1}
\psi(\zeta) = \sum_{s=0}^\infty \psi_s\zeta^s,\quad
\psis(\zeta) = \sum_{s=0}^\infty \psi_s\zeta^{-s-1},
\eeq
\begin{proposition}{\bf Bosonization Rule.}
\label{bosonization}
\beqa
\psi(\zeta)\ket{n+1}& = & \zeta^n\exp\left(
      -\sum_{\ell=1}^\infty \frac{\zeta^{-\ell}}{\ell}J_\ell\right)
      \ket{n}\nonumber\\
\psis(\zeta)\ket{n-1}& = & \zeta^{-n}\exp\left(
      \sum_{\ell=1}^\infty \frac{\zeta^{-\ell}}{\ell}J_\ell\right)
      \ket{n}.
\label{A.B.2}
\eeqa
\end{proposition}
We remark that although our fermions are `truncated' as in \Eq{A.B.1},
the bosonization formula has basically the same form as
for relativistic Dirac fermions\cite{DJKM}.

\noindent \proof
We first derive several lemmas.
\begin{lemma}{}
\beq\label{A.B.3}
\bra{\emptyset}\psi(\zeta_n)\cdots\psi(\zeta_1)\ket{n}
= \prod_{1\leq i<j\leq n} (\zeta_i-\zeta_j).
\eeq
\end{lemma}
\proof
In order to prove this identity, we have to observe that
the LHS of \Eq{A.B.3} is  a polynomial
of order $n-1$ for each $\zeta_i$,
which is  antisymmetric with respect to replacements
$\zeta_i\leftrightarrow \zeta_j$.
These two conditions already prove that the LHS
has to be proportional to
the Vandermonde determinant which appears on the RHS.  The coefficient
can be determined through a direct comparison of a particular term,
say, $\zeta_1^{n-1}\cdots\zeta_2$, of the both side.
    \qed

Next, we calculate the effect of  insertions of fermion
and current operators. They may be
 summarized  by  the following,
\begin{lemma}{}
\beqa
\frac{\bra{\emptyset}\psi(\zeta_n)
\cdots\psi(\zeta_1)\psi(\zeta)\ket{n+1}}
{\bra{\emptyset}\psi(\zeta_n)\cdots\psi(\zeta_1)\ket{n}}
& = & \prod_{i=1}^n (\zeta - \zeta_i),\label{A.B.4}\\
\frac{\bra{\emptyset}\psi(\zeta_n)
\cdots\psi(\zeta_1)\psis(\zeta)\ket{n-1}}
{\bra{\emptyset}\psi(\zeta_n)\cdots\psi(\zeta_1)\ket{n}}
& = & \left(\prod_{i=1}^n (\zeta - \zeta_i)\right)^{-1},\label{A.B.5}\\
\frac{\bra{\emptyset}\psi(\zeta_n)\cdots\psi(\zeta_1)J_\ell\ket{n}}
{\bra{\emptyset}\psi(\zeta_n)\cdots\psi(\zeta_1)\ket{n}}
& = & \sum_{i=1}^n \zeta^\ell_{i}.\label{A.B.6}
\eeqa
\end{lemma}

\noindent \proof
The first line, \Eq{A.B.4}, is a simple
consequence of the specific form of
the Vandermonde determinant, \Eq{A.B.3}.

\noindent The second line is proved
by using the Wick's theorem as follows,
$$
\begin{array}{l}
\bra{\emptyset}\psi(\zeta_n)\cdots\psi(\zeta_1)\psis(\zeta)\ket{n-1}\\
 =
\displaystyle{\sum_{i=1}^n\frac{(-1)^{i-1}}{\zeta - \zeta_i}
\bra{\emptyset}\psi(\zeta_n)\cdots
\psi(\zeta_{i+1})\psi(\zeta_{i-1})
\cdots\psi(\zeta_1)\ket{n-1}}\\
\displaystyle{
= \left(\sum_{i=1}^n\frac{1}{\zeta - \zeta_i}\prod_{j\neq i}
\frac{1}{(\zeta_i - \zeta_j)}\right)
\bra{\emptyset}\psi(\zeta_n)\cdots\psi(\zeta_1)\ket{n}}\\
\displaystyle{  =
\left(\prod_{i=1}^n\frac{1}{\zeta - \zeta_i}\right)
\bra{\emptyset}\psi(\zeta_n)\cdots\psi(\zeta_1)\ket{n}.}
\end{array}\nonumber
$$
Here, we use a anticommutation relation,
 $\left[\psis(\zeta),\psi(\zeta_i)\right]_+
= 1/(\zeta-\zeta_i)$, (assuming $|\zeta|>|\zeta_i|$)  and an identity,
$$\sum_{i=0}^n\left( \prod_{j=0, j\neq i}^n
\frac{1}{\zeta_i-\zeta_j}\right)
=0,$$ which can be proved by induction with respect to $n$.

\Eq{A.B.6} is a simple consequence of the commutation
relation $[J_\ell, \psi(\zeta)] = -\zeta^\ell \psi(\zeta).$
\qed

The third lemma in the following will give a direct relation
between fermion and current insertions,
\begin{lemma}{}
\beq\label{A.B.7}
\prod_{i=1}^n(\zeta-\zeta_i)=\zeta^n\exp\left(
-\sum_{\ell = 1}^\infty \frac{\zeta^{-\ell}}{\ell}
\left(\sum_{j=1}^n \zeta_j^\ell \right)\right).
\eeq
\end{lemma}
\proof
This fact is a straightfoward consequence of equality,
$$\zeta-\zeta_i = \zeta\exp\ln(1-\zeta/\zeta_i).$$
\qed
Combining the last two lemmas, we get the relations,
\beqa
\bra{\emptyset}\psi(\zeta_n)\cdots\psi(\zeta_1)\psi(\zeta)\ket{n+1}& =
& \zeta^n\bra{\emptyset}\psi(\zeta_n)\cdots\psi(\zeta_1)\exp\left(
      -\sum_{\ell=1}^\infty \frac{\zeta^{-\ell}}{\ell}J_\ell\right)
      \ket{n}\nonumber\\
\bra{\emptyset}\psi(\zeta_n)\cdots\psi(\zeta_1)\psis(\zeta)\ket{n-1}
& = & \zeta^{-n}
\bra{\emptyset}\psi(\zeta_n)\cdots\psi(\zeta_1)\exp\left(
      \sum_{\ell=1}^\infty \frac{\zeta^{-\ell}}{\ell}J_\ell\right)
      \ket{n}.
\nonumber
\eeqa
This  readily completes  the proof  of the
main theorem (\ref{bosonization}) of the present section.
Indeed,   the states $\bra{\emptyset}\psi(\zeta_n)\cdots\psi(\zeta_1)$,
span a complete set of $n$-fermion states. Thus any
$n$-particle state which is
orthogonal to all of them must vanish,
so that  Eqs.\ref{A.B.2} follow.
This completes our proof of the bosonization rule.
\qed

Finally,  we briefly describe how to
get
Eqs.\ref{3.3.17}  from \Eq{A.B.2}.  When we discuss the tau-functions,
we can make the replacements,
$$ J_\ell \longleftrightarrow \frac{\partial}{\partial z^{(\ell)}}$$
because we use the Hamiltonian $\exp(\sum_{i=0}^\infty J_iz^{(i)})$.
When we expand the exponential on the RHS of
\Eq{A.B.2} with respect to
the variable $\zeta$, we get the Schur polynomials
of ${\partial}/{\partial z^{(\ell)}}$ described in \Eq{3.3.18}.
In order to get the corresponding formulae for the anti-chiral sector,
one simply interchanges $\psi\leftrightarrow\psis$, and
kets $\leftrightarrow$ bras,   everywhere.

\subsection{W-geometry in the Matrix Model}
\label{A.3}

Last year, W-algebraic structures  have been exhibited\cite{FKN} in
the discrete formulation of the two-dimensional gravity,
i.e. in the matrix model.
The W-constraints, which are equivalent to the Schwinger-Dyson
equations, give an alternative
formulation of the Painlev\'e-Douglas
equations.  It was also discovered that the $A_\infty$ Toda
field theory governs the coupling-constant dependence of
the system.  Therefore, it will be interesting to clear out
the relation between the present work and the matrix model.
Actually, one can find a one-to-one correspondence between their
basic concepts.  Furthermore the free-fermion theory
which describes both models is the same non-relativistic one.
In the following we would like
to describe the ``dictionary" between them.
We restrict our attention to the two-matrix model for simplicity.
The generalization to other model does
not seem to imply any essential difficulties.

After one integrates over the angular part,
the partition function of the two-matrix model is reduced to
the integration over the eigenvalues\footnote{
For lack of letters, we gave up the consistency of  notations
between this particular appendix  and the rest of the article.
We hope that there will be no confusion.},
\beq\label{3.5.1}
Z([g],[\gb])=\int \prod_{\alpha=1}^N d\lambdab_\alpha
\Delta(\lambdab)e^{\sum_\alpha\bar{U}(\lambdab_\alpha)}
\int \prod_{\beta=1}^N d\lambda_\beta
\Delta(\lambda)e^{\sum_\beta U(\lambda_\beta)
+\lambda_\beta\lambdab_\beta}
\eeq
where $\Delta$ is the Vandermonde determinant and,
$$U(\lambda) = \sum_{\ell =1}^N g_{\ell}
               \lambda^\ell,\quad
  \bar{U}(\lambdab) = \sum_{\ell =1}^N \gb_{\ell}
               \lambdab^\ell.$$
This integral is  identical to the determinant,
\beqa\label{3.5.2}
Z([g],[\gb]) &=& \det \Gamma_N,\\
\label{3.5.3}
\Gamma_N & = & \left(
 \begin{array}{ccc}
  \Gamma_{0\0b} & \cdots & \Gamma_{N-1\0b}\\
  \vdots   &    ~   & \vdots                           \\
  \Gamma_{0 \overline{N-1}} & \cdots & \Gamma_{N-1{\overline{N -1}}}
 \end{array}\right),
\eeqa
where
\beqa\label{3.5.4}
\Gamma_{i\jb}& \equiv& (\lambdab^j,\lambda^i)\\
(\fb(\lambdab), f(\lambda)) & = & \int d\lambdab
\fb(\lambdab) e^{\bar{U}(\lambdab,\gb)}
\int d\lambda
f(\lambda)e^{U(\lambdab,g)+\lambda\lambdab}.\label{3.5.5}
\eeqa
The standard technique to solve this problem is to use the
orthogonal polynomials.  By carefully choosing their
coefficients, one can find polynomials of order $\ell$,
\beqa
P_\ell &=& \lambda^\ell
+{\cal C}_{\ell,\ell-1}\lambda^{\ell-s} +\cdots
= \sum_{s=0}^\ell {\cal C}_{\ell s} \lambda^s\nonumber\\
\bar{P}_\ell &=& \lambdab^\ell +
{\cal A}_{\ell-1\ell}\lambdab^{\ell-s} +\cdots
= \sum_{s=0}^\ell \lambdab^s {\cal A}_{s\ell}\label{3.5.6}
\eeqa
s.t. the inner products between them are diagonal,
\beq \label{3.5.7}
(\bar{P}_r, P_s ) = h_r\delta_{rs} \equiv {\cal B}^{-1}_{rs}.
\eeq
The partition function is then given by the product over $h$,
\beq \label{3.5.8}
Z([\gb][g]) = \prod_{r=0}^{N-1} h_r.
\eeq

We are using the specific notation in the above discussion
in order to make it clear  that the procedure above is
nothing but the  Gauss decomposition of the matrix $\Gamma$,
\beq\label{3.5.9}
\Gamma = {\cal C}^{-1} {\cal B}^{-1} {\cal A}^{-1}.
\eeq
This fact already proves that the two-matrix model is described by the
$A_N$-Toda theory.  These observations
were most explicitly claimed
in ref.\cite{Martinec}.   Comparing Eq.\ref{3.5.9} and
Eqs.\ref{2.3.13}--\ref{2.3.14}, it is obvious that the mathematical
structure that underlies  the matrix model and the moving
frame are  identical.  The correspondence is
$$
\begin{array}{ccc}
\mbox{ Moving~~ frame} \bfe,~~ \bbfe
&\LRA& \mbox{Orthogonal~~ polynomial} ~~P, ~\bar{P}.\\
\mbox{W frame}\>\> \bff^{(a)},~\bbff^{(b)}& \LRA&
 \mbox{Monomials}~~\lambda^a,~ \lambdab^b.\\
\eta_{i\jb}\>  (\mbox{Eq.\ref{2.1.4})} &\LRA& ~~(\Gamma_N)_{i\jb}.
\end{array}
$$

In order to get more informations on the eigenvalues $h$,
it is important to know the recursion
relation between the orthogonal polynomial,
\beq
\lambda P_\ell = \sum_{s=0}^{\ell+1}{\cal F}_{\ell s} P_s,
 \quad
\lambdab {\bar P}_\ell = \sum_{s=0}^{\ell+1}
\bar{P}_s{\cal G}_{s\ell}.
\label{3.5.10}
\eeq
These equations appeared in the moving-frame Eq.\ref{3.3.12}.
Hence the matrices ${\cal F}$, ${\cal G}$
in this equation should be identified
with the $F$, $G$ matrices of Eqs.\ref{3.3.2}--\ref{3.3.5}.
The orthogonal polynomials satisfy similar formula if we replace
$z^p$, $\zb^q$ by $g_p, \gb_q$.  This gives the lax pair equation
in two matrix model.

In this way, we see that the mathematical background of  matrix
models  and $CP^n$ W-surface are the same,  if we replace the
size of matrix from $n+1$ by  $N$.  In  matrix models,
we are interested in the limit $N\rightarrow \infty$.
It may be interesting to know if the global geometry of the
W-surface we observed in this
paper has some consequences in that  limit.
It might provide us new topological
insights into the structure of two dimensional
gravity.

%
%

\subsection{ Group-orbits of the fundamental weights}
\label{A.4}

In this paper, we have seen how $A_n$--Toda
field theory,
and the conformally reduced $A_n$--WZNW-dynamics
are equivalent to  embeddings of W-surfaces in $\CP$. Thus group
theory came out from the geometry. In the present appendix, we reverse
the viewpoint and start from K\"ahlerian manifolds that are coset
spaces, so that their group meaning is there from the
start. We shall see, how this group emerges again
when one turns the crank described above. This will clarify its
origin, and show a way of looking   for  embedding spaces,
whose W-surfaces are associated with the other classical Lie algebras.
In addition, the connection with Grassmannians, which has been so
important for section 4,  will emerge in a natural way.
\subsubsection{The general situation}
There are many coset spaces that are K\"ahlerian manifolds. At this
moment, the most interesting cases for us are the ones corresponding
to the fundamental representations. At first we recall some classical
results. The notation is standard. Consider a semi-simple lie
algebra $\gt$, in the Chevalley basis. $h_i$, are the Cartan
generators ($i$ runs form $1$ to $n=$ rank of $\gt$). Let
$E_{\pm j}$ be the step-operators associated with a set of primitive
roots. If the roots are noted $\vec \alpha_i$, the  fundamental
weights $\vec\lambda_i$  are such that
$\vec \lambda_i. \vec \alpha_j2/ \alpha_j^2 =\delta_{i,\,j }$.
We shall be interested in the cosets $G/H_j$ obtained by the action
of the maximally non-compact  Lie group associated with $\gt$, on
the
highest-weight state $|\lambda_j>$. $H_j$ is the stability group
of $|\lambda_j>$. It contains the  subgroup ${\cal N}_+$
generated by $E_{\alpha}$ with $\alpha \in \varphi_+$ ( where
$\varphi_+$ is the set of all positive roots) but is bigger. Indeed,
$E_{-\alpha} |\lambda_j>$ vanishes for $\alpha \in \varphi_+$ if
$\vec \lambda_j-\alpha$ is not a weight of the representation
associated with $\vec \lambda_j$. In general, $H_j$ is generated by the
Lie algebra
\begin{equation}
\label{4.7.2}
\htd_j \equiv \left \{ E_{\alpha}, \alpha \in \varphi_+;
E_{-\beta},  \beta \in \varphi_+, \lambda_j-\beta \>
\hbox{not a weight}; h_k, k\not= j\right \}.
\end{equation}
According to the common wisdom, one may parametrise $G/H_j$, at
least locally,  as
follows. Denote by  $\ct_j=\gt-\htd_j$ the set of generators
of $\gt$ that are not in $\htd_j$. This set is not closed as a Lie
 algebra,  but any element of
$G/H_j$ may be written as
\begin{equation}
\label{4.7.3}
\exp \left ( \sum_A X^A c_A\right ) |\lambda_j>, \quad
c_A \in \ct_j, \quad A=1,\,\cdots, \hbox {dim. of}\>
G/H_j.
\end{equation}
We shall be concerned with the  K\"ahlerian manifolds of coordinates
$X^A$ and $\Xb^{\Ab}$, with K\"ahler potential
\begin{equation}
\label{4.7.4}
{\cal K}_j=\ln <\lambda_j|
\exp \left ( -\sum_{\Ab} \Xb^{\Ab} c_{\Ab}^+\right )
\exp \left ( \sum_A X^A c_A\right ) |\lambda_j>.
\end{equation}
This geometrical situation is directly related with the general
solution of the $\gt$-Toda equations ($K^{(\gt)}_{ij}$ is the Cartan
matrix of $\gt$)
\begin{equation}
\label{4.7.5}
\partial \partialb \p_i=
-\exp\left (\sum_j K^{(\gt)}_{ij} \>\p_j\right )
\end{equation}
derived by \cite{LS}.
Indeed these authors have shown that the general solution of these
equations is of the form
\begin{equation}
\label{4.7.6}
e^{-\p_i}= <\lambda_i | \Mb^{-1}(\zb) M(z) |\lambda_i >
e^{-\xib_i^c(\zb) -\xi_i^c(z)}, \quad \hbox{where}
\end{equation}
\begin{eqnarray}
\label{4.7.7}
\partial M=M \sum_{j=1}^n s_j(z)E_{-j}&&, \quad
\partialb \Mb=\Mb \sum_{j=1}^n \sb_j(z)E_{j}, \nonumber \\
\xi_j^c(z)=\sum_j K^{(\gt)\,-1}_{ij} \ln s_j(z)&&,\quad
\xib_j^c(\zb)=\sum_j K^{(\gt)\,-1}_{ij} \ln \sb_j(\zb).
\end{eqnarray}
$s_j(z)$ and $\sb_j(\zb)$ are arbitrary chiral functions.
Comparing these last relations with Eq.\ref{4.7.4},
one sees that, for the W-surface
with coordinates $X^A(z)$, $\Xb^{\Ab}(\zb)$, such
that
\begin{eqnarray}
\label{4.7.8}
 M(z) |\lambda_i > e^{-\xi_j^c(z)}
&=&  \exp \left ( \sum_A X^A(z) c_A\right ) |\lambda_j>
\nonumber \\
\Mb(\zb) |\lambda_i >  e^{-\xib_j^c(\zb)}
&=&\exp \left ( \sum_{\Ab} \Xb^{\Ab}(\zb) c_{\Ab}\right ) |\lambda_j>,
\end{eqnarray}
$\p_i$ is the K\"ahler potential of the metric induced  in
the W-surface.
\subsubsection{The case of $A_n$}
The notations are the same as in section 2.3
\begin{equation}
\label{4.7.9}
h_i=\flat_i^+ \flat_i-\flat_{i+1}^+ \flat_{i+1},\quad
E_i=\flat_{i+1}^+ \flat_i,\quad E_{-i}=E_i^+,
\end{equation}
but here $i$ runs from $1$ to $n+1$
in agreement with the standardnotations. The
vector-space span by $\vec e_i$, $i=1$, $\cdots$, $n+1$,
$\vec e_i.\vec e_j=\delta_{i,j}$, and
\begin{equation}
\label{4.7.10}
\vec \alpha_i=\vec e_i-\vec e_{i+1}, \quad
\vec \lambda_i=\sum_{j=1}^i \vec e_j-
{1\over n+1} \sum_{j=1}^{n+1} \vec e_j.
\end{equation}
The highest-weight states are
\begin{equation}
\label{4.7.11}
|\lambda_j>= \flat_j^+ \flat_{j-1}^+ \cdots \flat_1^+ | 0 >
\end{equation}
\paragraph{The coset space associated with $\lambda_1$}
The above general scheme gives
\begin{eqnarray}
\label{4.7.12}
\htd_1&=&\left \{ {\cal N}_+\oplus \left\{ E_{-i}, i\not= 1\right
\}
\oplus \left\{ h_i, i\not= 1\right \} \right \}\nonumber \\
\ct_1&=&\left \{ \left \{ \flat_k^+\flat_1, k=2, \, \cdots, n+1\right \}
\oplus h_1 \right \}.
\end{eqnarray}
 $G/H_1$, which is  equal to $Sl(n+1)\bigl / Sl(n)\otimes Gl(1) $,
is parametrized by
\begin{equation}
\label{4.7.13}
e^{\Omega_1} \flat_1^+ | 0 >, \quad \hbox{with}\>
\Omega_1=\xi_1 h_1+\sum_{k=2}^{n+1} x^k \flat_k^+\flat_1.
\end{equation}
Straightforward computations give
\begin{eqnarray}
\label{4.7.14}
&&e^{\Omega_1} \flat_1^+ e^{-\Omega_1} =
e^{\xi_1} \left ( \flat_1^+ + \sum_{A=2}^{n+1} X^A \flat_A^+
\right )\nonumber \\
&&X^2=x^2\left ( {1-e^{-2\xi_1}\over \xi_1}\right ),\quad
X^A= x^A \left ({e^{\xi_1}-1\over \xi_1} \right ), \quad \hbox{for} \>
A \geq 3,
\end{eqnarray}
\begin{equation}
\label{4.7.15}
{\cal K}= \xi_1+\xib_1 +\ln \left ( 1+\sum_2^{n+1} X^A \Xb ^A
\right ).
\end{equation}
The $CP^n$ Fubini-Study metric comes out as expected. Solving
Eq.\ref{4.7.7} gives the coordinate of the W-surface
\begin{equation}
\label{4.7.16}
X^A=f^A(z)\equiv \int^z dx_1 s_1(x_1) \int^{x_1}dx_2s_2(x_2) \cdots
\int^{x_{A-2}}dx_{A-1}s_{A-1}(x_{A-1})
\end{equation}
\paragraph{Arbitrary $\lambda_k$}
In that case,
\begin{eqnarray}
\label{4.7.17}
\htd_k&=&\left \{ {\cal N}_+\oplus \left\{ \flat_p^+\flat_q,
q\leq p \leq k, \hbox{or} q >k \right \}
\oplus \left\{ h_i, i\not= k\right \} \right \}\nonumber \\
\ct_k&=&\left \{ \left \{ \flat_p^+\flat_q, q\leq k, p>k\right \}
\oplus h_k \right \}.
\end{eqnarray}
Writing an arbitrary point of $G/H_k$ as $\exp (\Omega_k)
|\lambda_k >$ ,  one may compute $f^{p,j}$, for $p\leq k$,
 by letting
\begin{equation}
\label{4.7.18}
e^{\Omega_k} \flat_p^+ e^{\Omega_k}= \sum_j f^{p,j} \flat_j^+,
\end{equation}
This gives
\begin{equation}
\label{4.7.19}
\exp (\Omega_k)
|\lambda_k >=(\sum f^{k,j_k} \flat_{j_k}^+)
\cdots   (\sum f^{1,j_1} \flat_{j_1}^+)| 0 >
\end{equation}
This is a redundant parametrisation, since this expression is
unchanged of one replaces $f^{p,j}$, $p=1,$ $\cdots$ $k$
by $\sum_1^k a^p_r f^{r,j}$ with $a$ an arbitrary $k\times k$
matrix. Thus $G/H_k$ coincides with the Grassmannian $G_{n+1, k}$
as is well known. The coordinates of the W-surfaces may be taken
to be $f^{p,j}=\partial^{(p-1)} f^{1,j}$ which coincides with the
associated embedding. This last form gives back the form of the
solution written in  \cite{BG1}.

\markright{References}

\end{document}